\documentclass[prb,showpacs,amsmath,amssymb,superscriptaddress,twocolumn,floatfix]{revtex4-1}

\usepackage{graphicx}

\def\unit#1{\ \mathrm{#1}}
\def\Rea{\mbox{Re }}
\def\Ima{\mbox{Im }}
\def\ve{\varepsilon}
\def\veeff{\ve_{\mathrm{eff}}}
\def\atan{\mbox{atan }}
\def\asin{\mbox{asin }}
\def\Tr{\mbox{Tr }}

\begin{document}

\title{Systematic study of magnetic linear dichroism and birefringence
  in (Ga,Mn)As}

\author{N. Tesa\v rov\'a}
\affiliation{Faculty of Mathematics and Physics, 
  Charles University in Prague, Praha, CZ-121 16, Czech Republic}

\author{T. Ostatnick\'y}
\affiliation{Faculty of Mathematics and Physics, 
  Charles University in Prague, Praha, CZ-121 16, Czech Republic}

\author{V. Nov\'ak}
\affiliation{
Institute of Physics, ASCR, $v.~v.~i.$, 
Cukrovarnick\'a 10, CZ-16253 Praha 6, Czech Republic}

\author{K. Olejn\'\i{}k}
\affiliation{
Institute of Physics, ASCR, $v.~v.~i.$, 
Cukrovarnick\'a 10, CZ-16253 Praha 6, Czech Republic}

\author{J. \v Subrt}
\affiliation{Faculty of Mathematics and Physics, 
  Charles University in Prague, Praha, CZ-121 16, Czech Republic}

\author{C.T. Ellis}
\affiliation{Department of Physics, University at Buffalo--SUNY,
Buffalo, New York 14260, USA}

\author{A. Mukherjee}
\affiliation{Department of Physics, University at Buffalo--SUNY,
Buffalo, New York 14260, USA}

\author{J.~Lee}
\affiliation{Department of Physics, University at Buffalo--SUNY,
Buffalo, New York 14260, USA}

\author{G.M. Sipahi}
\affiliation{Instituto de F\'\i{}sica de S\~ao Carlos, 
Universidade de S\~ao Paulo, CP 369, 13560-970, S\~ao Carlos, SP, Brazil}
\affiliation{Department of Physics, University at Buffalo--SUNY,
Buffalo, New York 14260, USA}

\author{J. Sinova}
\affiliation{Department of Physics, Texas A\& M University,
College Station, Texas 77843-4242, USA}
\affiliation{
Institute of Physics, ASCR, $v.~v.~i.$, 
Cukrovarnick\'a 10, CZ-16253 Praha 6, Czech Republic}

\author{J. Hamrle}
\affiliation{Department of Physics and Nanotechnology Centre,
Technical University of Ostrava, 17. listopadu 15,
CZ-70833 Ostrava-Poruba, Czech Republic}

\author{T. Jungwirth}
\affiliation{
Institute of Physics, ASCR, $v.~v.~i.$, 
Cukrovarnick\'a 10, CZ-16253 Praha 6, Czech Republic}
\affiliation{
School of Physics and Astronomy, University of Nottingham,
Nottingham NG7 2RD, United Kingdom}

\author{P. N\v emec}
\affiliation{Faculty of Mathematics and Physics, 
  Charles University in Prague, Praha, CZ-121 16, Czech Republic}

\author{J. \v Cerne}
\affiliation{Department of Physics, University at Buffalo--SUNY,
Buffalo, New York 14260, USA}

\author{K. V\'yborn\'y}
\affiliation{
Institute of Physics, ASCR, $v.~v.~i.$, 
Cukrovarnick\'a 10, CZ-16253 Praha 6, Czech Republic}
\affiliation{Department of Physics, University at Buffalo--SUNY,
Buffalo, New York 14260, USA}

\date{Aug26, 2013}

\begin{abstract}
Magnetic linear dichroism and birefringence in (Ga,Mn)As epitaxial
layers is investigated by measuring the polarization plane rotation of
reflected linearly polarized light when magnetization lies in the
plane of the sample. We report on the spectral dependence of the
rotation and ellipticity angles in a broad energy
range of $0.12-2.7\unit{eV}$ for a series of optimized samples
covering a wide range on Mn-dopings and Curie temperatures and find a
clear blue shift of the dominant peak at energy exceeding the host
material band gap. These results are discussed in the general context
of the GaAs host band structure and also within
the framework of the $k\cdot p$ and mean-field kinetic-exchange model
of the (Ga,Mn)As band structure.  We find a semi-quantitative agreement
between experiment and theory and discuss the role of disorder-induced
non-direct transitions on magneto-optical properties of (Ga,Mn)As.
\end{abstract}

% 75.47.-m  Magnetotransport phenomena; materials for magnetotransport
%           (for spintronics, see 85.75.-d; see also 72.15.Gd, 73.50.Jt,
%           73.43.Qt, and 72.25.-b in transport phenomena)

\maketitle

\section{Introduction}

Among optical spectroscopies, differential methods based on the
birefringence or the dichroism, i.e., sensitive to differences in
refractive indices between two optical modes, can give more information on
material electronic structure than absorption
measurements.\cite{Ferre:1984_a}  For instance, the absorption
coefficient $\alpha(\omega)$ in the dilute magnetic
semiconductor\cite{Jungwirth:2006_a} (DMS) (Ga,Mn)As is essentially
featureless\cite{Burch:2004_a} at frequencies $\omega$ close to
$E_g/\hbar$ (the band gap energy, $E_g\approx 1.52\unit{eV}$ for GaAs)
while the same material in the same frequency range exhibits a strong
peak in polarization plane rotation caused by the magnetic linear
dichroism and birefringence.\cite{Kimel:2005_a} At the same time, any
type of magnetism-induced dichroism or birefringence depends on the
ferromagnetic splitting of the bands (related to
saturated magnetization $\vec{M}$) and manganese-doped DMSs like
(Ga,Mn)As offer the unique possibility of tuning the strength of
magnetism by varying the Mn content $x_{\mathrm{nom}}$ over a broad
range. Studying the trends in magneto-optical spectra across a series
of samples with increasing Mn doping and comparing them to model
calculations allows to microscopically relate the individual spectral
features to the electronic structure of the (Ga,Mn)As material.

\begin{figure}
\begin{tabular}{cc}
\hskip-.5cm\includegraphics[scale=0.22]{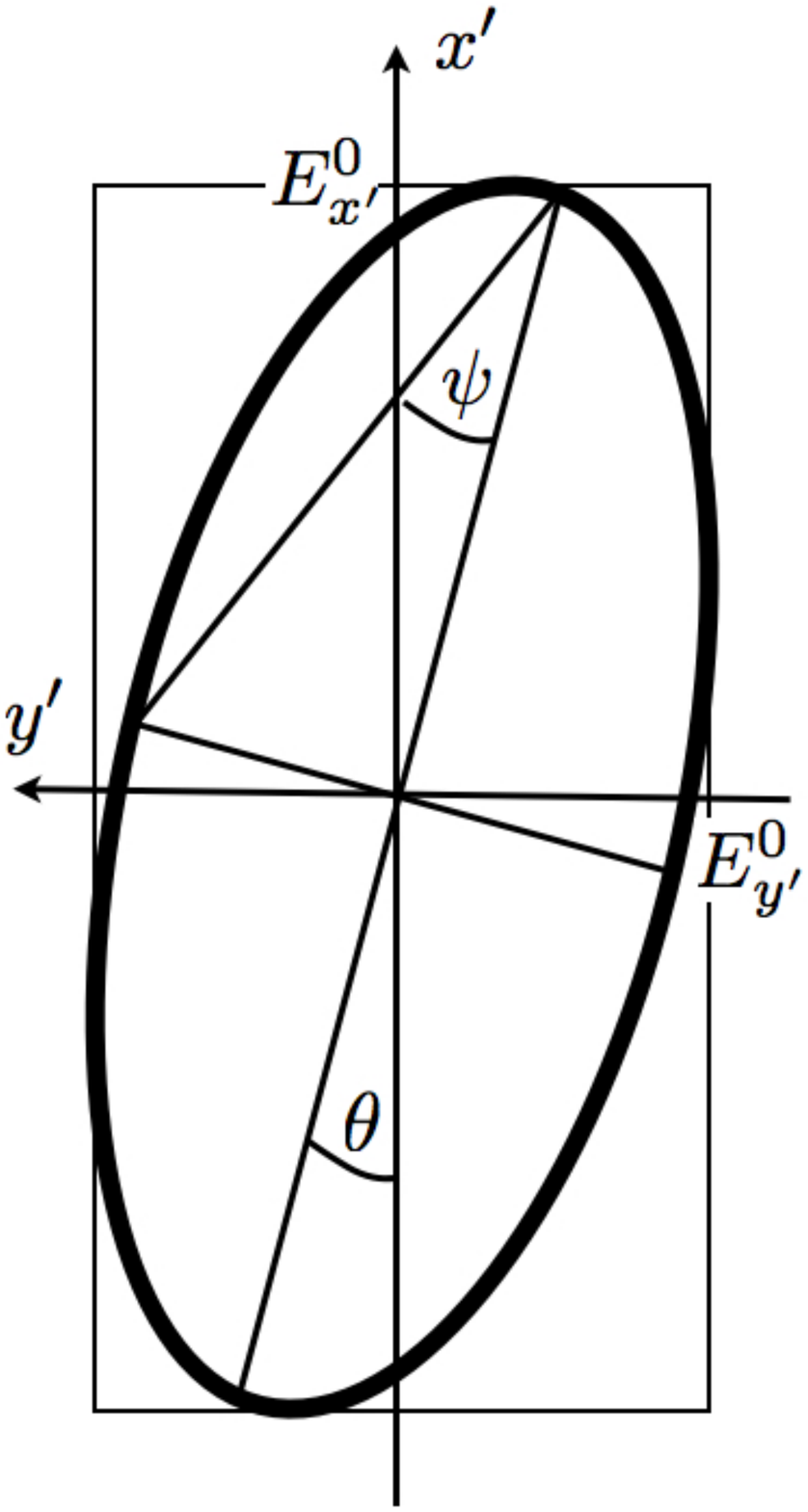} &
\hskip-1.0cm\includegraphics[scale=0.3]{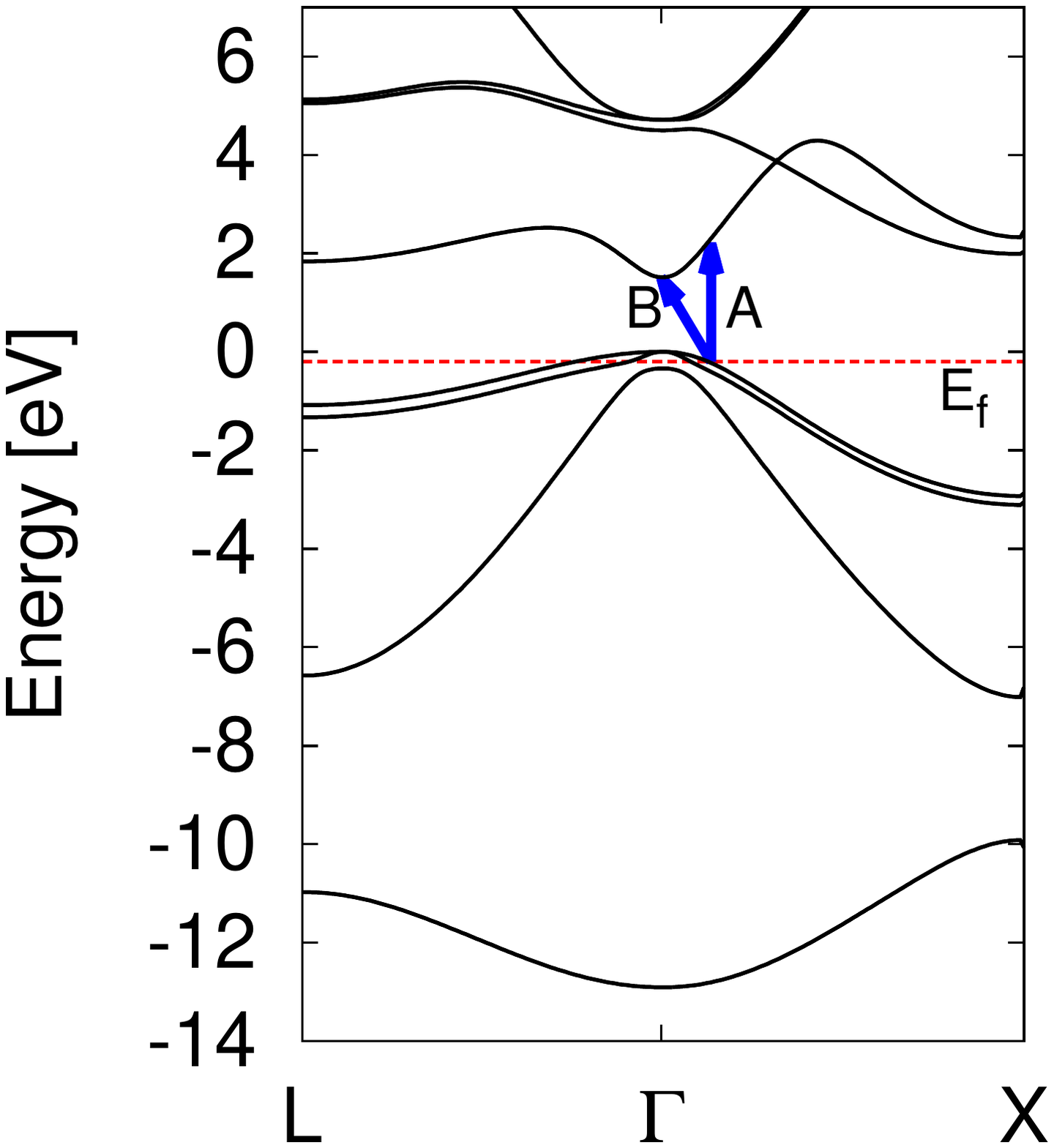}\\
(a) & \hskip-3cm (b)
\end{tabular}
\caption{(a) Measured magneto-optical quantities. Originally linearly
  polarized beam becomes elliptically polarized after interaction with
  the sample. Its ellipticity is characterized by angle $\psi$ and
  rotation of the major axis is $\theta$. (b) GaAs host band structure
  with Fermi level $E_f$ typical for our Mn-doped samples. Arrows
  indicate direct (A) and non-direct (B) transitions from the Fermi surface
  to the conduction band.}
\label{fig-01}
\end{figure}

Polarization-resolved magneto-optical effects appear in a multitude of
geometries and setups which we review in more detail in
Sec.~II below. In terms of the leading order of the effect, they can
be divided into effects linear and quadratic in $\vec{M}$.  In both cases, an
incident light beam linearly polarized along $\hat{x}'$ turns into an
elliptically polarized one whose major axis is rotated with respect to
$\hat{x}'$ by an angle $\theta$. The degree of ellipticity is
characterized by another (typically also small) angle $\psi$. Both
angles are defined in Fig.~\ref{fig-01}a.  Effects linear (or more
generally odd) in $\vec{M}$ give $\theta(-\vec{M})=-\theta(\vec{M})$ and are
related,\cite{Kim:2007_a,Acbas:2009_a} for $\omega\to 0$, to the dc
anomalous Hall effect.\cite{Nagaosa:2010_a} These effects are more
commonly investigated as they are often simpler to experimentally
access. They are typically larger and
it is simpler to separate them from
magnetization-independent optical signals. On the other hand,
even effects (quadratic in the leading order of $\vec{M}$) with 
$\theta(-\vec{M})=\theta(\vec{M})$
appear in literature less frequently. For example, the Voigt effect
in reflection (see Sec.~II and Fig.~\ref{fig-02}) 
has first been reported as late as in 1990.\cite{Schafer:1990_a}
Yet, they offer an alternative probe into the electronic 
structure of the material distinct from what is probed in odd-in-$\vec{M}$
measurements. The effects even in $\vec{M}$ are related to the
anisotropic magnetoresistance\cite{Kokado:2012_a,Rushforth:2007_a} for
$\omega\to 0$ and they do not vanish in certain situations where the
effects odd in $\vec{M}$ do. For example in compensated
antiferromagnets, the magneto-optical effects even in $\vec{M}$ can
still be detected \cite{Kirilyuk:2010_a} because contributions from
the two spin-sublattices with opposite spin orientations do not
cancel. As a probe into the antiferromagnetic
order,\cite{Kimel:2004_a} magneto-optical effect in the visible and
infrared range, such as the one described in this article, does not
rely on large-scale facilities as in the case of neutron diffraction
or x-ray Voigt effect.\cite{Mertins:2001_a} 

The magneto-optical effects odd in $\vec{M}$ have been extensively
explored in 
(Ga,Mn)As.\cite{Jungwirth:2006_a,Ando:1998_a,Jungwirth:2010_b,Acbas:2009_a}
While the visible\cite{Ando:1998_a,Jungwirth:2010_b}
range provides information on transitions between valence and
conduction bands which are relatively less sensitive to the spin-orbit
interaction effects, infra-red\cite{Acbas:2009_a} spectra enable to
explore transitions within valence bands. Quadratic (even in
$\vec{M}$) magneto-optical
response of (Ga,Mn)As is an alternative probe into its electronic
structure. In analogy with the dc anisotropic magnetoresistance, it crucially 
depends on the spin-orbit interaction in the whole spectral range.
Previous experiments have focused on measurements of the even in
$\vec{M}$ magneto-optical effects in selected (Ga,Mn)As samples
without studying their spectral dependence\cite{Moore:2003_a} or
limiting themselves to the visible spectral
range.\cite{Kimel:2005_a,Tesarova:2012_c} Here, we report measurements
in a spectral range of 0.12 to 2.7~eV and study systematically the
magneto-optical spectra even in $\vec{M}$ across a series of optimized
(Ga,Mn)As materials spanning a broad Mn-doping range summarized in
Tab.~\ref{tab-01} below.

Section~II is dedicated to a brief overview of magneto-optical effects
and clarification of the terminology that is not coherent across the
literature.\cite{Ferre:1984_a,Kimel:2005_a} Our experimental data are
presented in Section~III and we compare them in Section~IV to a
kinetic-exchange model\cite{Abolfath:2001_a}-based calculations of ac
permittivity that allow us to determine $\theta(\omega)$ and
$\psi(\omega)$. In Section~IV, we also discuss the complex individual
spectral features of $\theta(\omega)$ and clarify the role of linear
birefringence and dichroism (see also Appendix~D). Section~V concludes
the article.  In Appendix~A, we review theoretical description of
magneto-optical effects on the level of Maxwell's equations to which
the permittivity tensor is the input. Appendices B,~C and~D,
respectively contain additional experimental data, more details on the
transport calculation using the kinetic-exchange model, and details on
the optical part modelling, e.g. multiple reflections on the (Ga,Mn)As
epilayer.

\section{Overview of magneto-optical effects}

The purpose of this section is to recapitulate selected magneto-optical
effects, clarify the terminology and specify which of these effects is
considered in this article. The first magneto-optical
phenomenon was observed by Michael Faraday in 1846, followed by
another one found by John Kerr in 1877. They found that linearly
polarized light transmitted through (Faraday's discovery) or reflected
from (Kerr's discovery) a non-magnetic material subject to magnetic
field $\vec{B}$ has its polarization plane rotated. In their
experiments, the wavevector of the propagating light $\vec{k}$ was
parallel to $\vec{B}$. In 1899, Woldemar Voigt observed optical
anisotropy of a non-magnetic crystal for $\vec{k}\perp\vec{B}$ which
can also cause similar rotation of the polarization
plane. Historical overview of these and related discoveries can be
found in the introductory parts of
Refs.~\onlinecite{Ebert:1996_a,Ferre:1984_a}. As a matter of
definition, we will not include polarization-unrelated (or unresolved)
effects such as cyclotron resonance into our further
discussion.\cite{note1}

\begin{figure}
\leavevmode\kern-1cm\hbox{\includegraphics[scale=0.30]{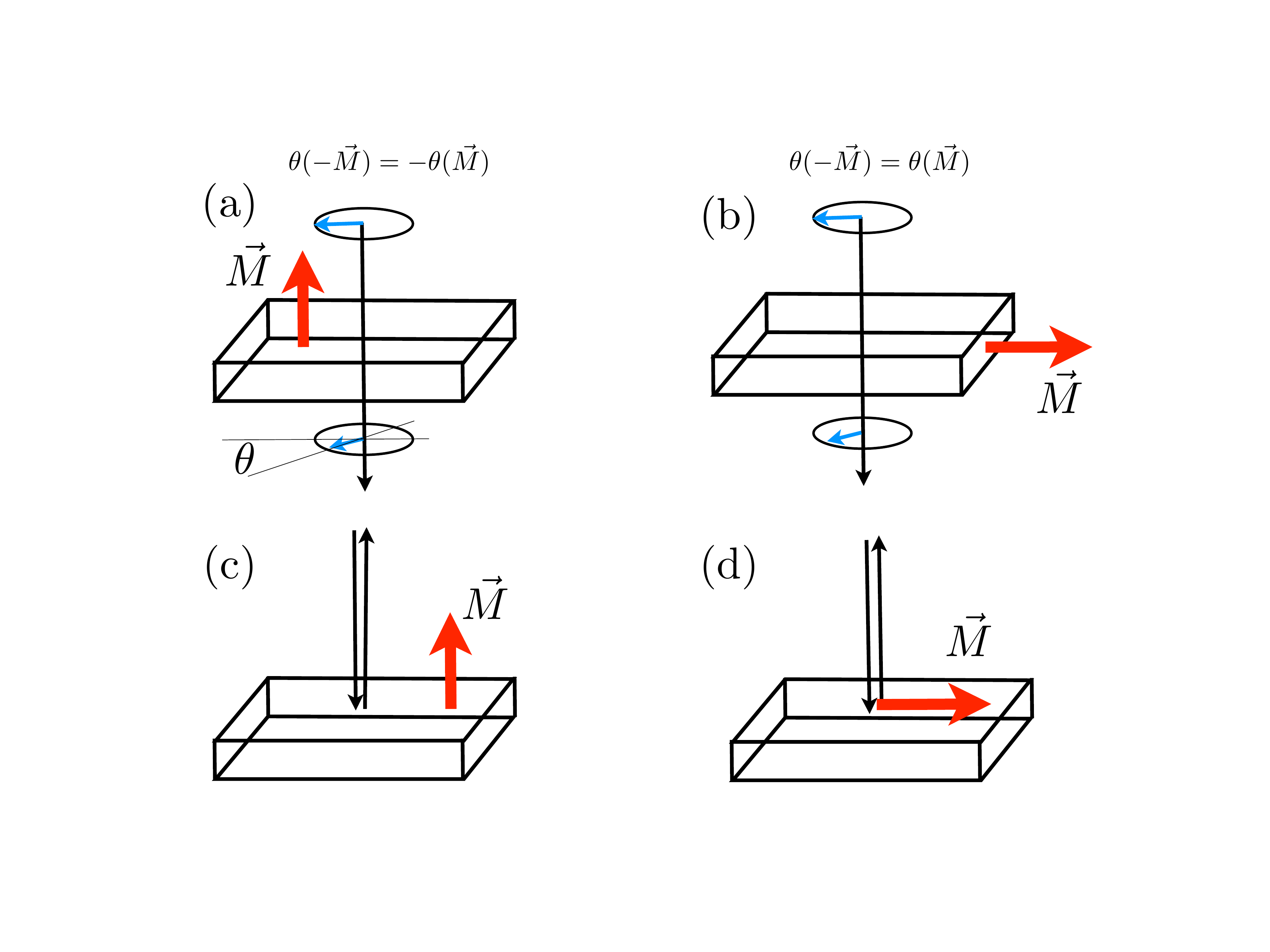}}
\caption{Selected magneto-optical effects. Polarization planes of incoming and
  outgoing beams are rotated by $\theta$ with respect to each other
  (possible ellipticity $\psi$ of the outgoing is not considered in
  these sketches).
  (a) Faraday effect, (b) Voigt effect, (c) Kerr effect, (d)
  Voigt effect in reflection.}
\label{fig-02}
\end{figure}

Analogous phenomena are found in magnetic materials where,
phenomenologically, $\vec{M}$ plays the same role as $\vec{B}$ in the
original observations of Faraday, Kerr and Voigt. Typical experiments
involve a slab or thin layer of the material and, for simplicity, let
us assume for now that it is not placed on a substrate and also
that $\vec{k}$ is perpendicular to the plane of the sample surface
(''normal incidence'').  Faraday and Kerr magneto-optical effects arise
for $\vec{M}\parallel\vec{k}$, i.e., out-of-plane magnetization while
Voigt effect occurs for in-plane magnetization
($\vec{M}\perp\vec{k}$).  As it has already been described above (see
Fig.~\ref{fig-01}a), the incident beam is linearly polarized and the Kerr
(Faraday or Voigt) effect are manifested in the rotation $\theta$
of the reflected (transmitted) beam polarization plane. 
Any of these effects will, in
general, be accompanied by a non-zero ellipticity characterized by
$\psi$ and both angles are sometimes combined into one complex
quantity, e.g. the complex Faraday angle $\theta_F$ in
Ref.~\onlinecite{Kim:2007_a}. The Voigt effect is even in $\vec{M}$
while the Faraday and Kerr effects are odd in $\vec{M}$. There is no
broadly accepted term for the quadratic (even-in-$\vec{M}$) 
magneto-optical effect in the 
reflection at normal incidence with in-plane $\vec{M}$, although 
sometimes it is called quadratic magneto-optical Kerr effect
(QMOKE),\cite{Buchmeier:2009_a} Hubert-Sch\"afer effect\cite{note9} 
or it is included in the ''reflection analogy to the Voigt 
effect''.\cite{Postava:1997_a}   We will adopt here the last
terminology. A schematic summary of the Faraday, Voigt and Kerr effects
and of the Voigt effect in reflection is shown in Fig.~\ref{fig-02}.

For other than the normal incidence, the Kerr effect is no longer
distinguished by $\vec{M}\parallel\vec{k}$ and it appears in several
variants. General magnetization $\vec{M}$ can now be decomposed into
out-of-plane component $M_\perp$ and in-plane components $M_L$ ($M_T$)
parallel (perpendicular) to the plane of incidence. The polar Kerr
effect, sometimes also called magneto-optical Kerr effect (MOKE), is
in the leading order linear in $M_\perp$ and it is the only effect
odd in $\vec{M}$ that does not vanish for $M_L=M_T=0$.  The longitudinal and
transversal Kerr effects depend on the in-plane components of
magnetization and to separate them from the Voigt effect in
reflection, the polarization signal dependence on the 
angle $\beta$ between $\vec{M}$ and
the polarization plane can be used. Unlike all three Kerr effects,
Voigt effect in reflection is proportional\cite{Postava:1997_a} to a
combination of $M_T^2, M_L^2$ and $M_LM_T$ which, at normal incidence,
combines into a $\sin 2\beta$ dependence.

In this work, we present a systematic spectral study of the Voigt effect in
reflection. As with
other magneto-optical phenomena, this effect includes rotation and
ellipticity measured in the beam after its interaction with the sample 
and from now on, we associate the terms ''rotation'' ($\theta$) and
''ellipticity'' ($\psi$) only with the Voigt effect in reflection
(unless explicitly stated otherwise). Both rotation and ellipticity are
related to complex refractive indices $n_\parallel$ and $n_\perp$ of two modes 
(see detailed explanation in Appendix~A) linearly polarized parallel
and perpendicular to $\vec{M}$. Rotation $\theta$ is caused both
by magnetic linear birefringence $\Delta \bar{n}\equiv \Rea
(n_\perp-n_\parallel)\not=0$ (MLB) and magnetic linear dichroism
$\Delta \bar{k}\equiv \Ima (n_\perp-n_\parallel)\not=0$ (MLD), an
illustrative example is given in Appendix~D. 
%More generally (not just
%in context of magneto-optical effects), birefringence 
%$\Delta \bar{n}\not=0$ can manifest itself in the
%well-known double-image refraction (propagation of ordinary and
%extraordinary rays through a slab of icelandic calcite described first by
%Rasmus Bartholin in 1669) while dichroism\cite{Ferre:1984_a} 
%$\Delta \bar{k}\not=0$ can effectively lead to mode filtering. Note
%that the term MLD is sometimes confusingly used\cite{Kimel:2005_a} both for 
%$\Delta \bar{k}\not=0$ and the nonzero difference in reflection
%coefficients of the two linearly polarized modes. 
We now
proceed to describe our experimental results of rotation and
ellipticity of the Voigt effect in reflection on (Ga,Mn)As samples.

\section{Experiment}

The samples used in our measurement are (Ga,Mn)As layers prepared by
optimized molecular-beam epitaxy growth and post-growth annealing
procedures\cite{Nemec:2012_b} with various nominal Mn doping ranging 
from $x_{\mathrm{nom}}=1.5\%$ to 14\% and cut into 4.5
by 5 mm chips. The basic material characteristics of our samples are 
listed in Tab.~\ref{tab-01}, additional information can be found in
the main text and supplementary information of
Refs.~\onlinecite{Jungwirth:2010_b},\onlinecite{Nemec:2012_b}.
All samples were grown on a GaAs substrate, producing a compressive
strain which favours an in-plane orientation of the easy axes
(EAs). The competition of in-plane cubic and uniaxial anisotropies
results in our (Ga,Mn)As films in two magnetic EAs tilted from the
[100] and [010] crystal axes towards the $[1\bar{1}0]$ in-plane
diagonal.\cite{Zemen:2009_a}  The tilt angle 
increases\cite{Nemec:2012_b} with increasing Mn-doping. The sample
substrates were wedged (1$^\circ$) to avoid spurious signals that
might appear due to the multiple reflections from the back side of
the substrate. In order to measure the
rotation and ellipticity angles $\theta$ and $\psi$ in a broad energy
range we developed a sensitive experimental technique which is
described in detail in Ref.~\onlinecite{Tesarova:2012_c}.  We use a Xe
lamp (0.33--2.7~eV) with a double prism CaF$_2$ monochromator and
discrete spectral lines from CO$_2$ (115--133~meV) and CO (215--232~meV)
lasers.\cite{Kim:2011_a}  Measurements are done in the reflection
geometry close to normal incidence ($\approx 6^\circ$ with respect
to the sample normal) whereas we assume that the polarization plane 
rotation due to the longitudinal Kerr effect is negligible.
The samples are mounted on a custom made
rotating sample holder attached to the cold finger which is cooled
down to 15~K. The holder enables a precise rotation of the
sample, and thus of the magnetization with respect to the incident
polarization using external magnetic field $\vec{B}$, which is
applied along a fixed in-plane direction.

Prior to the actual measurement of $\theta$ and $\psi$, the samples
are rotated so that one of the EAs is set parallel to
$\vec{B}$. Subsequent application of a moderately strong magnetic field
($B\equiv|\vec{B}|=0.6\unit{T}$) forces the magnetization to align
with this EA. After the magnetization is oriented along the EA
parallel to $\vec{B}$, the magnetic field is turned off and the sample
is rotated $45^\circ$ away from the field axis. The sample orientation
is kept fixed subsequently, and it is not changed during the
measurement of $\theta$ and $\psi$. The magneto-optical response of
the sample is measured using the polarization modulation technique at
base frequency $f=50\unit{kHz}$, where the reflected beam passes
through the photoelastic modulator (PEM).\cite{Sato:1981_a} The
optical axis of the PEM is oriented 45$^\circ$ with respect to the
magnetic field axis and the detected signals at $f$ and $2f$ are
proportional to ellipticity ($\psi$) and rotation ($\theta$) of the
reflected light polarization, respectively.\cite{Sato:1981_a,Kim:2011_a} In the
first step of the measurement, the polarization of the incident light
is set parallel with the magnetization orientation, so any non-zero
signal detected at $f$ or $2f$ is just background unrelated to
magneto-optical properties of the sample. In the second step we apply
$B\approx 0.6\unit{T}$ which rotates the magnetization to
$\beta=45^\circ$ relative to the incident beam polarization. In this
situation, the polarization components parallel and perpendicular to
magnetization experience different (complex) indices of refraction,
maximizing the rotation and ellipticity signals magnitude. The $\sin 2\beta$
dependence of $\theta$ has been checked (see Fig.~3d in
Ref.~\onlinecite{Tesarova:2012_c}). By taking a difference of $\theta$
(or $\psi$) between the first and second step, we obtain the pure
magneto-optical signal. This procedure replaces the commonly used 
$[\theta(\vec{M})-\theta(-\vec{M})]/2$ protocol for magneto-optical
phenomena odd in magnetization such as the Kerr effect.
We note, that in order to obtain the
correct sign and magnitude of $\theta$ and $\psi$, a calibration
procedure\cite{Kim:2007_a} has to be performed. Detailed description
of our experimental methods is given in Ref.~\onlinecite{Tesarova:2012_c}. 

Measured $\theta$ and $\psi$ for samples B,C,D,E,G of
Tab.~\ref{tab-01} are displayed in Fig.~\ref{fig-03}, while remaining two
samples are studied using a different technique and are discussed in
Appendix~B. Both rotation and ellipticity reach typically values of
several 0.1~mrad, show distinct spectral features in the studied range 
$\hbar\omega=115\unit{meV}$ to $2.7\unit{eV}$ and often change sign as a
function of radiation frequency
$\omega$. Such values are about an order of magnitude
smaller than the Kerr effect\cite{Acbas:2009_a} but still large enough
to use the Voigt effect in reflection as an efficient method to detect
in-plane component of the magnetization.\cite{Tesarova:2012_a} In the
more general context of magnetic materials, values of $\theta\approx
0.5$~mrad reported in Heusler alloys\cite{Hamrle:2007_a} are quoted
as\cite{Buchmeier:2009_a} ''record QMOKE values''. In agreement with
Kimel~et~al.\cite{Kimel:2005_a} who studied a single 
$x_{\mathrm{nom}}=2\%$ sample, we observe a peak in $\theta(\omega)$
exceeding $0.5$~mrad whose sign and position is consistent with this
earlier result. Compared to Ref.~\onlinecite{Kimel:2005_a}, we are now
able to follow spectral trends as $x_{\mathrm{nom}}$ is varied and we
discuss these in the following Section. Here, we only note that the
prominent peaks in $\theta(\omega)$ at $\hbar\omega\approx
1.7\unit{eV}$ shown in Fig.~\ref{fig-03} appear close to the peaks of
the Kerr effect\cite{Jungwirth:2010_b} and also the non-monotonic dependence 
of their height on $x_{\mathrm{nom}}$ is similar in both
magneto-optical effects. Finally, we remark that Voigt effect was also
measured in manganese-doped II-VI
materials. Ref.~\onlinecite{Kimel:2005_a} claims that magneto-optical
response even in $\vec{M}$ is ''drastically enhanced'' in (Ga,Mn)As
compared to that of (Cd,Mn)Te.\cite{Oh:1991_a}  While we do not
directly contradict this conclusion we find the comparison less
conclusive. Magneto-optical effects in a paramagnetic system
such as (Cd,Mn)Te are not spontaneous but must be induced by external
magnetic field, hence the spontaneous $\Delta \bar{n}$, $\Delta \bar{k}$
of (Ga,Mn)As must be compared to the proportionality constant
between $\Delta \bar{n}$, $\Delta \bar{k}$ and $B^2$ in
(Cd,Mn)Te. More importantly though, the transmission
measurements\cite{Oh:1991_a} are limited $\omega$ to sub-gap frequencies
where the signal is weaker and it is possible that the actual maximal
magneto-optical response of (Cd,Mn)Te would be comparable to that of
(Ga,Mn)As if we were comparing the parts of spectra that correspond to
each other.

\begin{table}
\begin{tabular}{c|lp{1.5cm}cp{.8cm}c}
& wafer$\quad$ & $x_{\mathrm{nom}}$ [\%] & $\quad x$ [\%] $\quad$ &
  $p$ $\unit{[nm^{-3}]}$ & $T_c$~[K] \\ \hline
    A&F010 &$\quad$ 1.5 & 1.0 & 0.15 & 29\\ 
    B&F002 &$\quad$ 3   & 1.8 & 0.66 & 77\\
    C&F020 &$\quad$ 5.2 & 3.6 & 1.08 & 132\\ 
    D&E115 &$\quad$ 7   & 5.5 & 1.41 & 159\\
    E&E122 &$\quad$ 9   & 6.9 & 1.55 & 179\\
    F&E079 &$\quad$ 12.5& 8.6 & $1.8^*$& 186\\
    G&F056 &$\quad$ 14  & 8.5 & $1.8^*$& 182\\
\end{tabular}
\caption{Basic sample parameters according to Tab.~I of
  Ref.~\onlinecite{Jungwirth:2010_b} (Supplemental
  Information). Asterisk indicates estimated value. Effective doping
  $x$ (which enters Eq.~(\ref{eq-11}) through the
  ferromagnetic splitting) is calculated from the measured saturated
  magnetization as explained in Appendix~C.}
\label{tab-01}
\end{table}

\begin{figure}
\vspace*{-1.3cm}
\begin{tabular}{c}
\includegraphics[scale=0.28]{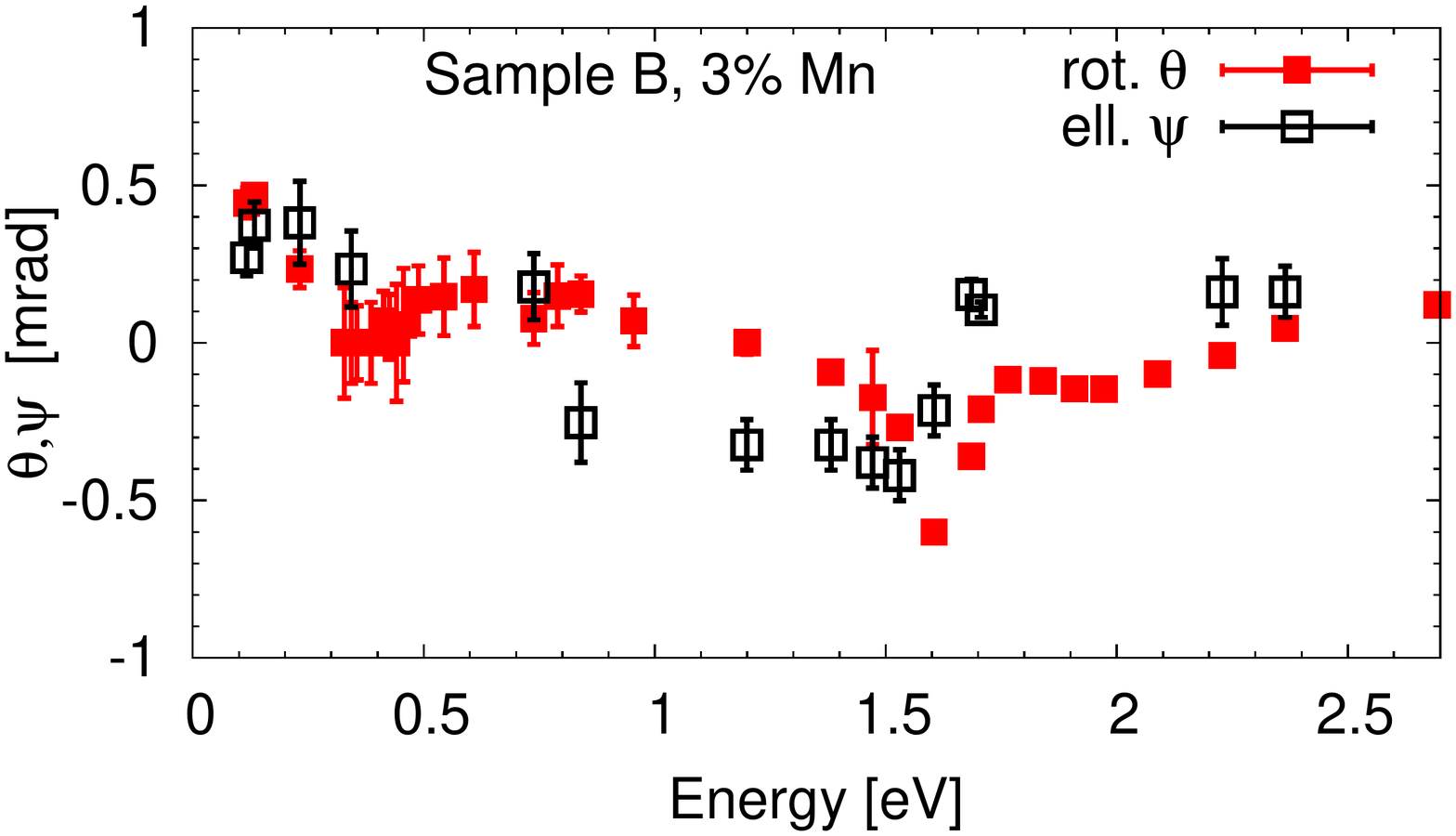} \\[-15mm]
\includegraphics[scale=0.28]{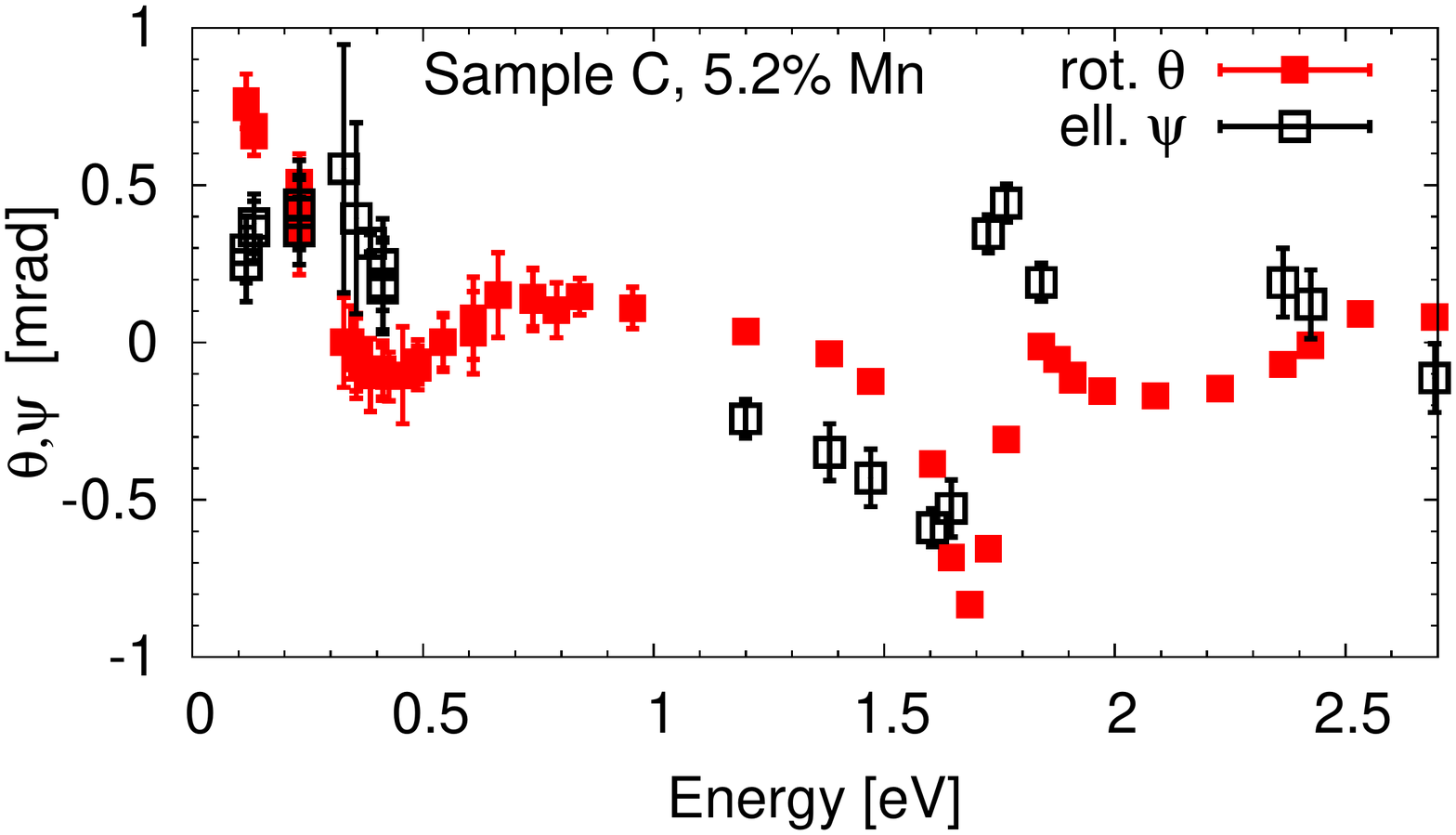} \\[-15mm]
\includegraphics[scale=0.28]{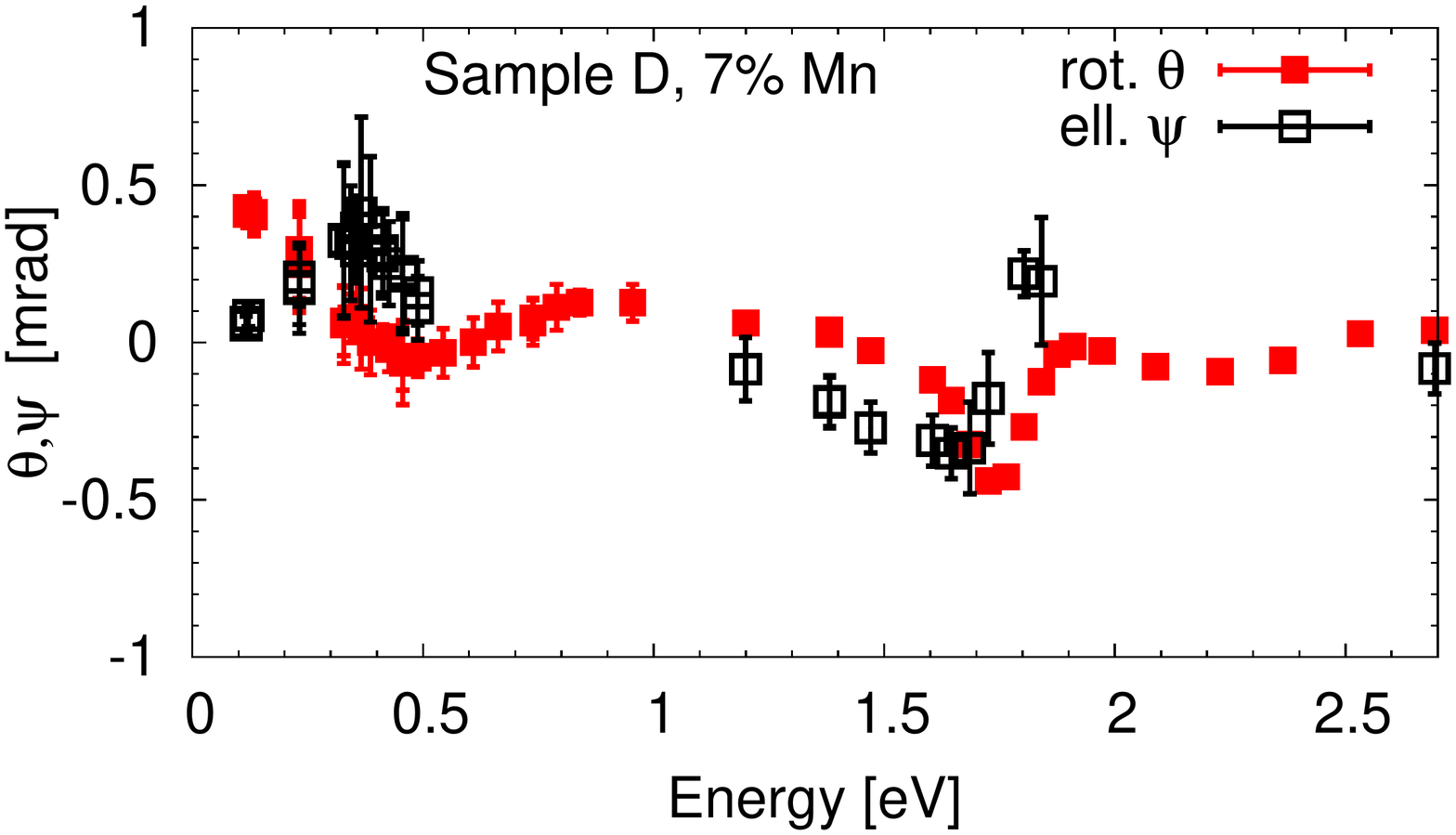} \\[-15mm]
\includegraphics[scale=0.28]{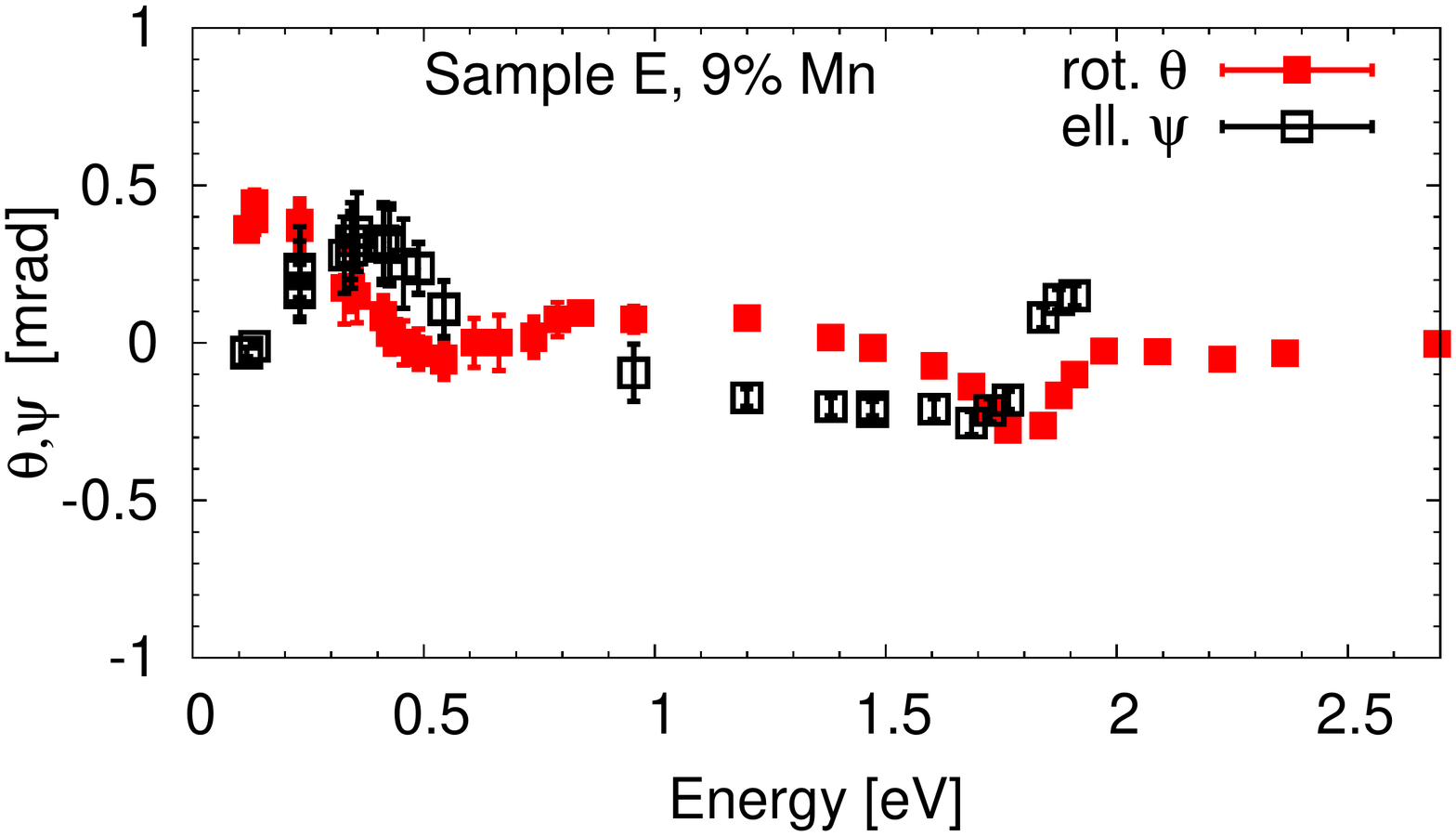} \\[-15mm]
\includegraphics[scale=0.28]{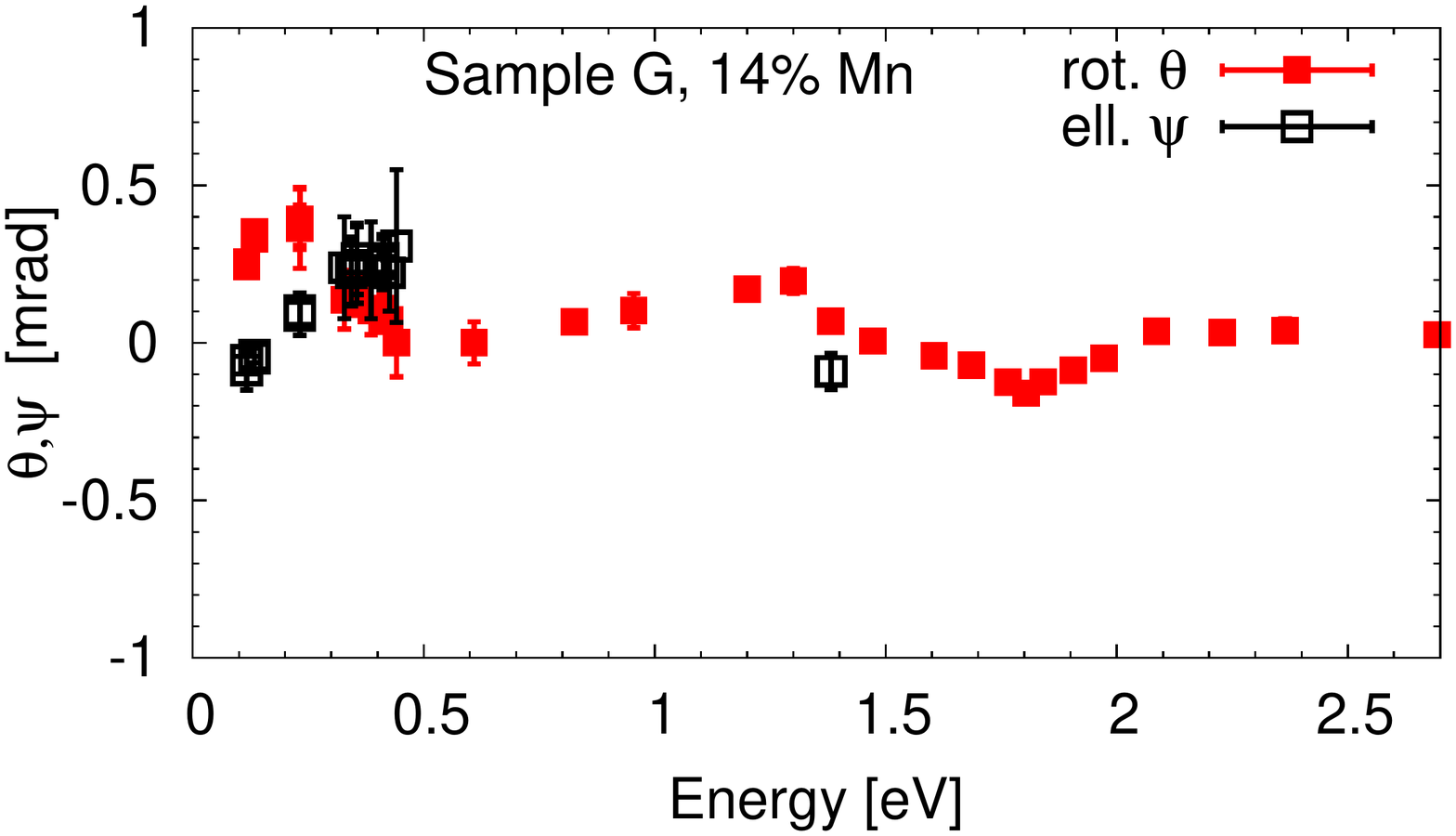}
\end{tabular}
\caption{Measured spectral dependence of the rotation $\theta$ 
and ellipticity $\psi$ for the Voigt effect in reflection. Manganese
doping levels indicated correspond to $x_{\mathrm{nom}}$.}
\label{fig-03}
\end{figure}

\section{Interpretation of the measured magneto-optical signals}

To understand observed spectral features in $\theta(\omega)$ and
$\psi(\omega)$ of the Voigt effect in reflection and their trends
across the set of samples, a model of the electronic bands close 
($\sim \hbar\omega$) to
the Fermi energy $E_f$ is needed. Any model having ambitions to yield
quantitative information on $\theta(\omega)$, $\psi(\omega)$ has to
start from a description of the (Ga,Mn)As electronic 
structure reflecting the GaAs host bands, exchange-splitting of the
bands in the ferromagnetic state of (Ga,Mn)As and the spin-orbit
coupling. Without the last two components, only positions of spectral
features in $\theta(\omega)$, $\psi(\omega)$ can  be anticipated but
not their shape and amplitude. GaAs host band structure in
Fig.~\ref{fig-01}(b), calculated by standard $spds^*$ tight-binding 
model\cite{Jancu:1998_a}, suggests that the prominent peak around
1.7~eV seen in $\theta(\omega)$ of~Fig.~\ref{fig-03} corresponds to
transitions between valence and conduction band. To analyze its
amplitude, we have to account for the combined effect of the
exchange-splitting and the spin-orbit interaction. Unlike other approaches
such as the quantum defect method\cite{Bebb:1969_a} used in
Ref.~\onlinecite{Chapler:2013_a} to analyze absorption spectra of
(Ga,Mn)As, the kinetic-exchange model of disordered carrier 
bands\cite{Dietl:2000_a,Abolfath:2001_a} which we briefly describe
below, naturally includes these two components. Apart from
successfully explaining the spectral trends in
absorption\cite{Jungwirth:2007_a} and of the Kerr effect in the visible 
range,\cite{Jungwirth:2010_b} this model therefore allows to calculate
$\theta(\omega)$, $\psi(\omega)$ of the Voigt effect in reflection which is
microscopically more constrained than the absorption or the visible
range Kerr effect. We show below that results of this model are in
semi-quantitative agreement with the measured data as in the
previously explored infrared Kerr effect\cite{Acbas:2009_a} which also
depends sensitively on the spin-orbit coupled exchange-split nature of
the valence band.

The path to theoretically evaluated $\theta(\omega)$, $\psi(\omega)$
involves three steps, the first of which is to obtain the band
structure $E_{\vec{k},a}$. Two of the aforementioned band structure
description components are
included in $\hat{H}_{KL}$ (host band structure and spin-orbit
interaction), the last component (ferromagnetic exchange-splitting) enters the
total Hamiltonian through kinetic-exchange parametrized by $J_{pd}$
($J_{sd}$) couplings between the dominantly $p$-like valence band
($s$-like conduction band) and Mn $d$-levels:
\begin{equation}\label{eq-11}
  \hat{H}=\hat{H}_{KL}+\frac{J_{pd}}{\mu_B}\vec{M}\cdot\hat{s}_h
  +\frac{J_{sd}}{\mu_B}\vec{M}\cdot\hat{s}_e + \hat{V}_{xc}.
\end{equation}
Here, $\hat{s}_{e/h}$ is the electron/hole spin operator, $\mu_B$ the
Bohr magneton and $\hat{V}_{xc}$ the correction due to many-body
effects which is important in heavily-doped
semiconductors as discussed below Eq.~(\ref{eq-15}). The choice of bands
included in the Kohn-Luttinger Hamiltonian~$\hat{H}_{KL}$ 
is dictated by the energy
range ($\hbar\omega$ up to 2.7~eV) in our experiments and $E_f$ 
of at most few 100~meV from the valence band top. As seen in
Fig.~\ref{fig-01}b, only conduction band, heavy holes (HH), light
holes (LH) and split-off band can be involved in optical excitations
from occupied to unoccupied states making up the total number of eight
bands in $\hat{H}_{KL}$. Parameters entering this $8\times 8$ matrix
are given in Appendix~C.
The magnetization $\vec{M}$ that determines the ferromagnetic
splitting in Eq.~(\ref{eq-11}) includes only the Mn magnetic moments,
hence the contribution of carrier spins must be subtracted from the saturation
magnetization. We use procedure described in Appendix~C 
below Eq.~(\ref{eq-16}).

Second step is to calculate the conductivity tensor components. We
take $\vec{M}\parallel\hat{x}$ and since $\sigma_{yz}$ has a
negligible\cite{note5} effect on $\theta$, we only need to determine
$\sigma_{xx}(\omega)$ and $\sigma_{zz}(\omega)$ which we henceforth
denote by $\sigma_\parallel$ and $\sigma_\perp$. They comprise of
intra- and inter-band contributions,
\begin{equation}\label{eq-12}
  \sigma_{\parallel/\perp}(\omega) = 
  \sigma_{\parallel/\perp}^{\mathrm{intra}}+
  \sigma_{\parallel/\perp}^{\mathrm{inter}}.
\end{equation}
where the former is simply taken as the Drude ac conductivity and the latter
is calculated from the Kubo linear-response formula whose input are the
energies $E_{\vec{k},a}$ and eigen-spinors $|n,\vec{k}\rangle$
obtained by numerical diagonalization of the $8\times 8$ 
Hamiltonian~(\ref{eq-11}). Formulae for both $\sigma^{\mathrm{intra}}$ and 
$\sigma^{\mathrm{inter}}$ are given in the Appendix~C. Complex
effective permittivity $\veeff$ then follows from Maxwell's equations
as discussed in Appendix~A. In textbooks, its two constituent terms
\begin{equation}\label{eq-10}
  \veeff \equiv \ve_0\left(\ve_b + \frac{i\sigma}{\omega\ve_0}\right).
\end{equation}
are usually ascribed to bound and free charges. This distinction is
certainly not a sharp one in the ac regime and more so at optical
frequencies. The ambiguity is naturally resolved by accounting for
all inter-band transitions between the eight selected bands in
$\sigma$ while all other processes, at lower as well as at larger
energies $\hbar\omega$, are included in the background
$\ve_b$. We include the intra-band transitions into $\sigma$ and adjust the
value of $\ve_b$ so that for intrinsic GaAs ($p=0$), $\veeff$
calculated using Eq.~(\ref{eq-10}) recovers the experimental ac
permittivity at optical frequencies\cite{Lautenschlager:1987_a} and it 
approaches $\ve_\infty=10.9$ in the $\omega\to 0$ limit.
Experimentally,\cite{Johnson:1969_a}
the permittivity of intrinsic GaAs approaches this value above 
the optical phonon resonance ($\hbar\omega\approx 30\unit{meV}$) which
is well below the lowest energies studied in our experiments.

Calculation of the rotation and ellipticity angles is the last 
step. Using the effective permittivity $\veeff^\parallel$ 
($\veeff^\perp$) obtained from $\sigma_\parallel$ ($\sigma_\perp$), we
calculate the refractive indices $n_\parallel$ and $n_\perp$ as
the square root of the permittivity (see also Appendix~D).
When multi-reflection effects on the sample-substrate interface are
neglected, we use Fresnel's formula
\begin{equation}\label{eq-05}
  r(n)=\frac{1-n}{1+n}
\end{equation}
to get reflection coefficients $r(n_\parallel)$ and $r(n_\perp)$ and
then calculate an auxiliary (complex-valued) quantity
\begin{equation}\label{eq-13}
  \chi = \frac{r(n_\parallel)-r(n_\perp)}{r(n_\parallel)+r(n_\perp)}.
\end{equation}
The rotation and ellipticity angles for the Voigt effect in reflection are
\begin{equation}\label{eq-14}
  \theta = \frac12 \atan \left(\frac{2\Rea\chi}{1-|\chi|^2}\right)
  \qquad
  \psi = \frac12 \asin \left(\frac{2\Ima\chi}{1+|\chi|^2}\right).
\end{equation}
The relationship between conductivities and actual experimentally
measured angles $\theta$, $\psi$ is thus markedly non-linear, yet as a
rough guide, the Voigt effect in reflection can be related to
$\sigma_\parallel-\sigma_\perp$ as explained in the simplified
situation pertaining to Eq.~(\ref{eq-09}) in Appendix~A and an example
of $\sigma_\parallel-\sigma_\perp$ is shown in Fig.~\ref{fig-11}. For most of
our calculations, we take the multi-reflections into account and use
Eq.~(\ref{eq-30}) instead of Eq.~(\ref{eq-05}). Discussion of their
importance is given in Appendix~D.

\begin{figure}
\begin{tabular}{c}
\includegraphics[scale=0.3]{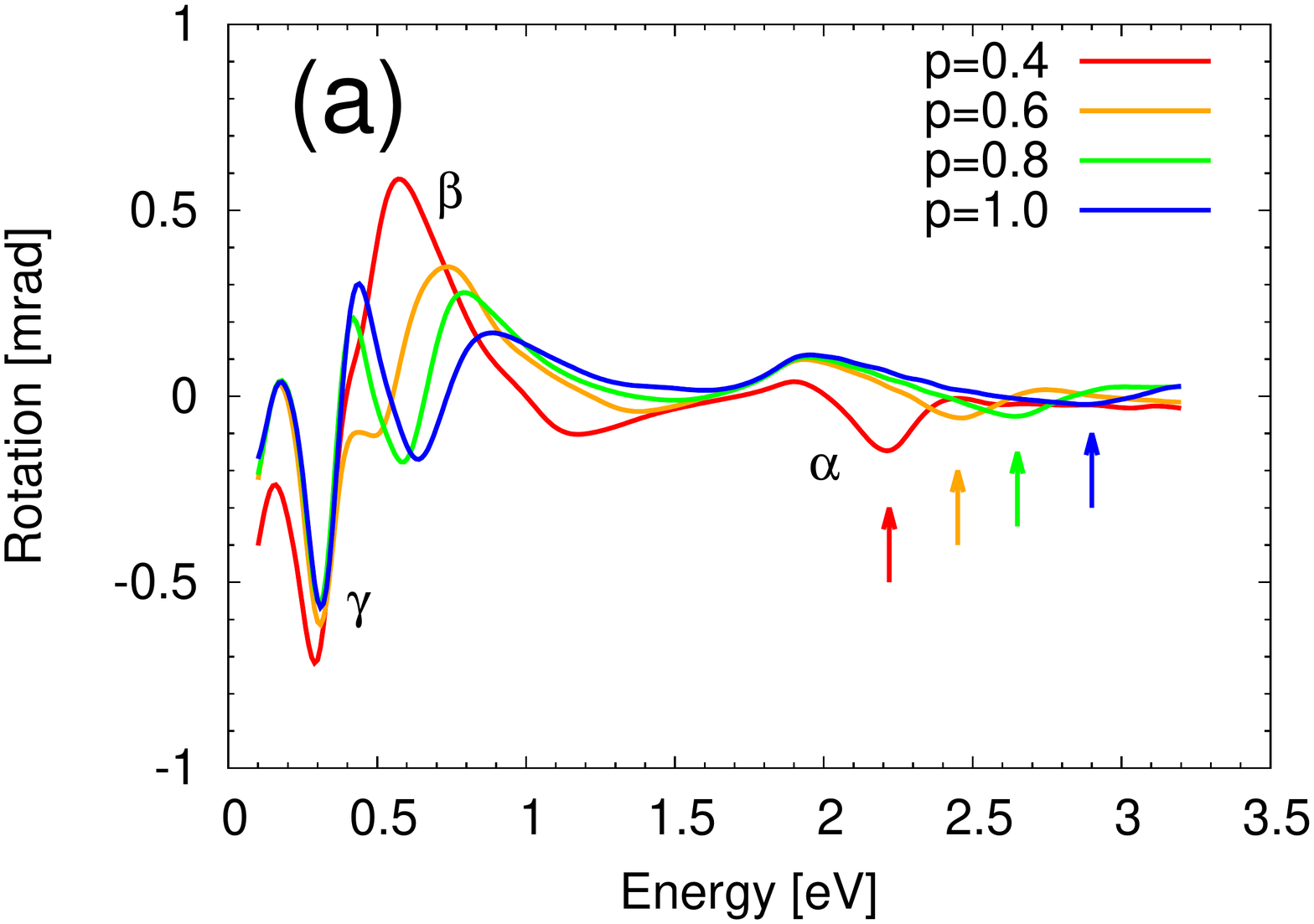} \\
\includegraphics[scale=0.3]{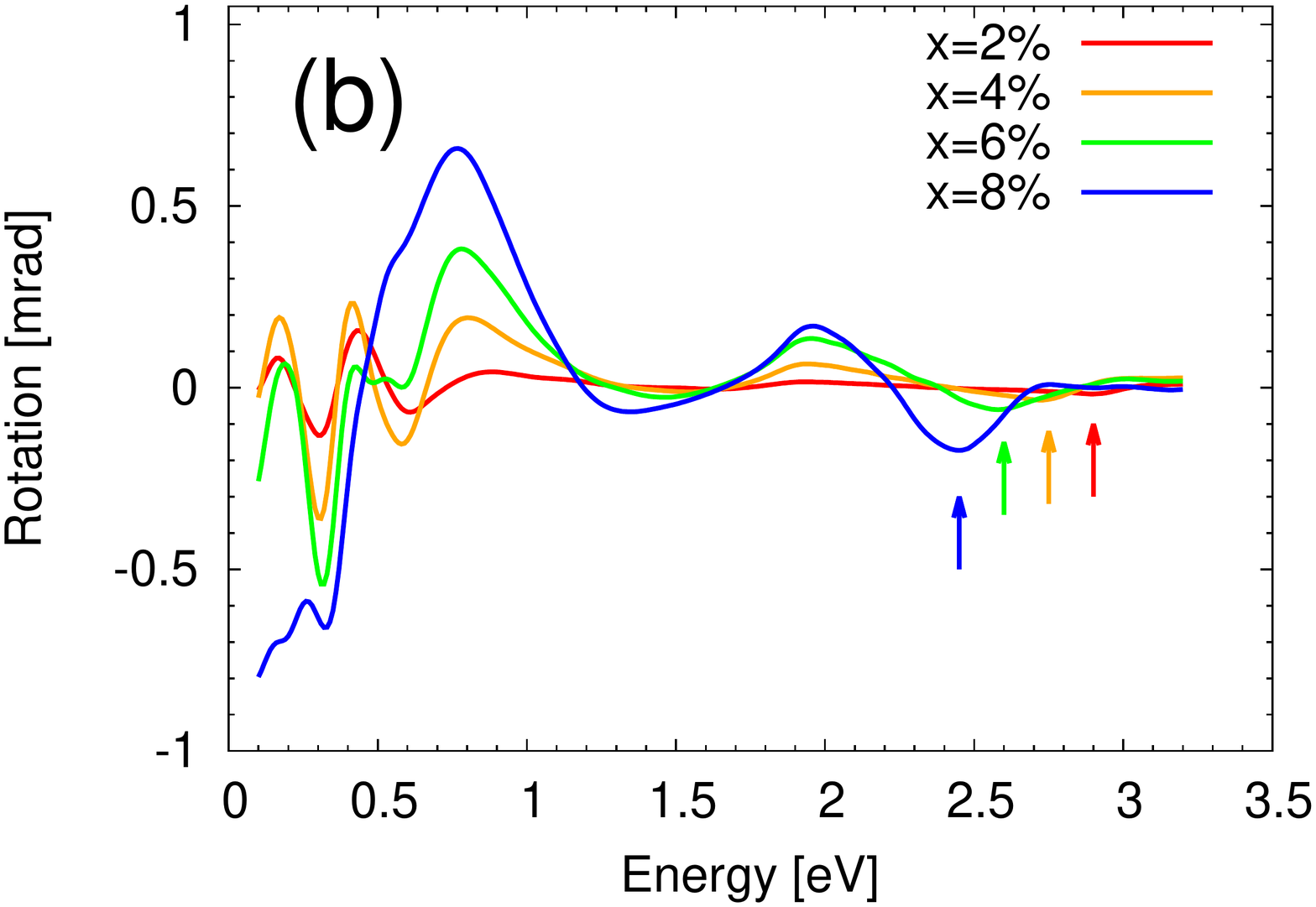} 
\end{tabular}
\caption{Calculated spectral dependence of the rotation angle $\theta$
with $\hat{V}_{xc}=0$ for a series of systems with (a) $x=5\%$ and 
varying $p$ (in$\unit{nm^{-3}}$) and (b) varying $x$ and constant 
$p=0.8\unit{nm^{-3}}$. Spectral features discussed in text are
labelled by Greek letters, arrows indicate the position of peak $\alpha$.}
\label{fig-04}
\end{figure}

To address experimental findings in Fig.~\ref{fig-03}, we calculate
$\theta(\omega)$ of Eq.~(\ref{eq-14}) in the range $\hbar\omega=0.1$
to $3.3$~eV. Such optical transition energies somewhat exceed the
range of applicability of our band structure model in
Eq.~(\ref{eq-11}); eight band $k\cdot p$ model does not describe the
conduction band bending between $\Gamma$ and $L$ points that can be
seen in Fig.~\ref{fig-01}(b). Transitions between valence and
conduction band in the $L$-point that will contribute to
$\sigma(\omega)$ at latest around $\hbar\omega=3$~eV are absent in our
model calculations. However, strong experimental magneto-optical
signals in Fig.~\ref{fig-03} all appear below $\hbar\omega=2$~eV and
$L$-point transitions should be unimportant for such energies (see
again Fig.~\ref{fig-01}b). In the calculations we consider an extended
range of $\hbar\omega$ up to $3.3\unit{eV}$ to show features which, as
we explain below, would be shifted in realistic materials with strong disorder
to lower energies.  The two basic sample parameters that enter our
model are the carrier (hole) density $p$ and the effective doping $x$,
see Eq.~(\ref{eq-16}), that determine primarily the Fermi level $E_f$
and ferromagnetic splittings in Eq.~(\ref{eq-11}), respectively. Given
the span of $p$ and $x$ in Tab.~\ref{tab-01}, we first show in
Fig.~\ref{fig-04} calculated $\theta(\omega)$ for fixed $x=5\%$ and
varying $p$ (panel a) and fixed $p=0.8\unit{nm^{-3}}$ and varying $x$
(panel b). The order of magnitude of $\theta(\omega)$ and its
structure agrees with experimental data in Fig.~\ref{fig-03}. We
discuss and compare the individual spectral features in more detail
below and for convenience, we label three of them by Greek letters
$\alpha$, $\beta$ and $\gamma$ as shown in Fig.~\ref{fig-04}a. We
begin our discussion by identifying the optical transitions which are
responsible for the individual spectral features. Before that, we just
briefly remark that $\theta(\omega)$ comprises both MLB and MLD
contributions as we demonstrate in a simple example
below Eq.~(\ref{eq-29}) in Appendix~D.
 
\begin{figure}
\begin{tabular}{c}
  \includegraphics[scale=0.3]{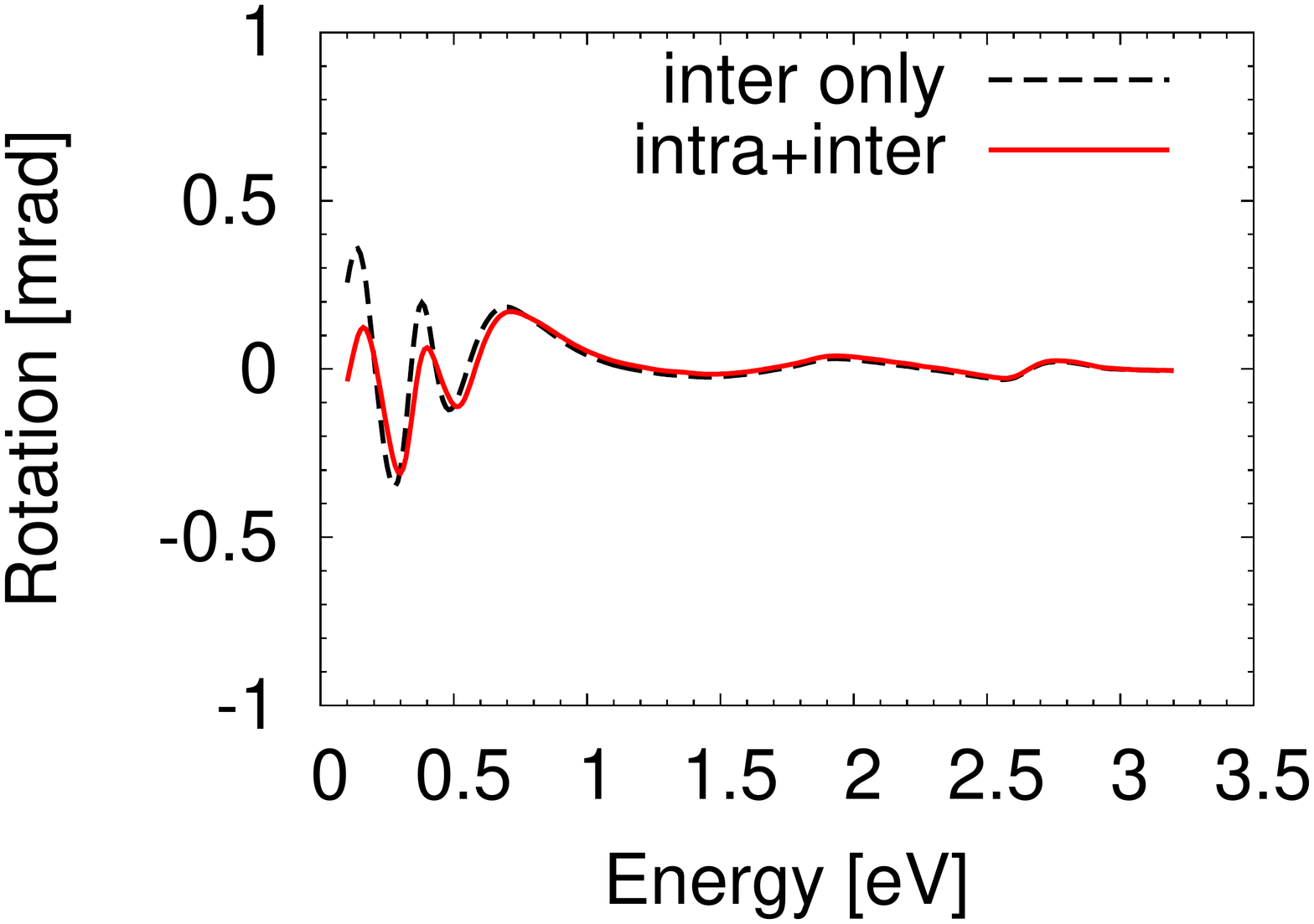} \\
  \includegraphics[scale=0.3]{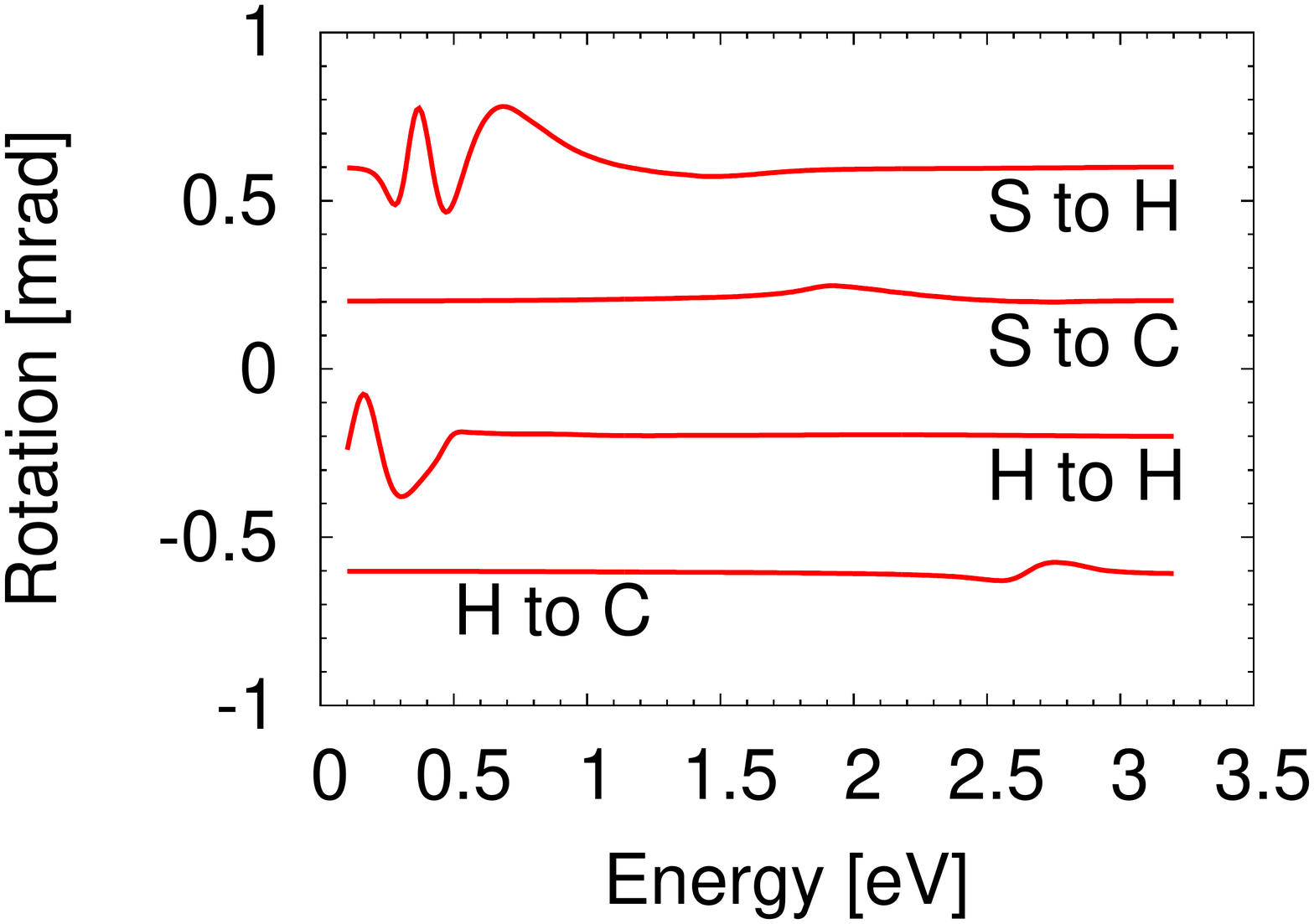}
\end{tabular}
\caption{Decomposition of $\theta(\omega)$ into individual terms
  appearing in the ac conductivity for $x=3\%, p=0.6\unit{nm^{-3}}$. 
  {\em Top:} $\theta(\omega)$ with and without intraband
  terms in Eq.~(\ref{eq-12}). {\em Bottom:} $\theta(\omega)$ calculated
  only using selected interband transitions in Eq.~(\ref{eq-19})
  (H=hole bands, C=conduction bands, S=split-off bands). Calculated 
  data in the
  bottom panel are vertically offset.} 
\label{fig-05}
\end{figure}

The conductivity that enters the rotation via reflection coefficients in
Eq.~(\ref{eq-13}) can be decomposed into contributions of
individual bands.  The relationship between $\theta$ and tensor components
of $\sigma$ is non-linear, yet it turns out that the individual
summands in $\sigma(\omega)$, see Eq.~(\ref{eq-19}), give rise to
well-defined structures in $\theta(\omega)$. The bottom panel in
Fig.~\ref{fig-05} demonstrates that peak $\alpha$ arises because of
optical transitions from the LH and HH bands (H) to the
conduction bands (C), peak $\beta$ is mostly due to transitions
between the split-off bands (S) and H while the intra-H transitions
underlie peak $\gamma$. The intra-band contribution to $\sigma$ has,
according to the top panel of Fig.~\ref{fig-05}, almost no
perceptible influence on the resulting $\theta(\omega)$, except for
the lowest energies ($\hbar\omega\approx 100$~meV). Sources of the
weak anisotropy of $\sigma^{\mathrm{intra}}$ are discussed in
Appendix~C below Eq.~(\ref{eq-18}).

The feature $\alpha$ seen consistently both in the model calculations
(Fig.~\ref{fig-04}) and experimental data (Fig.~\ref{fig-03}) is thus
largely due to optical transitions across the gap. Although our model
underestimates the effect of disorder in (Ga,Mn)As, this conclusion is
independent of the strength of disorder. Let us
take amorphous GaAs as an extreme example of a disordered system. 
Despite the completely destroyed translational symmetry, optical
properties such as photoemission spectra remain largely the same as
for perfect crystals. Important property of the amorphous material
underlying this similarity is the chemical bonding which is not very
different from the perfect crystal. Amorphous GaAs retains the so called Tauc
optical gap\cite{daSilva:2004_a,Tauc:1966_a} of the order of $E_g$.
The orbital character of states below the gap remains similar to the
perfect crystal and the main change\cite{note7} 
between the perfect crystal and amorphous material, using the language of the
former material, will be the presence of non-direct, $\vec{k}$
non-conserving, transitions in the amorphous material. We prefer the term
''non-direct'' to ''indirect'' to avoid confusion with phonon-mediated
transitions.  Since disorder generally
tends to reduce the gap\cite{daSilva:2004_a} and positions of peak
$\alpha$ in our experiments are consistently above the band-gap
in the perfect GaAs crystal ($E_g$), it implies that Fermi level in our
samples lies in the valence band as it is commonly
assumed.\cite{Jungwirth:2007_a} This conclusion is also supported by
a blue shift of peak $\alpha$ with increasing $x$, suggestive of the
Moss-Burstein shift.\cite{Moss:1954_a} From the point of view of our
model and with the help of the bottom panel of Fig.~\ref{fig-05}, peak
$\alpha$ in $\theta(\omega)$ arises from the direct transitions from
states at Fermi wavevector in H to states at the same wavevector in
C. Such transitions are shown by the vertical arrow labelled A in
Fig.~\ref{fig-01}b. Technically, the apparent conservation of
wavevector in a disordered system is a consequence of averaging over
impurity configurations (velocity operator matrix elements in 
Eq.~(\ref{eq-19}) are diagonal in $k$). To some extent, the
non-conservation of the wavevector is captured by the imaginary part
of the self-energy $\Gamma$ (see Appendix~C) but this is, strictly
speaking, only a correction justified in the weak-disorder 
regime. Given the relatively low sheet conductivities of our
samples,\cite{Jungwirth:2010_b} 
disorder corrections to Eqs.~(\ref{eq-19},\ref{eq-28})
may be sizable and the effect of non-direct transitions on
$\sigma^{\mathrm{inter}}$ larger than what is implied by Eq.~(\ref{eq-19}).

The direct transitions H$\to$C from the Fermi surface appear around
$\hbar\omega\approx E_g+|E_f|(1+m_H/m_C)$ where, for the sake of
illustration, we describe valence (conduction) bands by a single
effective mass $m_H$ ($m_C$).  This energy rapidly increases with
increasing $E_f$, i.e., with increasing hole density which in the
studied optimized (Ga,Mn)As samples is a monotonic increasing function
of $x_{\mathrm{nom}}$. This blue shift is so rapid for HHs
($m_H/m_C\sim 10$) that the corresponding spectral feature is even
out-of-range in Fig.~\ref{fig-04}. The actual transitions responsible
for peak $\alpha$ are those from the LHs to C
($m_H/m_C\sim 1$ by the order of magnitude) and even so, the blue
shift of the peak turns out to be much faster than what is observed
experimentally as we explain below (see also Fig.~\ref{fig-08}).  We
return to the non-direct transitions later and now discuss another
possible reason for the too high energies of the $\alpha$ peak in
Fig.~\ref{fig-04} as compared to the corresponding spectral feature in
the experimental Fig.~\ref{fig-03}.

Our (Ga,Mn)As samples are very heavily doped from the perspective of
traditional semiconductors, electron-electron interactions
can therefore appreciably contribute to the total 
energy.\cite{note6}  Indeed, the exchange energy per particle of free
spin-polarized electrons,
\begin{equation}\label{eq-15}
  E_x/N = -\frac{e^2}{4\pi\epsilon}\frac{3k_F}{4\pi}
\end{equation}
is of the order of 100~meV at carrier densities of the order of
$10^{21}\unit{cm^{-3}}$. The difficulty in evaluating the
exchange-correlation effects for the holes is in the presence of a
strong spin-orbit coupling. One possible approximative scheme is
discussed in Ref.~\onlinecite{Sipahi:1996_a}.  To assess the
qualitative effect of exchange energy on trends in the rotation
spectra of the Voigt effect in reflection, we use the following
scheme. We first disregard the correlation effects which are
small compared to exchange in Eq.~(\ref{eq-15}).  For
given $x$ and $p$, we first determine the band occupations by holes
$p_i$ ($\Sigma_i p_i=p$, $i=1,\ldots, 6$) as given by
Eq.~(\ref{eq-11}) with $V_{xc}=0$. For most of the considered dopings,
only the LH ($i=3,4$) and HH bands ($i=5,6$) are
occupied by holes. We next recalculate the corresponding densities
$p_{i}$ into Fermi wavevectors assuming isotropic dispersion and shift
the bands by $-E_x/N$ as given by Eq.~(\ref{eq-15}). Since $-E_x/N$ is
different for different bands, this procedure not only renormalizes
the Fermi level but also slightly changes $p_i$ and therefore we
iterate the procedure until we converge to a consistent set of $p_i$
and exchange shifts. Note that we neglected in this procedure the
exchange between bands $i\not=j$. To justify this approximation, at
least in part, we checked the spin-polarizations of individual
bands. For example, $x=3\%$, $p=0.6\unit{nm^{-3}}$ leads to
$p_{3,4,5,6}/p=0.02, 0.04, 0.33, 0.61$ and integral spin polarizations
$2\langle s\rangle_{3,4,5,6}=0.58, -0.14, 0.57, -0.91$. The
majority HHs are thus prevalent and nearly completely polarized,
hence their exchange interaction with holes in other bands will be
small and our estimate using Eq.~(\ref{eq-15}) with $k_F^3=6\pi^2 p_6$
should be a good approximation. On the other hand,
the exchange shifts for LH bands may contain sizable
corrections due to the neglected inter-band exchange and the values
52, 62, 126 and 154~meV thus serve only as a rough guide to assess
many-body effects on the magneto-optical spectra. These values are
similar to the band gap renormalization\cite{Zhang:2005_a} used
previously.\cite{Acbas:2009_a}  A commonly considered correction to
Eq.~(\ref{eq-15}) capturing part of the correlation effects is
logarithmic and weakly dependent on $p$ in our range of parameters. 
Appealing to the second term in Eq.~(36) of Ref.~\onlinecite{Sipahi:1996_a}
and the procedure described therein, we include it into our model through
a small constant shift of 6.5~meV (1.5~meV) for HH (LH) bands towards
the conduction bands. To summarize many-body corrections 
included in Eq.~(\ref{eq-11}), $\hat{V}_{xc}$ can be understood as a
single-particle operator that commutes with $\hat{H}$ and shifts the
selected bands as just described to account for exchange and partly
also correlations. This approximative treatment 
enhances the ferromagnetic splitting 
between minority and majority hole bands and also adds an additional
offset between the HH and LH bands.

\begin{figure}
\includegraphics[scale=0.32]{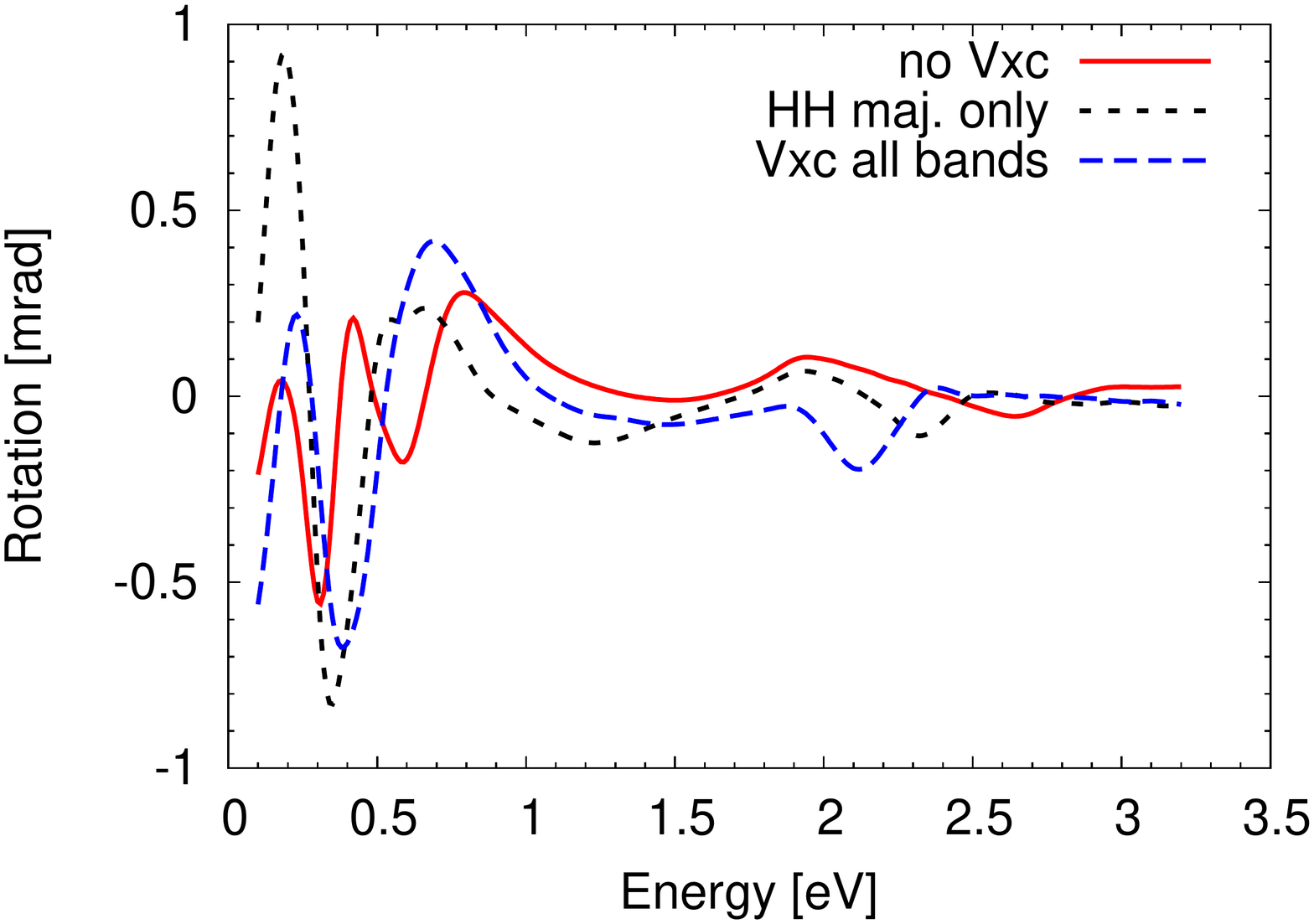}
\caption{Rotation $\theta(\omega)$ under various approximations 
  to $\hat{V}_{xc}$ for $x=5\%$,
  $p=0.8\unit{nm^{-3}}$. Note the position of the $\alpha$-peak:
  it appears at a relatively large energy for $\hat{V}_{xc}=0$ (''no
  Vxc'', corresponds to Fig.~\ref{fig-04}) and shifts to lower energies 
  when electron-electron exchange energy is taken into account. Adopting
  the approach of Ref.~\onlinecite{Acbas:2009_a} (''HH maj. only''), 
  we find the
  peak $\alpha$ around $2.3\unit{eV}$ and when applying the exchange
  shift to all bands (corresponds to Fig.~\ref{fig-07}), it
  shifts down to $2.1\unit{eV}$.}
\label{fig-06}
\end{figure}

\begin{figure}
\begin{tabular}{c}
\includegraphics[scale=0.3]{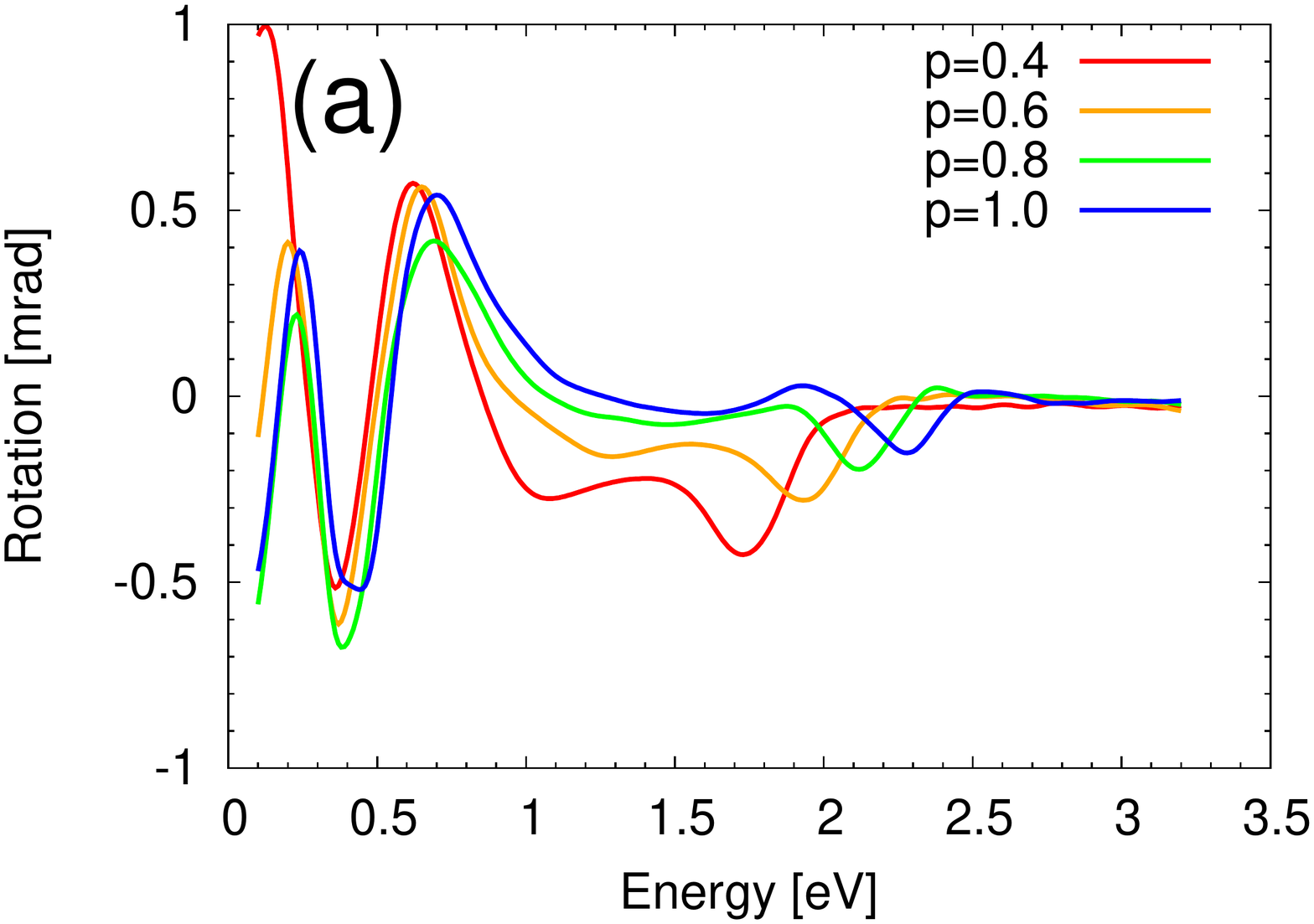} \\
\includegraphics[scale=0.3]{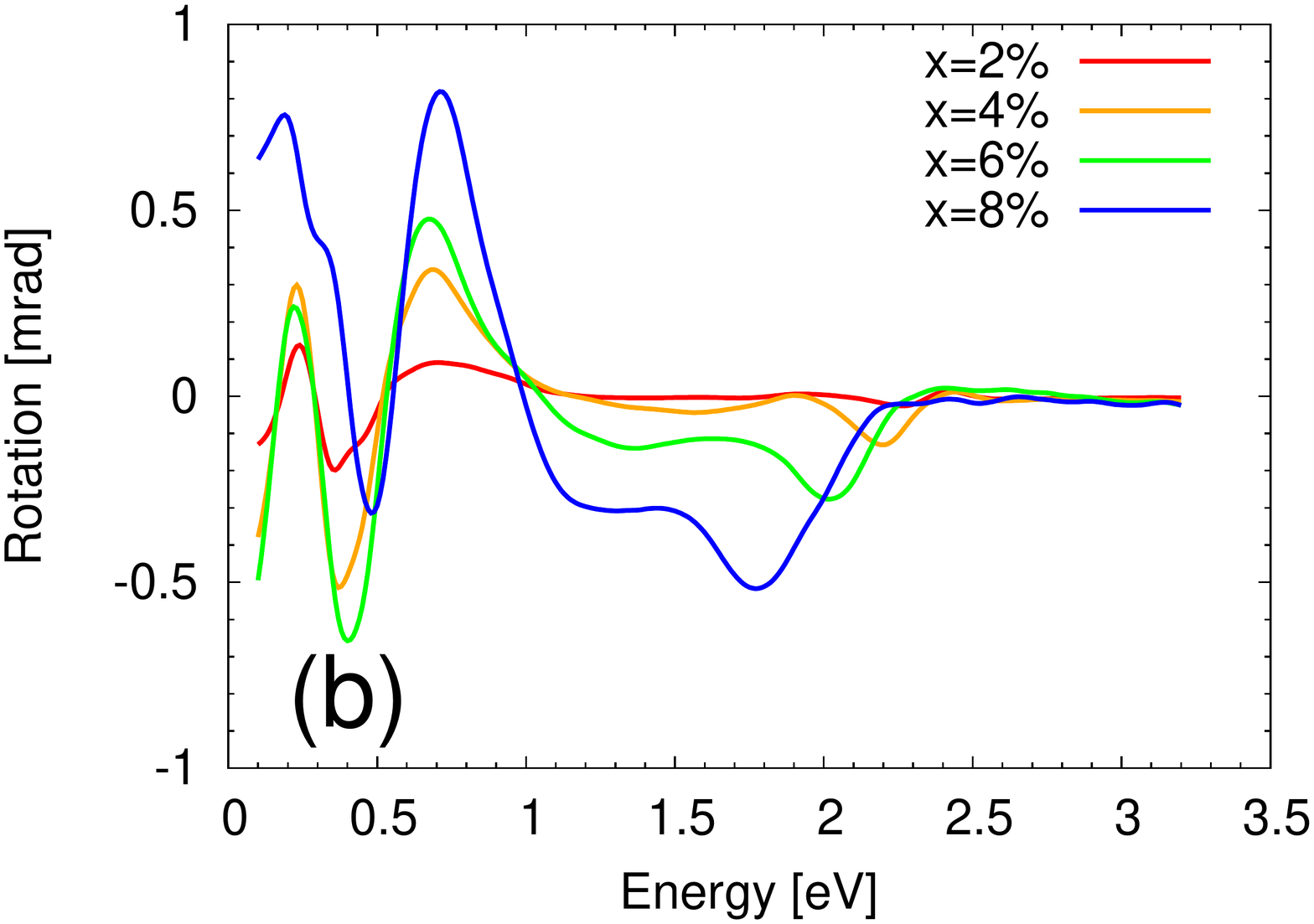}
\end{tabular}
\caption{Calculated spectral dependence of the rotation angle
  $\theta$, the same as Fig.~\ref{fig-04} but with $\hat{V}_{xc}$ taken into
  account as it is described below Eq.~(\ref{eq-15}).}
\label{fig-07}
\end{figure}

\begin{figure}
%\begin{tabular}{cc}
\includegraphics[scale=0.32]{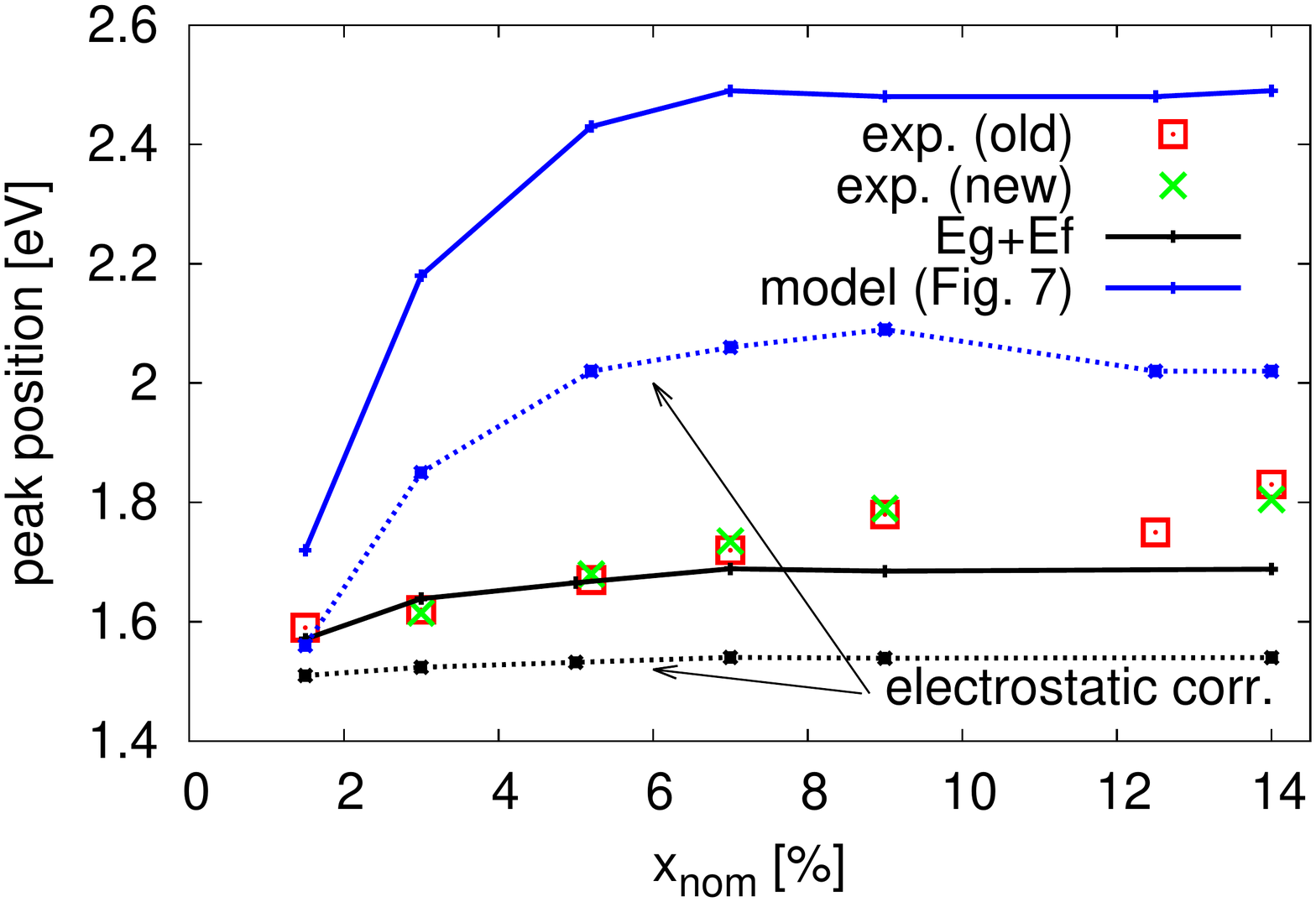}
%\end{tabular}
\caption{Positions of the $\alpha$-peak for the series of samples
  described by Tab.~\ref{tab-01}. Old and new experimental data
  correspond to results obtained by different techniques, peak
  positions were extracted from $\theta(\omega)$ data in
  Fig.~\ref{fig-03} (new) and from measurements below and above $T_c$
  described in Appendix~B (old). Model data (solid blue line) take
  into account direct transitions only, the extreme limit of
  non-direct transitions (solid black) corresponds to $E_g+E_f$ 
  where $V_{xc}$ is also included. Dotted lines show these two 
  limiting cases when electrostatic interaction with ionized 
  acceptors is considered.}\label{fig-08}
\end{figure}

Although the spectra with and without exchange-correction effects are
qualitatively similar, there are some notable differences. 
We observe a significant amplification and red shift of the peak
$\alpha$ by hundreds of meV, depending on the approximation as shown in
Fig.~\ref{fig-06}, and the double maximum structure around and above
$\hbar\omega=0.5\unit{eV}$ tends to merge into a single $\beta$-peak
structure. Absence of the double maximum in the range 0.5--1.2~eV in
experimental data of Fig.~\ref{fig-03} suggests that $\hat{V}_{xc}$
may be an important ingredient in the model. The peak $\alpha$ --- now
at smaller energies --- follows the same trends as with $\hat{V}_{xc}=0$:
it blue shifts with increasing $p$ and red shifts with increasing $x$ as shown
in Fig.~\ref{fig-07}. We checked these trends with the model 
of $\hat{V}_{xc}$ used previously by some of us for calculating
magneto-optical effects odd in magnetization\cite{Acbas:2009_a} 
where only majority HH band is exchange-shifted. To give an
impression, we display the corresponding spectrum as
the dotted line in Fig.~\ref{fig-06}. Finding the trends
independent of the approximation used for $\hat{V}_{xc}$, we proceed to use 
from now on the exchange shifts as described below Eq.~(\ref{eq-15})
which lead to $\theta(\omega)$ shown by dashed line in Fig.~\ref{fig-06}.

Even with the many-body band renormalization effects included, the
peak $\alpha$ still lies at considerably higher energy (above
$2\unit{eV}$) than what is observed experimentally (around
$1.7\unit{eV}$ in Fig.~\ref{fig-03}). We summarize its position in
Fig.~\ref{fig-08}. Experimental data from Fig.~\ref{fig-03} (crosses
in Fig.~\ref{fig-08}) and independent measurements described in
Appendix~B (empty boxes in Fig.~\ref{fig-08}) consistently show a slow
blue shift with increased nominal doping but the rate of this shift
with $x_{\mathrm{nom}}$ is much slower than what the mean-field
kinetic-exchange model predicts.  This is pointing to a shortcoming of
our electronic-structure model represented by Eq.~(\ref{eq-11}) where
the presence of Mn is effectively treated in a mean-field virtual-crystal
approximation. At the level of the Kubo formula in Eq.~(\ref{eq-19}),
our model only allows for direct transitions and below we continue the
discussion about how the experimental values of the peak $\alpha$
positions could possibly be explained by considering
non-direct transitions and electrostatic interaction between holes and
ionized acceptors.

Treating a realistic band-structure and disorder on equal footing is a
complicated task and we therefore discuss the effect of the latter
only qualitatively. It is important to keep in mind that the disorder
broadening $\Gamma$ which appears in the Kubo formula~(\ref{eq-19}) is
only a poor approximation to the non-conservation of wavevector
$\vec{k}$ in the strong-disorder case. Due to disorder in the crystal
caused primarily by random positions of Mn atoms substituting 
for the cations of the host
lattice, the Bloch theorem does not apply and $\vec{k}$ is not a good
quantum number. However, even in the extreme case of an amorphous
continuous covalent network discussed above, the valence and
conduction bands are largely preserved although the gap between them
may be smaller\cite{daSilva:2004_a} than the perfect-crystal value
$E_g$.  The lowest-energy optical transition would then appear close to
energy $E_f+E_g$ corresponding to the arrow labelled B in
Fig.~\ref{fig-01}b if we use the language of non-direct transitions
for the perfect-crystal band structure. In some sense, this could be
understood as calculating the band structure from Eq.~(\ref{eq-11})
and then replacing the matrix elements in Kubo formula~(\ref{eq-19})
by an expression that completely ignores $\vec{k}$ as opposed to
matrix elements diagonal in $\vec{k}$. Such estimate of the position
of peak $\alpha$ is its lower bound provided that we use the proper
value of $E_g$ reduced by the disorder.\cite{daSilva:2004_a} This
lower-bound property turns out to apply to our experimental data even
if we use the perfect-crystal value of $E_g$ reduced by $V_{xc}$ for
majority HH as it is shown in Fig.~\ref{fig-08} by the solid black curve. 
Admittedly, the experimental data are very close to this lower bound.

An effect that we have ignored in our discussion so far is the
electrostatic interaction between delocalized holes and ionized Mn
acceptors. The charge density of the latter is not constant as in the
jellium model and this causes an additional band-gap renormalization
that can be described by the real part of self-energy due to
hole-acceptor scattering.\cite{Zhang:2005_a}  We estimate it by 
Eq.~(5) of this reference with $g=1$ (full spin-polarization of the holes)
and HH effective mass of half the free electron 
mass as an additional shift of the valence bands towards the
conduction bands added to $\hat{V}_{xc}$. Positions of the peak $\alpha$
red shift by additional\cite{Jungwirth:2010_b} 
$\sim 300\unit{meV}$ as shown in
Fig.~\ref{fig-08} by the dotted blue line. For completeness, we
also show by dotted black line $E_g+E_f$ with $\hat{V}_{xc}$
included as well as the effect of band-gap renormalization due to the
ionized Mn acceptors described by Eq.~(5) in Ref.~\onlinecite{Zhang:2005_a}.
At this level, we can conclude that since the experimental data lie
approximately half-way between the lower and upper bounds delimited by
the dotted lines in Fig.~\ref{fig-08}, the non-direct transitions 
might play a significant role in optical transitions but band
renormalizations due to exchange-correlation and hole-acceptor electrostatic
interaction effects are also sizable. Quantitative modelling of experimental
magneto-optical data would require rigorous quantitative description 
of all these effects.

\begin{figure}
\begin{tabular}{c}
\includegraphics[scale=0.3]{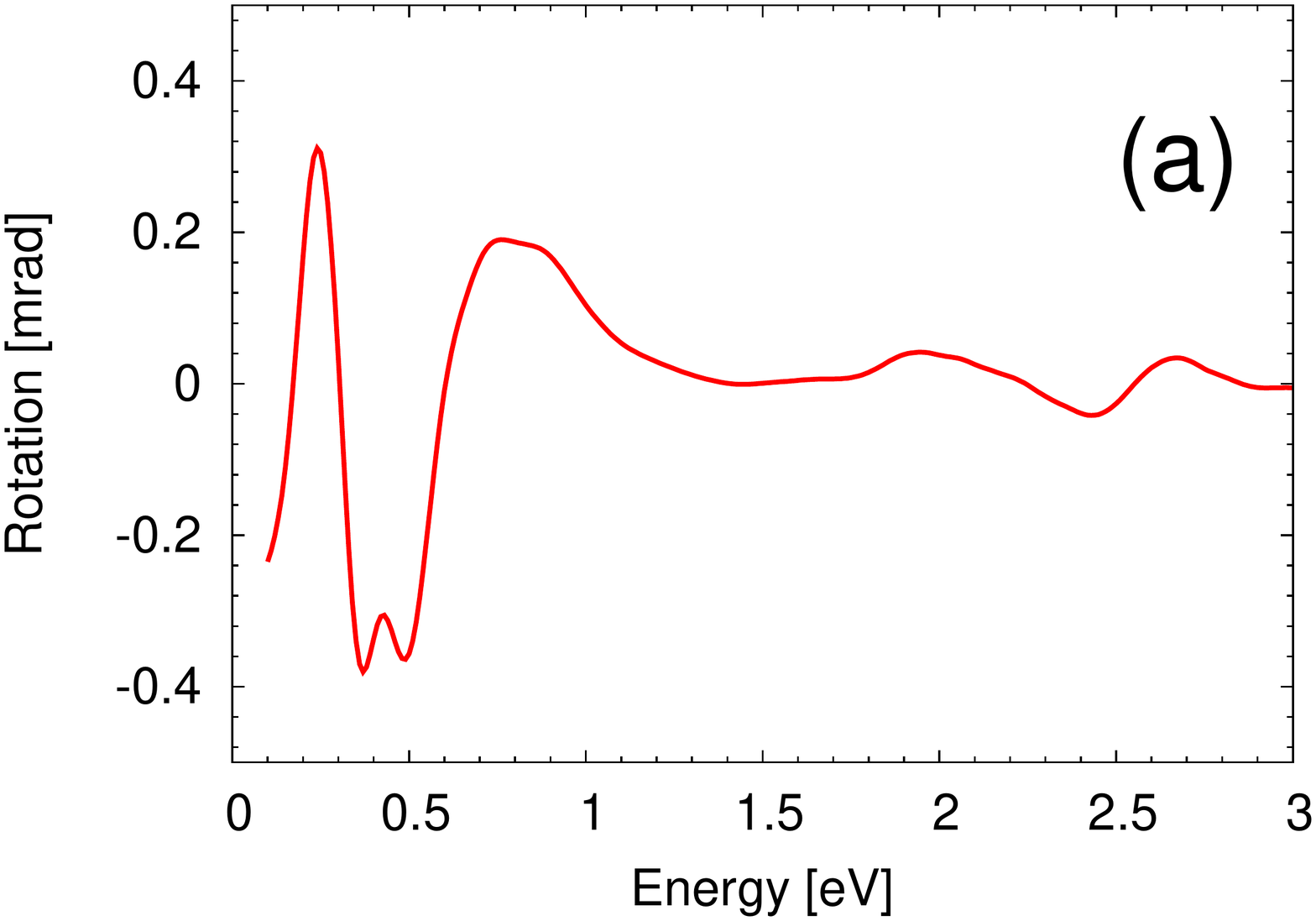} \\
\includegraphics[scale=0.3]{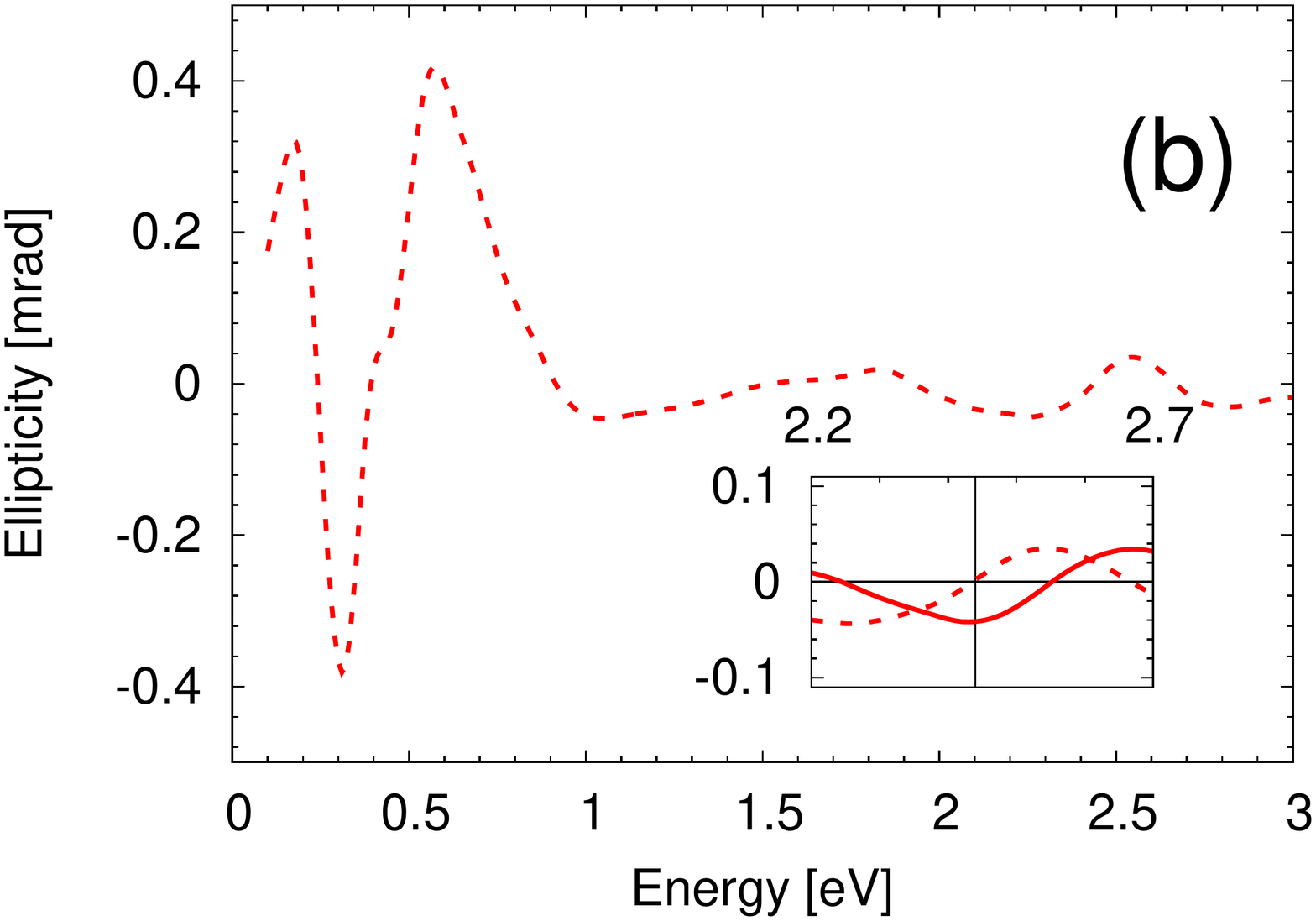}
\end{tabular}
\caption{Spectral dependence of (a) rotation and (b) ellipticity
  for $x=3.6\%$, $p=1.08$, with $\hat{V}_{xc}$ included. These
  parameters correspond to sample C ($x_{\mathrm{nom}}=5.2\%$). 
  The inset shows detail of rotation (solid) and ellipticity (dotted) 
  around peak $\alpha$.}
\label{fig-09}
\end{figure}

Returning to our model of direct transitions only, the experimental
feature where it performs relatively poorly are the peak amplitudes.
Although the predicted order of magnitude is correct (0.1~mrad for all
peaks $\alpha,\beta,\gamma$ in Figs.~\ref{fig-04},\ref{fig-07}), it is
notable that $\alpha$ is smaller than both $\beta$ and $\gamma$ while
the same peak $\alpha$ is by far the largest in experiments summarized in
Fig.~\ref{fig-03}. Smaller amplitude of peaks
$\beta,\gamma$ together with possibly larger linewidth in experiment
may be a consequence of disorder-induced non-direct transitions which
are not included in our model. Beyond the very crude treatment of
the chemical aspect of disorder in our model,\cite{note7}
we speculate that translational symmetry breaking
underlies the differences between the model and experiment also in the
case of peak $\alpha$. Our model shows that the amplitude
monotonically grows with $x$ across our set of samples.
% p.9 w06/mld/koresp/39/130303-mldpaper-figures.pdf
This is understandable since all magneto-optical effects must vanish
in the limit $x\to 0$ and actually, this decay is seen in experimental
data for samples C,B,A in Fig.~\ref{fig-10}b.  (We disregard the small
magneto-optical effects in non-magnetic GaAs which are present at
finite magnetic fields.\cite{Ando:1998_a}) On the other hand, disorder 
might play larger role at higher doping concentrations in metallic
samples sufficiently far from the metal-insulator transition
hence the decrease of peak $\alpha$ heights
for samples C,D,E. We note that the experimentally determined
extraordinarily large\cite{Buchmeier:2009_a} height of peak $\alpha$
($-0.86$~mrad) is about three times larger than the result of the
model based on Eq.~(\ref{eq-11}) (see Fig.~\ref{fig-09}a). It is
possible that a more refined choice of the scattering rates, instead
of a single parameter $\Gamma$ (see Appendix~C), could reduce this
difference but such analysis is beyond scope of this article.

We finally comment on the spectral dependence of ellipticity $\psi(\omega)$
which can also be readily calculated using Eq.~(\ref{eq-14}). Since
the ellipticity is experimentally somewhat more difficult to
access,\cite{Tesarova:2012_c} less data than for $\theta(\omega)$ is
available and we keep the following discussion short. Let us compare
Fig.~\ref{fig-09} to sample C in Fig.~\ref{fig-03}. Our model 
gives the correct
order of magnitude and functional shape of ellipticity related to the
peak $\alpha$ in rotation. The inset of Fig.~\ref{fig-09}b clearly
shows that Lorentzian peak $\alpha$ in rotation corresponds to an
anti-Lorentzian one in ellipticity which is found again in experimental data.
Regarding other spectral features in ellipticity, we find a minimum
close to $\hbar\omega=300\unit{meV}$ in Fig.~\ref{fig-09}b while no
such feature is observed experimentally. Given the complicated
structure of the valence bands and large shifts of the
$\alpha$-peak ascribed to disorder as discussed above, we will not
attempt to speculate on how this feature could be suppressed and only
note that the experimental magnitude of $\psi\approx 0.4\unit{mrad}$ is similar 
to calculations in Fig.~\ref{fig-09}b. Our model can therefore capture only
semi-quantitatively the major trends seen in the experimental
magneto-optical data.

\section{Conclusions}

Rotation $\theta(\omega)$ and ellipticity $\psi(\omega)$ measured for
the Voigt effect in reflection, a direct consequence of the magnetic
linear dichroism and birefringence, represent a much more sensitive
spectroscopic probe into the electronic structure of (Ga,Mn)As than,
for instance, unpolarized optical absorption experiments. Our measured
data are compatible with the previously published $\theta(\omega)$ on
selected samples and limited spectral range
and we investigate variations of the
spectra with manganese doping which influences both the
exchange-splitting and Fermi level. We confirm that $\theta$ at
energies exceeding the gap of the GaAs host can reach values larger
than so far reported in other ferromagnetic materials.
The corresponding peak is found to
blue shift with increasing manganese doping and we analyse this
trend using the $k\cdot p$ mean-field kinetic-exchange model.
We find that even with exchange-correlation band renormalization
effects taken into account,
this model yields appreciably larger energies at which this feature is
seen, compared to experiment, and we attribute this fact to the
neglected of non-direct transitions caused by the disorder. Apart from
this deficiency, the model correctly reproduces the structure of
experimental $\theta(\omega)$ and $\psi(\omega)$ ranging from
$112$~meV to $2.7$~eV, captures the sign of the peaks and,
semi-quantitatively, also their amplitude. A more quantitative
description of the measured magneto-optical spectra would require to
combine the modeling of the complex, spin-orbit coupled band structure
with a more detailed treatment of the strong disorder effects in
(Ga,Mn)As, as previously done, e.g., in the studies of unpolarized
absorption spectra.\cite{Yang:2003_b}

\section*{Acknowledgments}

We thank Jan Zemen and Pavel Motloch for providing tight-binding data
for the band structure in Fig.~\ref{fig-01} and Jiajun Li for
preliminary calculations of non-direct optical transitions in
disordered crystals. Communication with Rudolf Sch\"afer and
Vladim\'\i{}r Kambersk\'y helped to clarify terminology.  Thanks for
helpful discussions are also due to Carsten~A.~Ullrich, Alexander
Khaetskii, Andreas Dirks, Jong Han, Igor \v Zuti\'c, Florian Eich,
J\"org Wunderlich, Jan Kune\v s and very specially, to Jan Ma\v
sek. Support of the Academy of Sciences of the Czech Republic via
Praemium Academiae and funding from the ERC Advanced Grant 268066 is
gratefully acknowledged.  Work done at the University at Buffalo was
supported by NSF-DMR1006078, by the US Department of Energy, Office of
Basic Energy Sciences, Division of Materials Sciences and Engineering
under Award DE-SC0004890 and NSF ECCS-1102092.  We also acknowledge
funding by FAPESP (\#~2011/19333-4) and CNPq (\#~246549/2012-2), M\v
SMT (grant Nr. LM2011026), U.S. agencies through ONR-N000141110780,
NSF-DMR-1105512, Grant Agency of the Czech Republic through grant
No.~P204/12/0853 and Grant Agency of Charles University in Prague
through grant No.~443011.

\begin{appendix}

\section{Classical theory of magneto-optical effects}

Maxwell's equations allow to show how the magneto-optical effects 
described in Sec.~II follow from properties of the bulk magnetic
material. Inspired by the argumentation of Ref.~\onlinecite{Osgood:1998_a}, 
we review in this Appendix how MLD/MLB (or their circular 
counterparts), i.e. difference in 
imaginary/real parts of the refractive indices for two linearly
(circularly) polarized modes, is calculated from bulk ac conductivity
tensor of the material. Relation between these refractive indices and
the particular magneto-optical effects is also explained here using
simple examples and we refer the reader to Appendix~D for a
discussion of the more realistic relationship pertaining to our measurements.

Consider an electromagnetic wave $\vec{E}(\vec{r},t)=\vec{E}_0
e^{i(kz-\omega t)}$ propagating along $\vec{k}\parallel\hat{z}$. 
The non-zero ac conductivity $\sigma(\omega)$ and Maxwell
equations imply that
\begin{equation}\label{eq-01}
\nabla (\nabla\cdot\vec{E}) - \nabla^2 \vec{E}
  = -\mu\sigma \dot{\vec{E}} -\mu\ve\ddot{\vec{E}}
\end{equation}
which yields an equation for $\vec{E}_0$ whose solutions correspond to
propagating modes when $n^2\equiv(ck/\omega)^2>0$ where $c$ is the
light velocity.  Character of these modes depends on the form of the 
permeability $\mu$, permittivity $\ve$ and conductivity
$\sigma$ tensors. The right-hand-side of Eq.~(\ref{eq-01}) takes on
the form $\omega^2\mu\veeff \vec{E}(\vec{r},t)$ where $\veeff$ can be
written as $\ve_0 + i\sigma/\omega$ if we replace $\ve$ by vacuum
permittivity $\ve_0$ or as in Eq.~(\ref{eq-10}),
depending on how the ambiguity discussed below Eq.~(\ref{eq-10}) is
resolved. We obtain the modes by solving Eq.~(\ref{eq-01}) and
their $\vec{E}_0$ and refractive indices $n$ depend on the form of 
the effective permittivity tensor $\veeff$.
We now consider two examples related to
the magnetization-in-plane and out-of-plane magneto-optical
experiments discussed in Sec.~II. The permeability
$\mu$ will from now on be considered a scalar equal to the vacuum
permeability % p.139 in Sugano/Kojima
and a material of cubic
symmetry will be assumed whose index of refraction in the absence of
magnetization equals $n_0=\sqrt{\veeff/\ve_0}$.

In the first example, $\vec{M}\parallel\hat{z}$ which 
implies\cite{note3} an effective permittivity tensor of the form
\begin{equation}\label{eq-02}
  \veeff = \ve_0
  \left(\begin{array}{ccc}
    \ve_{xx} & \ve_{xy} & 0 \\
    -\ve_{xy} & \ve_{xx}& 0 \\
    0 & 0 &\ve_{zz}
  \end{array}\right)
\end{equation}
with dimensionless components $\ve_{ij}$. The eigenmodes obtained by
solving Eq.~(\ref{eq-01}) are two circularly polarized waves with
\begin{equation}\label{eq-03}
  \left(\begin{array}{c}E_x^0\\ E_y^0\\ E_z^0\end{array}\right)
  \propto  \left(\begin{array}{c}1\\ i\\ 0\end{array}\right)\mbox{ and }
  \left(\begin{array}{c}1\\-i\\ 0\end{array}\right)
\end{equation}
for $n^2_+=\ve_{xx}+i\ve_{xy}$ and $n^2_-=\ve_{xx}-i\ve_{xy}$.  
Let us now explain how the Faraday effect arises in such a
situation, sketched in Fig.~\ref{fig-02}a.
Consider a slab of a magnetic material of thickness $d$
described by $\veeff$ in Eq.~(\ref{eq-02}) placed in vacuum, assume
normal incidence and, for simplicity, the absence of reflections on
the vacuum-sample surface. An incoming linearly polarized wave with 
$\vec{E}_0=(E_x^0,0,0)$ will propagate through the sample in two
circularly polarized modes at different group velocities. Under the
additional (typically satisfied) assumption $|n_\pm-n_0|\ll n_0$, we can
conclude using 
\begin{equation}\label{eq-04}
  n_+ - n_- \approx \frac{n_+^2-n_-^2}{2n_0} = \frac{i\ve_{xy}}{n_0}
\end{equation}
that the polarization plane of the outgoing wave will be rotated by 
$\theta\approx -(d\omega /c) \Ima \ve_{xy}/n_0$. In this
transmission geometry (and under the simplifying assumption on surface
reflections), the Faraday rotation is directly related to magnetic
circular birefringence (MCB) while magnetic circular dichroism (MCD)
will make the outgoing wave elliptically polarized ($\psi\not=0$).

The (polar) Kerr effect is obtained by considering reflection off an
interface between a semi-infinite magnetic material and vacuum.  The
Fresnel formula~(\ref{eq-05}) for the reflection coefficient $r$ at
normal incidence (ratio of outgoing to incident beam's $E_x^0$) reads
$r=(1-n)/(1+n)$ and applying it to $n_\pm$ defined below
Eq.~(\ref{eq-03}), we obtain $r_\pm$ for the two circularly polarized
modes. For $r_+/r_-=ae^{i\xi}$ (with $a,\xi$ real), the originally
linearly polarized wave will be reflected as elliptically polarized
(unless $a=1$) with the major axis rotated by $\theta=\xi$ (see
Fig.~\ref{fig-01}). In a general case, it is not possible to link MCB
alone directly to the rotation and unlike with the Faraday effect,
both MCB and MCD will influence $\theta$ because the relation between $r$
and $n$ is non-linear. An illustrative example of this is given in
Appendix~D.

In the second example $\vec{M}\parallel\hat{x}$, which implies the same form of
$\veeff$ as in Eq.~(\ref{eq-02}) up to a permutation of indices:
\begin{equation}\label{eq-06}
  \ve_0
  \left(\begin{array}{ccc}
    \ve_{xx} & 0 & 0 \\
    0 & \ve_{zz} & \ve_{yz} \\
    0 &-\ve_{yz} & \ve_{zz}
  \end{array}\right).
\end{equation}
Solving Eq.~(\ref{eq-01}) for $\vec{E}_0$ gives
\begin{equation}\label{eq-07}
  \left(\begin{array}{c}E_x^0\\ E_y^0\\ E_z^0\end{array}\right)
  \propto  \left(\begin{array}{c}1\\ 0\\ 0\end{array}\right)\mbox{ and }
  \left(\begin{array}{c}0\\1\\ \ve_{yz}/\ve_{zz}\end{array}\right)
\end{equation}
with refractive indices $n_\parallel^2=\ve_{xx}$ and
$n_\perp^2=\ve_{zz}[1+(\ve_{yz}/\ve_{zz})^2]$. Voigt rotation
(after transmission through a slab of the magnetic material as
sketched in Fig.~\ref{fig-02}b) is related to 
\begin{equation}\label{eq-08}
  n_\parallel-n_\perp 
  \approx\frac12 n_0(\ve_{xx}-\ve_{zz}-\frac{\ve_{yz}^2}{\ve_{zz}})
\end{equation}
in analogy to Eq.~(\ref{eq-04}) and polarization plane rotation in the
Voigt effect in reflection
(assuming $\beta=\pi/4$ and $b=r_\parallel/r_\perp$ real for
simplicity) follows from 
\begin{equation}\label{eq-09}
  \tan\theta = \frac{1-b}{1+b} \approx \frac{n_0}{2(n_0^2-1)}
  (\ve_{xx}-\ve_{zz}-\frac{\ve_{yz}^2}{\ve_{zz}}).
\end{equation}
For other mutual positions of $\vec{M}$ and polarization plane,
$\theta$ will follow the $\sin 2\beta$ dependence as mentioned in
Sec.~II. In particular, when incident beam polarization is parallel or
perpendicular to $\vec{M}$, light in the magnetic material travels
simply as the first or second mode in~(\ref{eq-07}) and the
polarization remains unchanged. 

\begin{figure*}
\begin{tabular}{cc}
\includegraphics[scale=0.3]{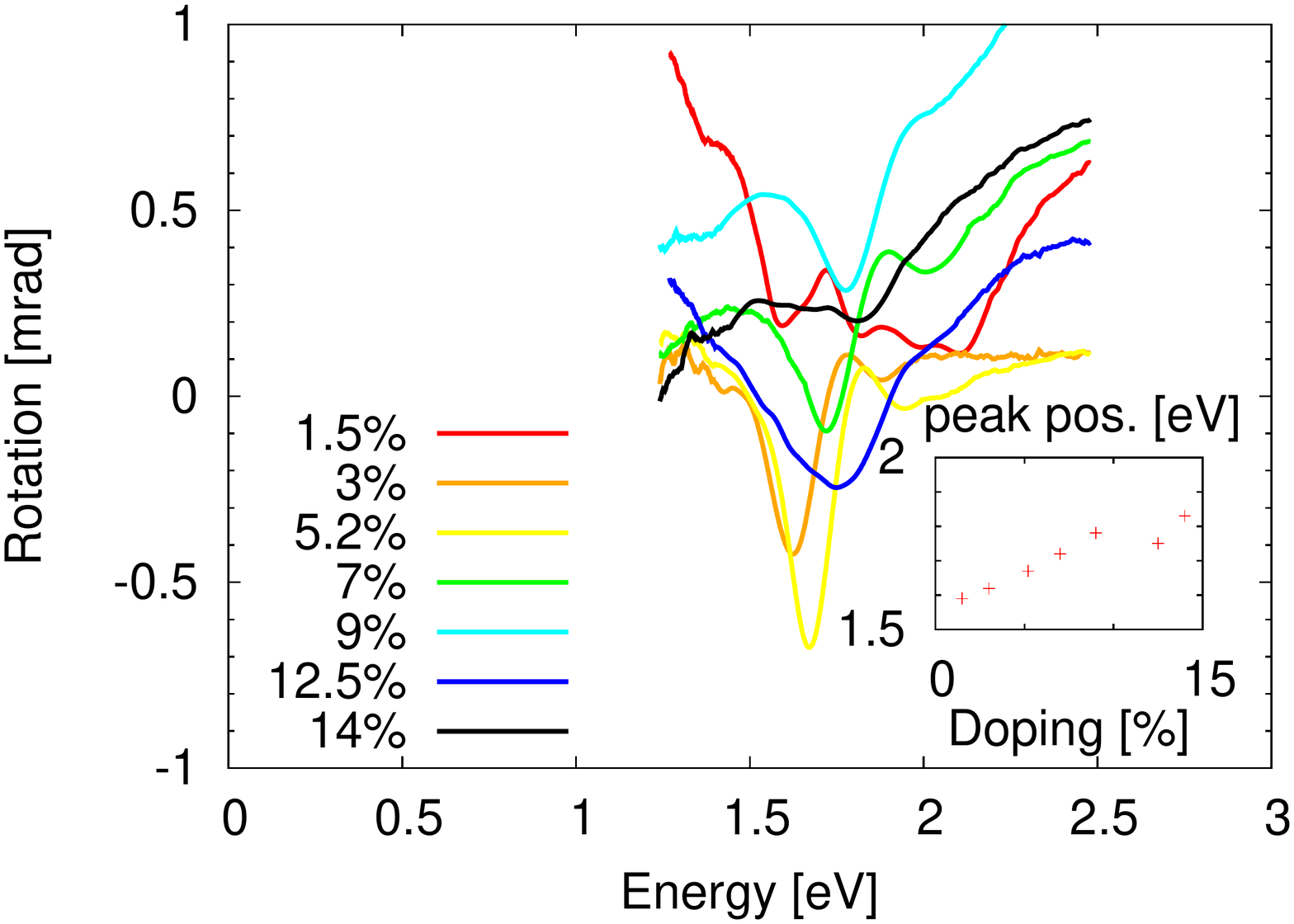} &
\includegraphics[scale=0.3]{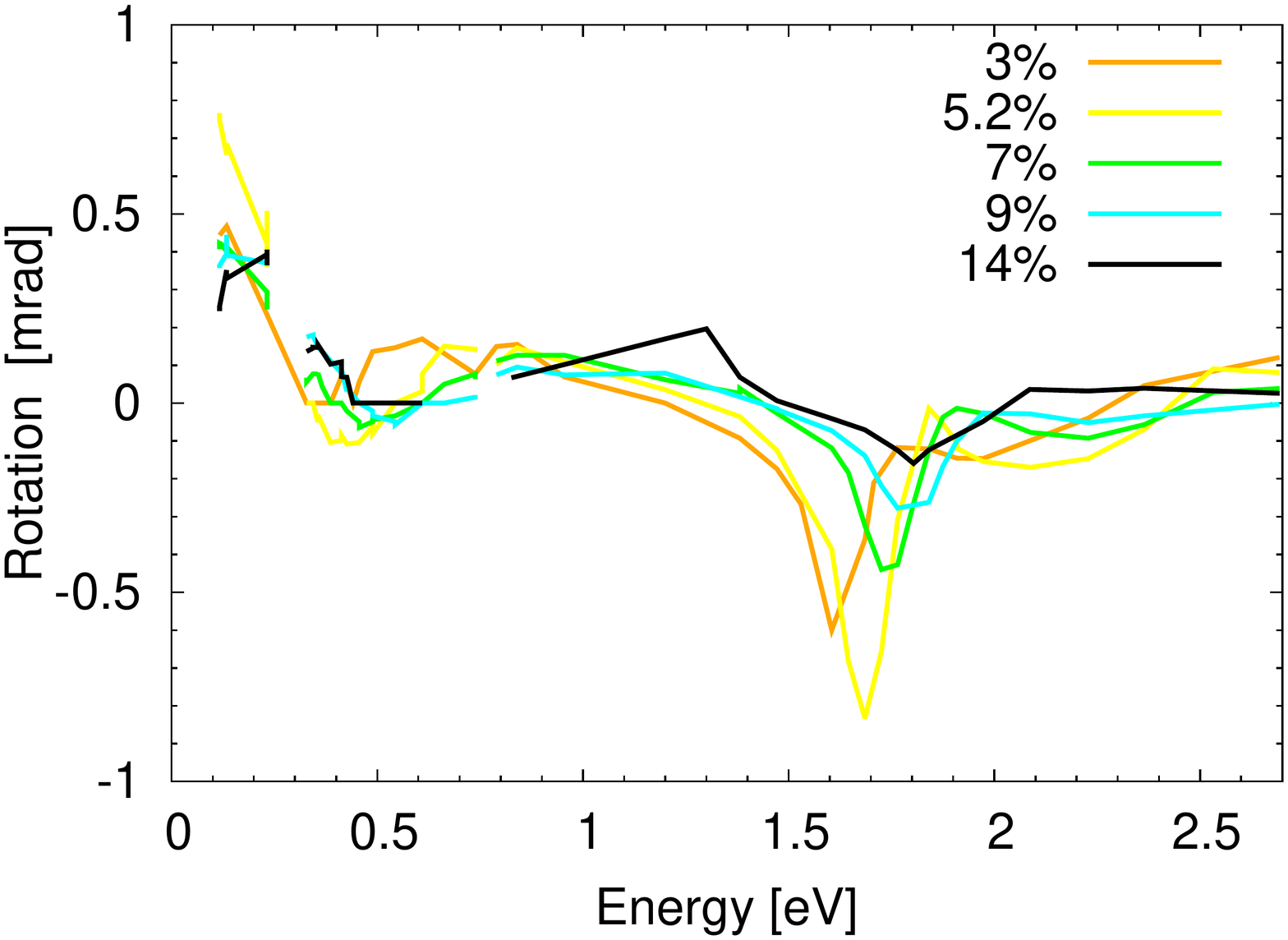} \\
(a) & (b)
\end{tabular}
\caption{Overview of the measured rotation angle $\theta$ for samples
  A--G using (a) the subtraction of data above $T_c$ and (b) the in situ
  rotation of $\vec{M}$. Inset in panel (a) shows the position in eV of
  the peak $\alpha$ as a function of Mn doping. All Mn concentrations
  indicated are $x_{\mathrm{nom}}$ (see Tab.~\ref{tab-01}).}
\label{fig-10}
\end{figure*}

With these two examples at hand, we can make several
observations. Recall that we have always considered the normal
incidence here. The in-plane magnetization leads to magneto-optical
effects even in magnetization, $\theta(\vec{M})=\theta(-\vec{M})$, as
stated in Sec.~II.  In Eq.~(\ref{eq-08}), $\ve_{xx}-\ve_{zz}$ is even
in $\vec{M}$ owing to the Onsager relations, and $\ve_{yz}^2$ is even
because $\ve_{yz}(\vec{M})$ is odd.\cite{note4} Next, we can see that
a non-zero difference between $n_{\parallel}$ and $n_\perp$ in
non-dissipative systems ($\Ima n_{\parallel}=\Ima n_\perp=0$), a
circumstance that could be called ''pure MLB'', causes rotation in the
Voigt effect in reflection.  However, as soon as $n_{\parallel}$ and
$n_\perp$ are complex, both MLB and MLD will influence $\theta$
because of the non-linear dependence of $r$ on $n$ in
Eq.~(\ref{eq-05}). We again refer to the illustrative example given in
Appendix~D. Similar statement holds about ellipticity of the Voigt
effect in reflection.  Some confusion can arise because of different
terminology used in the literature: Ref.~\onlinecite{Ferre:1984_a}
relates MLB to the real part of refractive indices while
Ref.~\onlinecite{Kimel:2005_a} to the real part of the reflection
coefficients.  We find the former terminology more appropriate because
it is generic for both reflection and transmission
coefficients. Independent of the terminology, it is safe to state that
different complex refractive indices $n_\parallel$ and $n_\perp$ cause
$\theta\not=0$, $\psi\not=0$ in both transmission and reflection
experiments. Nonzero $n_\parallel - n_\perp$ arises due to difference
in diagonal components of $\veeff$ or nonzero $\ve_{yz}$, as seen in
Eq.~(\ref{eq-08}). Since $\ve_{yz}^2/\ve_{zz}$ is in our case
negligible,\cite{note5} one can conclude that the Voigt effect in
reflection or in transmission is (via MLB and MLD) primarily driven by
the difference of diagonal components of $\sigma(\omega)$
corresponding to directions parallel and perpendicular to $\vec{M}$,
i.e., by the ac anisotropic magnetoresistance.

We conclude this Appendix by explaining the relationship between
terminology used in this article (components of the effective
permittivity tensor $\veeff$) and the notation of ''quadratic
magneto-optic tensor components''\cite{Birss:1964, Visnovsky:1986_a} used
elsewhere.\cite{Postava:2002_a, Bhagavantam:1966, Hamrlova:2013_a} The
basic conceptual difference between the two approaches is whether
$\vec{M}$ is kept fixed and different polarizations of light are
considered (the former approach) or vice versa (the latter approach).
An advantage of the latter approach is its aptitude to describe the ac
analogy of crystalline anisotropic magnetoresistance
components\cite{Ranieri:2008_a} which we completely ignore in this
article, motivated by their smallness in the dc
limit.\cite{Rushforth:2007_a} We expand the effective permittivity
tensor into a Taylor series in powers of the magnetization Cartesian
components $M_k$:
\begin{equation}\label{eq-31}
\varepsilon_{ij} 
 = \varepsilon_{ij}^{(0)}+K_{ijk}M_k+G_{ijkl}M_kM_l+\ldots 
\end{equation}
where $\varepsilon_{ij}^{(0)}$ is the part independent on
magnetization, $K_{ijk}$ and $G_{ijkl}$ are rank three and
four tensors and the last two are sometimes 
also called linear and quadratic magneto-optical
tensors. They represent the parts of the permittivity tensor which are
linear and quadratic in magnetization, respectively.

The form of $K_{ijk}$ and $G_{ijkl}$ depends on the symmetry of the
crystal\cite{Visnovsky:1986_a} as well as on the orientation of
principal crystal axis with respect to $xyz$-axis in which the
permittivity tensor is expressed.\cite{Hamrlova:2013_a} In case of
cubic crystals with point symmetry (crystal classes 23=T, m3=T$_h$,
432=O, $\overline{4}$3m=T$_d$ and m3m=O$_h$) where $\langle
100\rangle$, $\langle 010\rangle$ and $\langle 001\rangle$ are
parallel with $x$, $y$ and $z$-axis, respectively, following three
statements hold.  Non-magnetic part of the permittivity tensor is
constant, $\varepsilon_{ij}^{(0)}=\delta_{ij}\varepsilon^{(0)}$, where
$\delta_{ij}$ is the Kronecker delta. The third rank tensor $K_{ijk}=
\gamma_{ijk} K$ where $\gamma_{ijk}$ is Levi-Civita symbol. The 
rank four tensor $G_{ijkl}$ can be written in matrix form
as\cite{Visnovsky:1986_a}
\begin{equation}
\label{eq-32}
\left( \hskip-.5mm
\begin{array}{c}
\varepsilon_{xx}^{(2)}\\
\varepsilon_{yy}^{(2)}\\
\varepsilon_{zz}^{(2)}\\
\varepsilon_{yz}^{(2)}\\
\varepsilon_{zx}^{(2)}\\
\varepsilon_{xy}^{(2)}
\end{array}
\hskip-.5mm\right) \hskip-1mm = \hskip-1mm \left(
\begin{array}{cccccc}
G_{11} & G_{12} & G_{12} & 0 & 0 & 0 \\
G_{12} & G_{11} & G_{12} & 0 & 0 & 0 \\
G_{12} & G_{12} & G_{11} & 0 & 0 & 0 \\
0 & 0 & 0 & 2G_{44} & 0 & 0 \\
0 & 0 & 0 & 0 & 2G_{44} & 0 \\
0 & 0 & 0 & 0 & 0 & 2G_{44} 
\end{array}
\right)\hskip-1mm \left( \hskip-.5mm
\begin{array}{c}
M_x^2 \\
M_y^2 \\
M_z^2\\
M_yM_z\\
M_zM_x\\
M_xM_y
\end{array}
\hskip-.5mm\right)
\end{equation}
where $\ve_{ij}^{(2)}=G_{ijkl}M_kM_l$.
In the case of an isotropic material, the number of free 
parameters is further reduced because $2G_{44}=G_{11}-G_{12}$. 

In our analysis in Sec.~IV, magnetization was always oriented along
$x$-axis and combining Eq.~(\ref{eq-32}) and~(\ref{eq-31}), we arrive
at
\begin{equation}\label{eq-33}
 \varepsilon=
 \left(
 \begin{array}{ccc}
   \varepsilon^{(0)} + G_{11} M_x^2 & 0 & 0 
   \\
   0 & \varepsilon^{(0)} + G_{12} M_x^2 & K M_x
   \\
   0 & -K M_x & \varepsilon^{(0)} + G_{12} M_x^2
 \end{array}
 \right).
\end{equation}
Repeating the analysis leading to Eq.~(\ref{eq-08}), we now find
$n_\parallel-n_\perp=(G_{11}-G_{12}-K^2/\varepsilon^{(0)})M_x^2$.

\section{Additional experimental data}

%%% fig-10 a few lines above...

As it is discussed in Sec.~III and in Ref.~\onlinecite{Tesarova:2012_c},
a difficulty related to the measurement of magneto-optical
phenomena even in $\vec{M}$ is that the actual signal cannot be separated from
background simply by subtracting the results in magnetic field $B$ and
$-B$. One possible approach is to subtract results at $T>T_C$
from those at the low temperature of interest.
Phenomena related to magnetism are suppressed at $T>T_C$ and the
remaining signal stemming from the experimental apparatus is still often
large, see Fig.~3b in Ref.~\onlinecite{Tesarova:2012_c}.
Our measurements using this technique are
summarized in Fig.~\ref{fig-10}a (note that with this technique, we
measured also samples A and F not available Fig.~\ref{fig-03}). 
Even between two measurements of the same sample, inferred
$\theta(\omega)$ may be offset because of temperature-dependence in
optical properties of measurement setup elements. On the other hand,
the technique\cite{Tesarova:2012_c} of in situ rotating $\vec{M}$ is
free of these artefacts as it is apparent from the $\theta(\omega)$
data of Fig.~\ref{fig-03} summarized in Fig.~\ref{fig-10}b. The peak
positions (given in the inset of the panel a) agree well between the
two methods --- compare the two sets of experimental data in
Fig.~\ref{fig-08}.

\section{Microscopic model}

This appendix contains detailed information about the model of
(Ga,Mn)As electronic structure embodied in Eq.~(\ref{eq-11}), its
parameters and the Kubo formula used to calculate conductivity tensor
components entering Eq.~(\ref{eq-12}).

Individual samples are primarily characterized by the Mn doping $x$
(fraction of Ga atoms substituted by Mn) and total hole density $p$.
The former is taken as $x=N_{\mathrm{Mn}}a_l^3/4$ where
$a_l=0.565325\unit{nm}$ is the GaAs lattice constant and
$N_{\mathrm{Mn}}$ is the density of Mn atoms. Since Mn substituting
for a Ga atom is a single acceptor, it follows $p=N_{\mathrm{Mn}}$ and
$|\vec{M}|=5\mu_B N_{\mathrm{Mn}}$ in the ideal case (for the moment,
we neglect magnetic moment of the holes, included in Eq.~(\ref{eq-16})
below). However, compensating impurities (e.g. As antisites or Mn
atoms in interstitial position) will reduce both $p$ and magnetization
$|\vec{M}|$. These two quantities therefore have to be determined
independently by measurement as it is done in Fig.~10 and Tab.~I of
the Supplemental Information in
Ref.~\onlinecite{Jungwirth:2010_b}. For our article, the nominal
doping $x_{\mathrm{nom}}$ serves only as a convenient ''label'' of the
samples summarized in Tab.~\ref{tab-01}. We take $p$ directly from
Ref.~\onlinecite{Jungwirth:2010_b} and using the values of $M_{sat}$
from the same source, we calculate the effective doping
\begin{equation}\label{eq-16}
  x=\frac{M_{sat}a_l^3}{8(S_{\mathrm{Mn}}+S_{\mathrm{carr}})\mu_B}
\end{equation}
which is also given in Tab.~\ref{tab-01}. The Mn magnetic moment 
$S_{\mathrm{Mn}}=5/2$ dominates $M_{sat}$, carriers contribute by a
smaller part and we take $S_{\mathrm{carr}}=-0.25$ because the
(incompletely polarized\cite{Piano:2010_a}) hole spins are oriented
antiparallel to those of the Mn. Using this $x$, we calculate
$M=|\vec{M}|$ in Eq.~(\ref{eq-11}) as $8x S_{\mathrm{Mn}}\mu_B/a_l^3$.
Note that Eq.~(\ref{eq-16}) basically expresses the notion that in
annealed metallic samples there are approximately $4.5$~Bohr magnetons
per manganese atom.\cite{Jungwirth:2005_a}

Our $\hat{H}_{KL}$ in Eq.~(\ref{eq-11}) is the eight-band
Kohn-Luttinger Hamiltonian identical to the corresponding block in Eq.~(2) of 
Ref.~\onlinecite{Hankiewicz:2004_a}. We use GaAs Luttinger parameters
$\gamma_{1/2/3}=6.98/2.06/2.93$ together with
$\Delta_{SO}=341\unit{meV}$, $E_g=1.519\unit{eV}$,
$E_P=2m_0P^2/\hbar^2=24.8\unit{eV}$, $m_c^*=0.067m_0$ where $m_0$ is
the electron vacuum mass. The middle two terms in
Eq.~(\ref{eq-11}) describe the ferromagnetic splitting in our model.
When $\vec{M}||\hat{x}$, as we always assume in our calculations, they
combine into an $8\times 8$ matrix $h \hat{m}$ where
\begin{equation}\label{eq-17}
  \hat{m}=\left(\begin{array}{cccccccc}
    0&0&\frac{\sqrt3}{2}&0&\frac{\sqrt3}{\sqrt2}&0&0&0 \\
    0&0&1&\frac{\sqrt3}2&\frac{-1}{\sqrt2}&0&0&0 \\
    \frac{\sqrt3}2&-1&0&0&0&\frac{-1}{\sqrt2}&0&0\\
    0&\frac{\sqrt3}2&0&0&0&-\frac{\sqrt3}{\sqrt2}&0&0\\
    \frac{\sqrt3}{\sqrt2}&\frac1{\sqrt2}&0&0&0&\frac12&0&0\\
    0&0&\frac1{\sqrt2}&\frac{\sqrt3}{\sqrt2}&-\frac12&0&0&0\\
    0&0&0&0&0&0&0&\eta\\
    0&0&0&0&0&0&\eta&0\\
  \end{array}
  \right)
\end{equation}
with $\eta=J_{sd}/J_{pd}$ and the prefactor $h=J_{pd} M/\mu_B $. 
The kinetic-exchange couplings are $J_{pd}=55\unit{meV\cdot nm^{3}}$
and $J_{sd}=-9.2\unit{meV\cdot nm^{3}}$. By diagonalizing $\hat{H}$ of
Eq.~(\ref{eq-11}) in each $\vec{k}$-point of a suitably chosen mesh
around the $\Gamma$-point of the Brillouin zone, we obtain band
dispersions $E_{a\vec{k}}$ and corresponding spinors $|a,\vec{k}\rangle$. 

These two ingredients can be used to calculate the conductivity tensor
components in Eq.~(\ref{eq-12}) whose intraband part
\begin{equation}\label{eq-18}
  \sigma_{jj}^{\mathrm{intra}} = 
  \sum_n \frac{\sigma_0^{j,n}(1+i\hbar\omega/\Gamma)}{1+(\hbar\omega/\Gamma)^2}
\end{equation}
contains only the diagonal matrix elements of the velocity operator
$\hat{v}_j$ ($j=x,y,z$ denotes its Cartesian component) appearing in
the dc Drude conductivity $\sigma_0^{j,n}$ along given direction in
the $n$-th band. Relaxation times corresponding to
$\Gamma=100\unit{meV}$ are assumed to be $n$ and
$\vec{k}$-independent. Off-diagonal components
$\sigma_{ij}^{\mathrm{intra}}$ are not calculated, since they
contribute only little to the Voigt effect in reflection.\cite{note5}
Due to the combined effect of the ferromagnetic splitting (keep in
mind that $\vec{M}||\hat{x}$) and spin-orbit interaction, there is a
small difference between $\sigma_0^{x,n}$ and
$\sigma_0^{z,n}$. Additionaly, Eq.~(\ref{eq-18}) does not take into
account anisotropy induced by external magnetic field which is used in
experiments to control $\vec{M}$. Both effects lead to a small
anisotropy in $\sigma_{ii}^{\mathrm{intra}}$ which we estimated to
have only negligible effect on the resulting spectra of
$\theta(\omega)$ and $\psi(\omega)$.

Off-diagonal matrix elements 
$v_j^{ab\vec{k}}=\langle a,\vec{k}|\hat{v}_j|b,\vec{k}\rangle$ enter
the interband part of Eq.~(\ref{eq-12}) for which we use
\begin{equation}\label{eq-19}
\begin{split}
    \sigma_{jl}^{\mathrm{inter}}
  =
  -\frac{i \hbar e^2}{V} \sum_{\vec{k},a,b} &
  \big(f(E_{a\vec{k}})-f(E_{b\vec{k}})\big)\times \\ &\hskip-10mm
\frac{v_j^{ab\vec{k}}v_l^{ba\vec{k}}}{(E_{a\vec{k}}-E_{b\vec{k}}+i\Gamma)
  (E_{a\vec{k}}-E_{b\vec{k}}-\hbar\omega+i\Gamma)}
\end{split}
\end{equation}
where $f(E)$ is the Fermi-Dirac distribution function that contains
the Fermi level $E_f$ determined from the total hole concentration $p$
and $V$ is the system volume. In the remainder of this Appendix, we
show how the linear-response conductivity of a non-interacting system in
Eqs.~(\ref{eq-18},\ref{eq-19}) can be derived from the quantum
mechanical analogue of Liouville's theorem
\begin{equation}\label{eq-20}
\frac{\partial\hat{\rho}(t)}{\partial{t}}=
\frac{1}{i\hbar}[\hat{\mathbb{H}}(t),\hat{\rho}(t)].
\end{equation}
Here, $\hat{\rho}(t)$ is the density matrix and
$\hat{\mathbb{H}}(t)=\hat{H}+\hat{H}'(t)$ the total single-electron 
Hamiltonian of an
originally unperturbed system ($\hat{H}$) subject to a small
perturbation $\hat{H}'(t)$. We loosely follow Appendix of
Ref.~\onlinecite{Kolorenc:2002_a} where only the dc ($\omega\to 0$) limit
is considered. The derivation below is conceptually close to that of
Sec.~4 in Ref.~\onlinecite{Allen:2006_a} and remarks to the more
general context of Kubo formula can be found in that reference.

Perturbation of interest to us will be a weak monochromatic 
electric field $\vec{E}(\vec{r},t)=\vec{E}_0 
e^{i(\vec{q}\cdot\vec{r}-\omega t)}$ whose wavevector $\vec{q}$ is
small in the sense $|\vec{q}|a_l\ll 1$ and we put it equal to zero. 
Using vector potential in the
Coulomb gauge to describe this field, $\vec{A}=\vec{E}/(i\omega)$, the
perturbation in the linear order of $|\vec{E}_0|$ is  
$\hat{H}'(t)=e/(i\omega)\vec{E}(t) \cdot \hat{\vec{v}}$ where 
$\hat{\vec{v}}\equiv\hat{\vec{p}}/m_0$. Note that in this convention,
the current operator in Eq.~(\ref{eq-23}) is {\em not} 
proportional to $\hat{\vec{v}}$. 

Eq.~(\ref{eq-20}) can be solved by separating time dependent perturbative term
from Hamiltonian using
$$
  \frac{\partial}{\partial{t}}e^{i\hat{H}t/\hbar}\hat{\rho}(t)
  e^{-i\hat{H}t/\hbar}=\frac{1}{i\hbar} 
e^{i\hat{H}t/\hbar} [\hat{H}'(t),\hat{\rho}(t)]e^{-i\hat{H}t/\hbar}.
$$
The result
\begin{equation}\label{eq-21}
\hat{\rho}(t) = \hat{\rho}(t_0) + 
   \frac{1}{i\hbar}\int_{t_0}^t dt' \, e^{i\hat{H}(t'-t)/\hbar} 
   [\hat{H}'(t'),\hat{\rho}(t)] e^{-i\hat{H}(t'-t)/\hbar}.
\end{equation}
is exact and it can be evaluated iteratively. However, as our goal is
to calculate linear response of the system to $\vec{E}_0$ 
only the lowest order from the Dyson series is considered.
Then, variation of the density matrix 
$\delta\hat{\rho}(t)=\hat{\rho}(t)-\hat{\rho}_0$ from its equilibrium
value $\hat{\rho}_0\equiv \hat{\rho}(t_0)$ is given by
\begin{equation}\label{eq-22}
\begin{split}
\delta\hat{\rho}(t) = 
 \frac{1}{i\hbar}\int_{t_0}^t dt' \, e^{i\hat{H}(t'-t)/\hbar}
 [\hat{H}'(t'),\hat{\rho}_0] e^{-i\hat{H}(t'-t)/\hbar}
\end{split}
\end{equation}
and the current $\langle \hat{\vec{J}}\rangle =
\Tr \{(\delta\hat\rho)\, \hat{\vec{J}}\} + 
\Tr \{\hat{\rho}_0\hat{\vec{J}}\}\equiv \vec{J}_1+\vec{J}_2$.
Linear-response conductivity is then straightforwardly
$\sigma_{i j}=\partial \langle \hat{J}_i\rangle
/\partial E_{j}$, where $i,j = x,y,z$ and $E_j$ ($\hat{J}_i$) is $j$
the Cartesian component of $\vec{E}(t)$ ($\hat{\vec{J}}$).
The current operator in Coulomb gauge
\begin{equation}\label{eq-23}
\hat{\vec{J}}=-\frac{e}{m V}\frac{d}{dt}\hat{\vec{x}}
=
-\frac{e}{m V i \hbar}[\hat{\vec{x}},\mathbb{\hat{H}}]
=
-\frac{e}{m V}(\hat{\vec{p}}+e\vec{A})
\end{equation}
implies non-zero $\vec{J}_2$ because of the $\vec{A}$ 
term in Eq.~(\ref{eq-23}).
This $\vec{J}_2=i n e^2 / (m \omega) \vec{E}$ is often referred to as
diamagnetic or gauge current; $n\equiv \Tr \hat{\rho}_0$ is the total 
density of eletrons. Below, we show that this term which is divergent
in the dc ($\omega\to 0$) limit drops out and turn 
our attention to $\vec{J}_1$.
Here, the $\vec{A}$ term in Eq.~(\ref{eq-23}) can be omitted in the
linear response since $\delta\hat{\rho}(t)$ also contains a factor of $E_j$.
The rest of $\vec{J}_1$, called paramagnetic current, gives
\begin{equation}\label{eq-24}
-\frac{me}{V}\Tr \{(\delta \hat{\rho})\, \hat{p}_i \} =
\frac{e^2 E_j}{V \hbar \omega}\int_{0}^\infty d\tau \,e^{i\omega \tau}
\,\Tr \{\hat{v}_i^I(\tau)[\hat{v}_j,\hat{\rho}_0] \}
\end{equation}
where $\hat{v}_i^I(\tau)=e^{-i\hat{H} t/\hbar}\hat{v_i}e^{i\hat{H}
t/\hbar}$, $\tau\equiv t-t'$ and $t_0$ is set to $-\infty$. Using 
the invariance of trace to cyclic permutations of operators
inside it, the conductivity reads
\begin{equation}\label{eq-25}
\sigma_{i j}=\frac{e^2}{V \hbar \omega}
\int_{0}^\infty d\tau \,e^{i(\omega+i\Gamma/\hbar)\tau}
\Tr \{\hat{\rho}_0[\hat{v}_i^I(\tau),\hat{v}_j] \}
+\frac{ine^2}{m\omega}\delta_{ij}.
\end{equation}
Positive $\Gamma$ in the exponential ensures convergence and in clean
systems, it can be set to zero at the end of the calculation.
Without further discussing this step
here,\cite{note8} we replace this auxiliary $\Gamma$ by the estimated
spectral broadening (taken to be 100~meV as already mentioned).

With the knowledge of the complete set of eigenstates,
$\hat{H}|a,\vec{k}\rangle=E_{a\vec{k}}|a,\vec{k}\rangle$, conductivity
of Eq.~(\ref{eq-25}) can be rewritten\cite{Kolorenc:2002_a} as
\begin{equation}\label{eq-26}
\frac{i e^2}{V \omega}
\sum_{a, b, \vec{k}}\left(f_{a\vec{k}}-f_{b\vec{k}}\right)
\frac{v_i^{ab\vec{k}}v_j^{ba\vec{k}}}{\hbar(\omega+i \Gamma/\hbar)-
(E_{b\vec{k}}-E_{a\vec{k}})}
+\frac{ine^2}{m\omega}\delta_{ij}
\end{equation}
where $f_{a\vec{k}}\equiv f(E_{a\vec{k}})$ is the Fermi-Dirac function
(encoded in $\hat{\rho}_0$). The first term diverges as $\omega \to 0$
but the divergent part can be separated using identity
$1/\hbar \omega (\hbar \omega+x)=[1/\hbar\omega -1/(\hbar \omega + x)]/x$ and 
the first of these two terms precisely cancels 
the second term in Eq.~(\ref{eq-25})
which stems from the gauge current $\vec{J}_2$. The Kubo formula for
conductivity is therefore
\begin{equation}\label{eq-27}
\sigma_{i j}=-\frac{i \hbar e^2}{V}
\sum_{a,b}\left(\frac{f_{a\vec{k}}-f_{b\vec{k}}}{E_{a\vec{k}}-
                                           E_{b\vec{k}}+i\Gamma}\right) 
\frac{v_i^{ab\vec{k}}v_j^{ba\vec{k}}}{\hbar\omega+i \Gamma-
                                      (E_{b\vec{k}}-E_{a\vec{k}})}
\end{equation}
which is identical to Eq.~(\ref{eq-19}). In a perfect crystal, $\vec{k}$
is a good quantum number and only dipole transitions between empty and
filled bands are allowed. In other words, dipole matrix element
($v^{ab}$) is diagonal with respect to $\vec{k}$. In such a case,
$\Gamma\to 0$ limit can easily be taken and conventional expression for optical
conductivity in semiconductors and insulators results. To model
(Ga,Mn)As which is strongly disordered, we take a finite value of
$\Gamma$ as stated below Eq.~(\ref{eq-18}). 

Note that Eq.~(\ref{eq-27}) 
contains only interband ($a\not=b$) terms.
To derive the intraband conductivity $\sigma_{jl}^{\mathrm{intra}}$,
more careful treatment of the $|\vec{q}|\to 0$ limit is required. 
We arrive, for $a=b$, at a formula similar to Eq.~(\ref{eq-27})
where the first fraction after the summation symbol
is replaced\cite{Allen:2006_a} by 
$\partial f_{a\vec{k}}/\partial E_{a\vec{k}}$ and 
\begin{equation}\label{eq-28}
\sigma_{j j}^\textrm{intra}
=\frac{i \hbar e^2}{V}\sum_{a,\vec{k}}
\left(-\frac{\partial f_{a\vec{k}}}{\partial E_{a\vec{k}}}\right)
\frac{|v_j^{aa\vec{k}}|^2}{\hbar\omega+i \Gamma}
=\sum_{a}\frac{\sigma_0^{j,a}}{1-i \hbar \omega/ \Gamma}
\end{equation}
where $\sigma_{0}^{j,a}$ is the dc conductivity of band $n$ as in
Eq.~(\ref{eq-18}). If $\hbar/\Gamma$ is replaced by relaxation time
$\tau$, this turns into a more familiar form of the Drude formula,
giving $\Gamma$ a straightforward physical interpretation.
Eq.~(\ref{eq-28}) is often written in terms of single-particle Green's
functions, which is useful for perturbative treatment of disorder. 
% This leads to Bastin formula and Kubo-Greenwood formula.

\begin{figure}
\begin{tabular}{cc}
\includegraphics[scale=0.16]{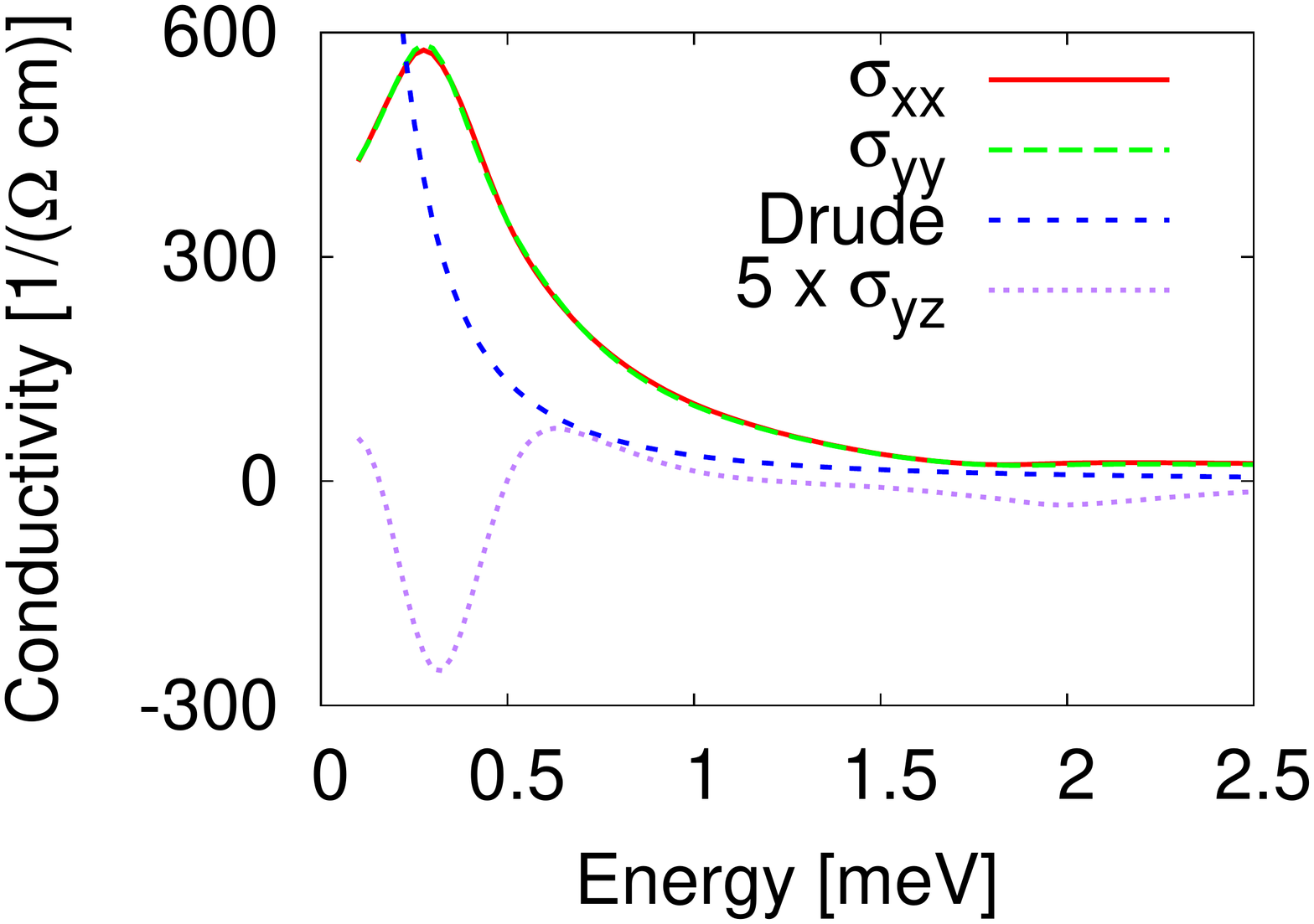} &
\hskip-5mm\includegraphics[scale=0.16]{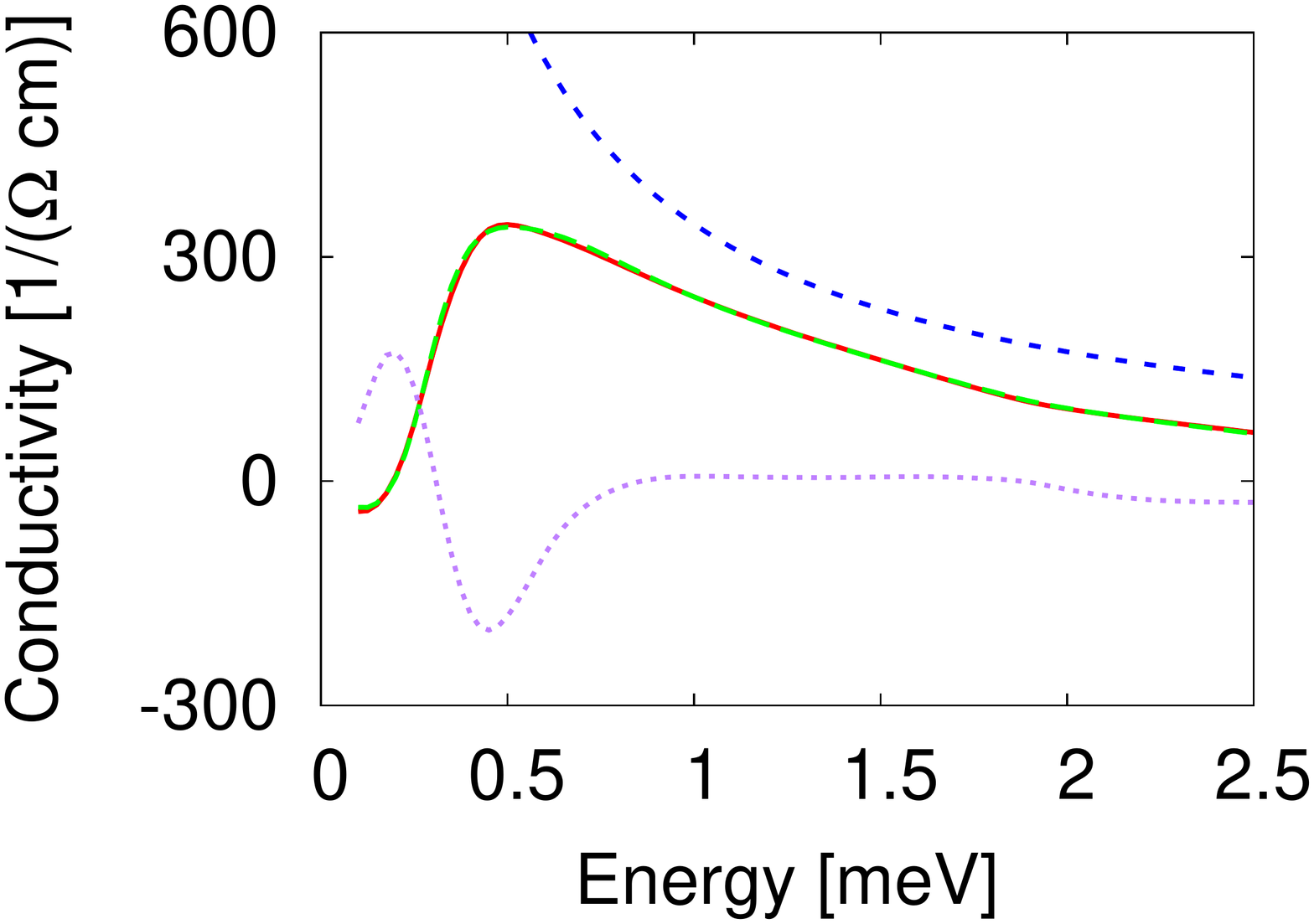} \\
(a) & (b) \\
\includegraphics[scale=0.16]{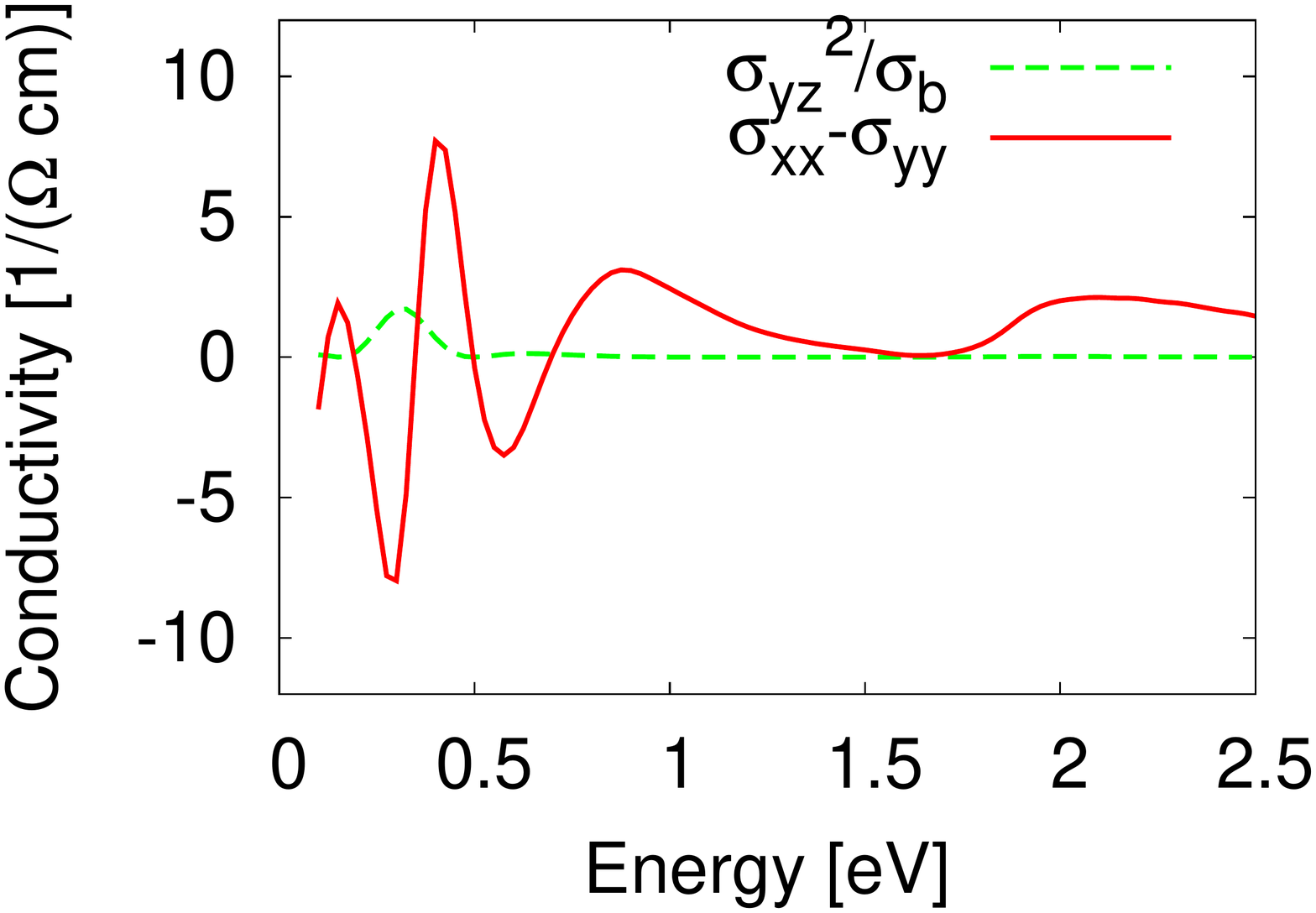} &
\hskip-5mm\includegraphics[scale=0.16]{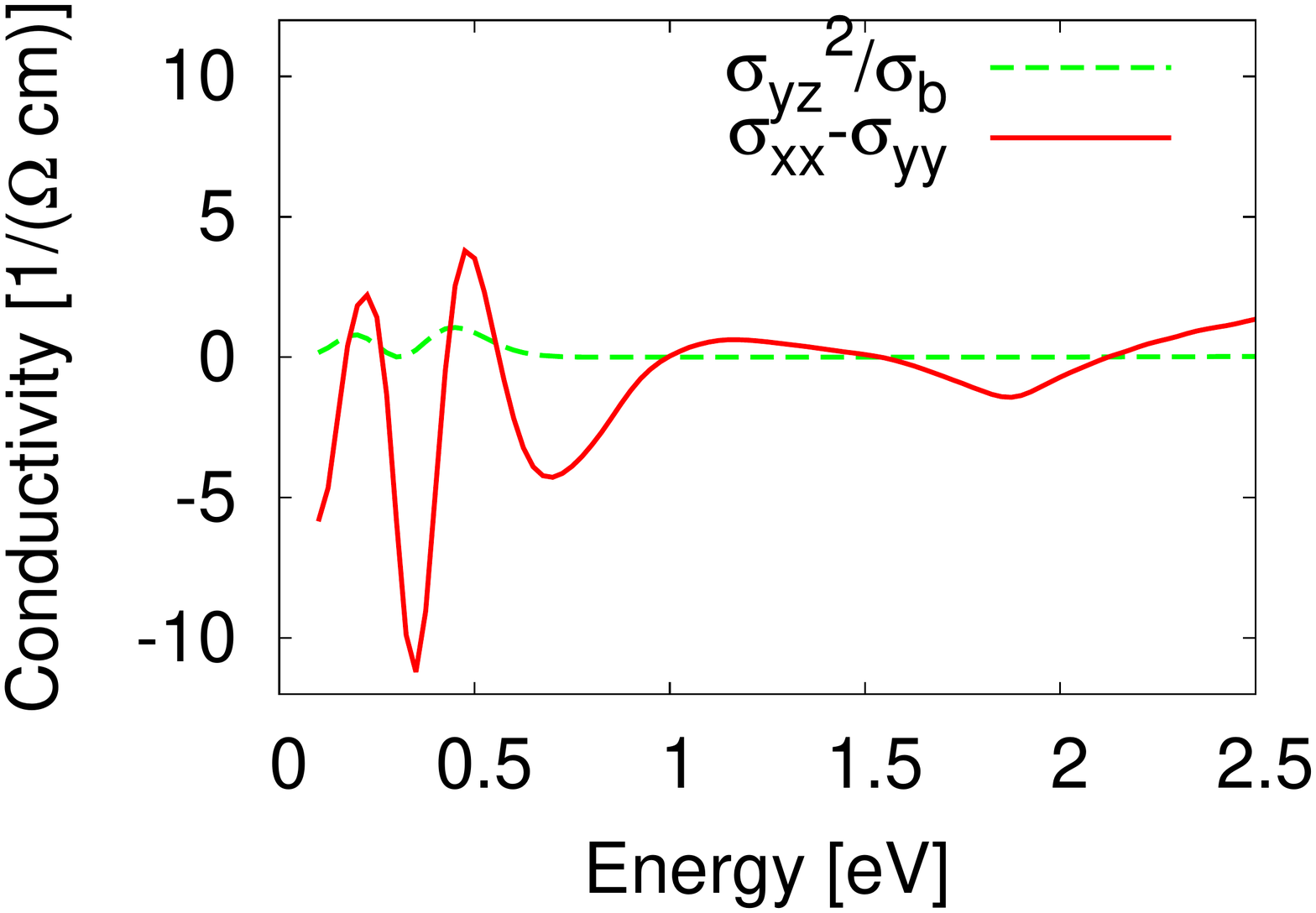} \\
(c) & (d)
\end{tabular}
\caption{Conductivities $\sigma_{xx}$, $\sigma_{yy}=\sigma_{zz}$ 
  and $\sigma_{yz}$  
  (magnetization along $\hat{x}$) corresponding to $p=1\unit{nm^{-3}}$ and
  $x=5\%$.
  (a) Real and (b) imaginary part of the interband conductivities
  according to Eq.~(\ref{eq-19}), intraband (Drude) part is
  also shown. Note that the off-diagonal conductivity is magnified by
  a factor of $5$. (c/d) Real/imaginary part of the difference between
  diagonal components. The relatively small
  $(\Rea \sigma_{yz})^2/\sigma_b$ and $(\Ima \sigma_{yz})^2/\sigma_b$ 
  with $\sigma_b=1500\unit{(\Omega\cdot cm)^{-1}}$ are also shown in
  panels (c) and (d).}
\label{fig-11}
\end{figure}

\section{From conductivity to the Voigt effect in reflection}

Typical $\sigma_{\parallel},\sigma_{\perp}$ as of Eq.~(\ref{eq-12})
are shown in Fig.~\ref{fig-11}(a,b) (real and imaginary parts). Angle $\theta$
(and similarly $\psi$) is according to Eq.~(\ref{eq-09}) related to their
difference $\sigma_{\parallel}-\sigma_\perp$ and we therefore also plot
this quantity. Some spectral features of Fig.~\ref{fig-04}a can be seen
in Fig.~\ref{fig-11}c ($\Rea \sigma_{\parallel}-\sigma_\perp$) and
Fig.~\ref{fig-11}d ($\Ima \sigma_{\parallel}-\sigma_\perp$) but
their relationship is not straightforward. 

Once the optical conductivities $\sigma_\parallel(\omega)$ and
$\sigma_\perp(\omega)$ are known, effective permittivity and 
refractive indices can be calculated using Eq.~(\ref{eq-10})
\begin{equation}\label{eq-29}
n_\parallel^2=\mu\veeff^\parallel=\mu(\ve_b+\frac{i\sigma_{||}}\omega),\qquad
n_\perp^2 \approx \mu\veeff^\perp=\mu(\ve_b+\frac{i\sigma_\perp}\omega)
\end{equation}
where $\mu$ is the relative permeability which we take $\mu=1$.
The $\ve_{yz}^2/\ve_{zz}$ term contributing to $n_\perp$
according to Eq.~(\ref{eq-08}) can be neglected:\cite{note5}
diagonal components of permittivity are dominated by the background
$\ve_b$ and this large value causes $\sigma_{yz}/\sigma_b^2$ 
(with $\sigma_b=1500\unit{(\Omega\cdot cm)^{-1}}$ appropriate 
for $\ve_b\approx 10.9$) to be small compared to $\sigma_\parallel-
\sigma_\perp$ as shown in the lower panels of Fig.~\ref{fig-11}.
Since both real and imaginary parts of $n_\parallel$ and $n_\perp$
differ, meaning that both (magnetic linear) birefringence and
dichroism is present in our system, let us consider an illustrative
example of how MLD and MLB individually influence the resultant 
$\theta$. Assume that $\ve_{xx}=11.60+0.70i$ and
$\ve_{zz}=11.61+0.71i$; this is inspired by values in Fig.~\ref{fig-11}
and it would correspond to $\sigma\approx
100\unit{(\Omega\cdot cm)^{-1}}$ at $\hbar\omega=1$~eV and
$|\sigma_{xx}-\sigma_{zz}|$ of the order of $1\unit{(\Omega\cdot cm)^{-1}}$.
The rotation $\theta\approx \Rea \chi$, as given by Eq.~(\ref{eq-14}),
will be $0.151\unit{mrad}$ if reflection on an infinitely thick
(Ga,Mn)As layer is considered, corresponding to Eq.~(\ref{eq-05}). Now
consider pure MLD situation: $\ve_{xx}=11.60+0.70i$ and
$\ve_{zz}=11.60+0.71i$ would give $\theta=0.009\unit{mrad}$. On
the other hand, $\ve_{xx}=11.60+0.70i$ and $\ve_{zz}=11.61+0.70i$ (pure
MLB) results in $\theta=0.142\unit{mrad}$. It is clear that both MLD
and MLB can significantly contribute to the spectra of rotation of the
Voigt effect in reflection.

Taking into account the effect of the substrate as in Fig.~3 and
Eq.~(5) of Ref.~\onlinecite{Kim:2007_a}, (Ga,Mn)As refractive index 
$n$ leads to the reflection coefficient
\begin{equation}\label{eq-30}
  r(n) = 
\frac{(n_s-1)\cos(kd) -i(n-n_s/n)\sin(kd)}{(n_s+1)\cos(kd)-i(n+n_s/n)\sin(kd)}.
\end{equation}
Using Eqs.~(\ref{eq-29},\ref{eq-30}) we get $r(n_\parallel)$ and
$r(n_\perp)$ that can be inserted into Eq.~(\ref{eq-13}) and we
finally obtain the rotation and ellipticity $\theta$, $\psi$.
Multiple reflections in a (Ga,Mn)As layer are taken into
account in Eq.~(\ref{eq-30}),
the complex $k=n\omega/c$, the layer has a finite thickness $d$
and it is sandwiched between vacuum and GaAs substrate with refractive
indices $1$ and $n_s=\sqrt{\ve_b+i\sigma_{\mathrm{GaAs}}/\omega\ve_0}$, 
respectively. As it was explained below Eq.~(\ref{eq-10}), we use
$\omega$-dependent $\ve_b$ which, together with intrinsic GaAs ac conductivity
$\sigma_{\mathrm{GaAs}}(\omega)$ calculated from Eq.~(\ref{eq-19}),
reproduces experimentally known %\cite{Lautenschlager:1987_a} permittivity
refractive index of GaAs. This seemingly over-cautious method of
determining $n_s$ is important for maintaining the consistency of our optical
model in Eq.~(\ref{eq-30}). It guarantees that in the $x\to 0$, $p\to
0$ limit applied to our (Ga,Mn)As layer, reflection from the
layer/substrate interface will be zero.

\begin{figure}
\includegraphics[scale=0.32]{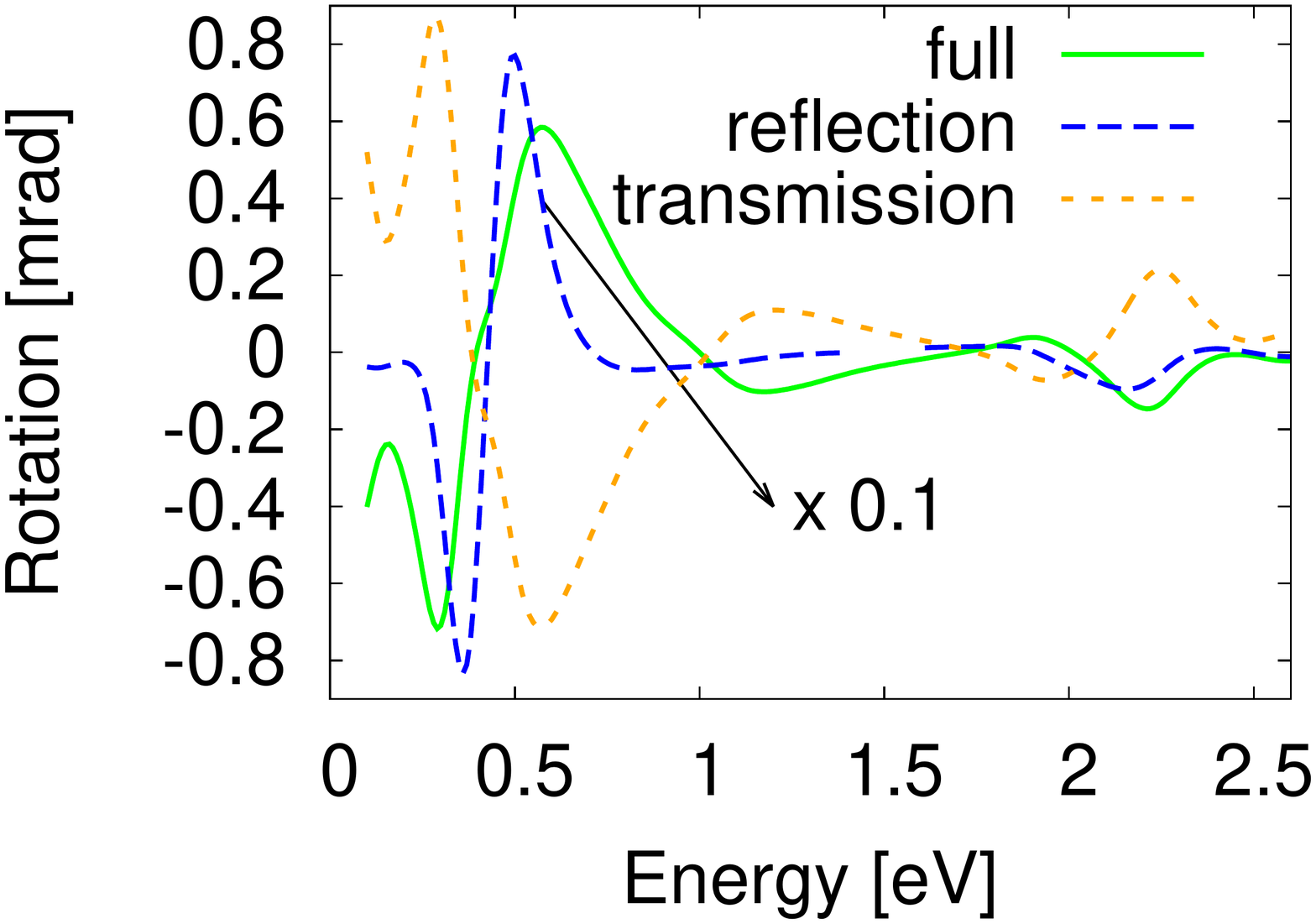} 
\caption{Rotation for a sample with $p=0.4\unit{nm^{-3}}$, $x=5\%$ and
$d=20\unit{nm}$ in three situations: transmission (i.e. Voigt effect),
reflection from a thick layer ($d\to\infty$) and geometry of our
measurements (labelled 'full'). Note that the dashed curve is
downscaled by a factor of 10 in the range $\hbar\omega<1.5\unit{eV}$.}
\label{fig-12}
\end{figure}

Indeed, multilayer optical properties significantly influence the
final form of the spectra (experimentally, thickness-dependence of 
$\theta$ at $\hbar\omega=1.58$~eV was studied by Al-Qadi 
et~al. in Ref.~\onlinecite{Qadi:2012_a}). Fig.~\ref{fig-12} shows 
that differences
between the transmission and reflection Voigt effect 
experiments could be significant, yet the
spectral features (and their position in particular) remain to some
extent unaffected. For example, peak $\alpha$ is somewhat suppressed
in pure reflection (dashed curve in Fig.~\ref{fig-12}) that would
correspond to an experiment with a thick layer ($d\to\infty$). Features
$\beta$, $\gamma$ in the sub-gap energy range would, however, be order
of magnitude larger in pure reflection. A hypothetical experiment
measuring transmission (including multireflections)
through a thin
($d=20\unit{nm}$) layer would give the Voigt effect as shown by
the dotted curve where, roughly speaking, the spectrum only changes
the overall sign. Multiple reflections between the
substrate-sample and air-sample interfaces substantially modify the spectra 
although their effect may even be
somewhat exaggerated in our model. Based on an estimate 
$\Rea\sigma_{xx}=\Ima\sigma_{xx}=200\unit{(\Omega\cdot cm)^{-1}} $ at
$\hbar\omega=1\unit{eV}$ (compare Fig.~\ref{fig-11}), we obtain index
of refraction $n\approx 1.2$ and therefore a large reflection
coefficient at the substrate/sample interface. With absorption
coefficients $\alpha(\omega)\sim 10000\unit{cm^{-1}}$, i.e. 
$\alpha d\approx 0.01\ll 1$, the wave will be (in our model) able 
to travel many times through the sample. Experimental comparison of
the effect in samples with different thicknesses however suggests that
both $\alpha d$ is larger and the sample/substrate contrast is lower
($n/n_s$ closer to one).

\end{appendix}

\def\urlprefix{}
\def\url#1{}\bibliography{lit}

\begin{thebibliography}{67}
\expandafter\ifx\csname natexlab\endcsname\relax\def\natexlab#1{#1}\fi
\expandafter\ifx\csname bibnamefont\endcsname\relax
  \def\bibnamefont#1{#1}\fi
\expandafter\ifx\csname bibfnamefont\endcsname\relax
  \def\bibfnamefont#1{#1}\fi
\expandafter\ifx\csname citenamefont\endcsname\relax
  \def\citenamefont#1{#1}\fi
\expandafter\ifx\csname url\endcsname\relax
  \def\url#1{\texttt{#1}}\fi
\expandafter\ifx\csname urlprefix\endcsname\relax\def\urlprefix{URL }\fi
\providecommand{\bibinfo}[2]{#2}
\providecommand{\eprint}[2][]{\url{#2}}

\bibitem[{\citenamefont{Ferre and Gehring}(1984)}]{Ferre:1984_a}
\bibinfo{author}{\bibfnamefont{J.}~\bibnamefont{Ferre}} \bibnamefont{and}
  \bibinfo{author}{\bibfnamefont{G.~A.} \bibnamefont{Gehring}},
  \bibinfo{journal}{Rep. Prog. Phys.}
  \textbf{\bibinfo{volume}{47}}, \bibinfo{pages}{513} (\bibinfo{year}{1984}).


\bibitem[{\citenamefont{Jungwirth
  et~al.}(2006{\natexlab{a}})\citenamefont{Jungwirth, Sinova,
  Ma\ifmmode~\check{s}\else \v{s}\fi{}ek, Ku\ifmmode~\check{c}\else
  \v{c}\fi{}era, and MacDonald}}]{Jungwirth:2006_a}
\bibinfo{author}{\bibfnamefont{T.}~\bibnamefont{Jungwirth}},
  \bibinfo{author}{\bibfnamefont{J.}~\bibnamefont{Sinova}},
  \bibinfo{author}{\bibfnamefont{J.}~\bibnamefont{Ma\ifmmode~\check{s}\else
  \v{s}\fi{}ek}},
  \bibinfo{author}{\bibfnamefont{J.}~\bibnamefont{Ku\ifmmode~\check{c}\else
  \v{c}\fi{}era}}, \bibnamefont{and} \bibinfo{author}{\bibfnamefont{A.~H.}
  \bibnamefont{MacDonald}}, \bibinfo{journal}{Rev. Mod. Phys.}
  \textbf{\bibinfo{volume}{78}}, \bibinfo{pages}{809}
  (\bibinfo{year}{2006}{\natexlab{a}}).

\bibitem[{\citenamefont{Burch et~al.}(2004)\citenamefont{Burch, Stephens,
  Kawakami, Awschalom, and Basov}}]{Burch:2004_a}
\bibinfo{author}{\bibfnamefont{K.~S.} \bibnamefont{Burch}},
  \bibinfo{author}{\bibfnamefont{J.}~\bibnamefont{Stephens}},
  \bibinfo{author}{\bibfnamefont{R.~K.} \bibnamefont{Kawakami}},
  \bibinfo{author}{\bibfnamefont{D.~D.} \bibnamefont{Awschalom}},
  \bibnamefont{and} \bibinfo{author}{\bibfnamefont{D.~N.} \bibnamefont{Basov}},
  \bibinfo{journal}{Phys. Rev. B} \textbf{\bibinfo{volume}{70}},
  \bibinfo{pages}{205208} (\bibinfo{year}{2004}).

\bibitem[{\citenamefont{Kimel et~al.}(2005)\citenamefont{Kimel, Astakhov,
  Kirilyuk, Schott, Karczewski, Ossau, Schmidt, Molenkamp, and
  Rasing}}]{Kimel:2005_a}
\bibinfo{author}{\bibfnamefont{A.~V.} \bibnamefont{Kimel}},
  \bibinfo{author}{\bibfnamefont{G.~V.} \bibnamefont{Astakhov}},
  \bibinfo{author}{\bibfnamefont{A.}~\bibnamefont{Kirilyuk}},
  \bibinfo{author}{\bibfnamefont{G.~M.} \bibnamefont{Schott}},
  \bibinfo{author}{\bibfnamefont{G.}~\bibnamefont{Karczewski}},
  \bibinfo{author}{\bibfnamefont{W.}~\bibnamefont{Ossau}},
  \bibinfo{author}{\bibfnamefont{G.}~\bibnamefont{Schmidt}},
  \bibinfo{author}{\bibfnamefont{L.~W.} \bibnamefont{Molenkamp}},
  \bibnamefont{and} \bibinfo{author}{\bibfnamefont{T.}~\bibnamefont{Rasing}},
  \bibinfo{journal}{Phys. Rev. Lett.} \textbf{\bibinfo{volume}{94}},
  \bibinfo{pages}{227203} (\bibinfo{year}{2005}).

\bibitem[{\citenamefont{Kim et~al.}(2007)\citenamefont{Kim, Acbas, Yang,
  Ohkubo, Christen, Mandrus, Scarpulla, Dubon, Schlesinger, Khalifah
  et~al.}}]{Kim:2007_a}
\bibinfo{author}{\bibfnamefont{M.-H.} \bibnamefont{Kim}},
  \bibinfo{author}{\bibfnamefont{G.}~\bibnamefont{Acbas}},
  \bibinfo{author}{\bibfnamefont{M.-H.} \bibnamefont{Yang}},
  \bibinfo{author}{\bibfnamefont{I.}~\bibnamefont{Ohkubo}},
  \bibinfo{author}{\bibfnamefont{H.}~\bibnamefont{Christen}},
  \bibinfo{author}{\bibfnamefont{D.}~\bibnamefont{Mandrus}},
  \bibinfo{author}{\bibfnamefont{M.~A.} \bibnamefont{Scarpulla}},
  \bibinfo{author}{\bibfnamefont{O.~D.} \bibnamefont{Dubon}},
  \bibinfo{author}{\bibfnamefont{Z.}~\bibnamefont{Schlesinger}},
  \bibinfo{author}{\bibfnamefont{P.}~\bibnamefont{Khalifah}},
  \bibnamefont{et~al.}, \bibinfo{journal}{Phys. Rev. B}
  \textbf{\bibinfo{volume}{75}}, \bibinfo{pages}{214416}
  (\bibinfo{year}{2007}).

\bibitem[{\citenamefont{Acbas et~al.}(2009)\citenamefont{Acbas, Kim, Cukr,
  Nov\'ak, Scarpulla, Dubon, Jungwirth, Sinova, and Cerne}}]{Acbas:2009_a}
\bibinfo{author}{\bibfnamefont{G.}~\bibnamefont{Acbas}},
  \bibinfo{author}{\bibfnamefont{M.-H.} \bibnamefont{Kim}},
  \bibinfo{author}{\bibfnamefont{M.}~\bibnamefont{Cukr}},
  \bibinfo{author}{\bibfnamefont{V.}~\bibnamefont{Nov\'ak}},
  \bibinfo{author}{\bibfnamefont{M.~A.} \bibnamefont{Scarpulla}},
  \bibinfo{author}{\bibfnamefont{O.~D.} \bibnamefont{Dubon}},
  \bibinfo{author}{\bibfnamefont{T.}~\bibnamefont{Jungwirth}},
  \bibinfo{author}{\bibfnamefont{J.}~\bibnamefont{Sinova}}, \bibnamefont{and}
  \bibinfo{author}{\bibfnamefont{J.}~\bibnamefont{Cerne}},
  \bibinfo{journal}{Phys. Rev. Lett.} \textbf{\bibinfo{volume}{103}},
  \bibinfo{pages}{137201} (\bibinfo{year}{2009}).

\bibitem[{\citenamefont{Nagaosa et~al.}(2010)\citenamefont{Nagaosa, Sinova,
  Onoda, MacDonald, and Ong}}]{Nagaosa:2010_a}
\bibinfo{author}{\bibfnamefont{N.}~\bibnamefont{Nagaosa}},
  \bibinfo{author}{\bibfnamefont{J.}~\bibnamefont{Sinova}},
  \bibinfo{author}{\bibfnamefont{S.}~\bibnamefont{Onoda}},
  \bibinfo{author}{\bibfnamefont{A.~H.} \bibnamefont{MacDonald}},
  \bibnamefont{and} \bibinfo{author}{\bibfnamefont{N.~P.} \bibnamefont{Ong}},
  \bibinfo{journal}{Rev. Mod. Phys.} \textbf{\bibinfo{volume}{82}},
  \bibinfo{pages}{1539} (\bibinfo{year}{2010}).


\bibitem{Schafer:1990_a} R. {Sch\"afer} and A. Hubert,
  phys. stat. sol. (a) \textbf{118}, 271 (1990).

\bibitem[{\citenamefont{Mertins et~al.}(2001)\citenamefont{Mertins, Oppeneer,
  Kune\ifmmode~\check{s}\else \v{s}\fi{}, Gaupp, Abramsohn, and
  Sch\"afers}}]{Mertins:2001_a}
\bibinfo{author}{\bibfnamefont{H.-C.} \bibnamefont{Mertins}},
  \bibinfo{author}{\bibfnamefont{P.~M.} \bibnamefont{Oppeneer}},
  \bibinfo{author}{\bibfnamefont{J.}~\bibnamefont{Kune\ifmmode~\check{s}\else
  \v{s}\fi{}}}, \bibinfo{author}{\bibfnamefont{A.}~\bibnamefont{Gaupp}},
  \bibinfo{author}{\bibfnamefont{D.}~\bibnamefont{Abramsohn}},
  \bibnamefont{and}
  \bibinfo{author}{\bibfnamefont{F.}~\bibnamefont{Sch\"afers}},
  \bibinfo{journal}{Phys. Rev. Lett.} \textbf{\bibinfo{volume}{87}},
  \bibinfo{pages}{047401} (\bibinfo{year}{2001}).

\bibitem[{\citenamefont{Kokado et~al.}(2012)\citenamefont{Kokado, Tsunoda,
  Harigaya, and Sakuma}}]{Kokado:2012_a}
\bibinfo{author}{\bibfnamefont{S.}~\bibnamefont{Kokado}},
  \bibinfo{author}{\bibfnamefont{M.}~\bibnamefont{Tsunoda}},
  \bibinfo{author}{\bibfnamefont{K.}~\bibnamefont{Harigaya}}, \bibnamefont{and}
  \bibinfo{author}{\bibfnamefont{A.}~\bibnamefont{Sakuma}},
  \bibinfo{journal}{J. Phys. Soc. Jap.}
  \textbf{\bibinfo{volume}{81}}, \bibinfo{pages}{024705}
  (\bibinfo{year}{2012}).

\bibitem[{\citenamefont{Rushforth et~al.}(2007)\citenamefont{Rushforth,
  V\'yborn\'y, King, Edmonds, Campion, Foxon, Wunderlich, Irvine,
  Va\ifmmode~\check{s}\else \v{s}\fi{}ek, Nov\'ak et~al.}}]{Rushforth:2007_a}
\bibinfo{author}{\bibfnamefont{A.~W.} \bibnamefont{Rushforth}},
  \bibinfo{author}{\bibfnamefont{K.}~\bibnamefont{V\'yborn\'y}},
  \bibinfo{author}{\bibfnamefont{C.~S.} \bibnamefont{King}},
  \bibinfo{author}{\bibfnamefont{K.~W.} \bibnamefont{Edmonds}},
  \bibinfo{author}{\bibfnamefont{R.~P.} \bibnamefont{Campion}},
  \bibinfo{author}{\bibfnamefont{C.~T.} \bibnamefont{Foxon}},
  \bibinfo{author}{\bibfnamefont{J.}~\bibnamefont{Wunderlich}},
  \bibinfo{author}{\bibfnamefont{A.~C.} \bibnamefont{Irvine}},
  \bibinfo{author}{\bibfnamefont{P.}~\bibnamefont{Va\ifmmode~\check{s}\else
  \v{s}\fi{}ek}}, \bibinfo{author}{\bibfnamefont{V.}~\bibnamefont{Nov\'ak}},
  \bibnamefont{et~al.}, \bibinfo{journal}{Phys. Rev. Lett.}
  \textbf{\bibinfo{volume}{99}}, \bibinfo{pages}{147207}
  (\bibinfo{year}{2007}).

\bibitem[{\citenamefont{Kirilyuk et~al.}(2010)\citenamefont{Kirilyuk, Kimel,
  and Rasing}}]{Kirilyuk:2010_a}
\bibinfo{author}{\bibfnamefont{A.}~\bibnamefont{Kirilyuk}},
  \bibinfo{author}{\bibfnamefont{A.~V.} \bibnamefont{Kimel}}, \bibnamefont{and}
  \bibinfo{author}{\bibfnamefont{T.}~\bibnamefont{Rasing}},
  \bibinfo{journal}{Rev. Mod. Phys.} \textbf{\bibinfo{volume}{82}},
  \bibinfo{pages}{2731} (\bibinfo{year}{2010}).

\bibitem[{\citenamefont{Kimel et~al.}(2004)\citenamefont{Kimel, Kirilyuk,
  Tsvetkov, Pisarev, and Rasing}}]{Kimel:2004_a}
\bibinfo{author}{\bibfnamefont{A.~V.} \bibnamefont{Kimel}},
  \bibinfo{author}{\bibfnamefont{A.}~\bibnamefont{Kirilyuk}},
  \bibinfo{author}{\bibfnamefont{A.}~\bibnamefont{Tsvetkov}},
  \bibinfo{author}{\bibfnamefont{R.~V.} \bibnamefont{Pisarev}},
  \bibnamefont{and} \bibinfo{author}{\bibfnamefont{T.}~\bibnamefont{Rasing}},
  \bibinfo{journal}{Nature} \textbf{\bibinfo{volume}{429}},
  \bibinfo{pages}{850} (\bibinfo{year}{2004}).

\bibitem[{\citenamefont{Ando et~al.}(1998)\citenamefont{Ando, Hayashi, Tanaka,
  and Twardowski}}]{Ando:1998_a}
\bibinfo{author}{\bibfnamefont{K.}~\bibnamefont{Ando}},
  \bibinfo{author}{\bibfnamefont{T.}~\bibnamefont{Hayashi}},
  \bibinfo{author}{\bibfnamefont{M.}~\bibnamefont{Tanaka}}, \bibnamefont{and}
  \bibinfo{author}{\bibfnamefont{A.}~\bibnamefont{Twardowski}},
  \bibinfo{journal}{J. Appl. Phys.} \textbf{\bibinfo{volume}{83}},
  \bibinfo{pages}{6548} (\bibinfo{year}{1998}).

\bibitem[{\citenamefont{Jungwirth et~al.}(2010)\citenamefont{Jungwirth,
  {Horodysk\'{a}}, {Tesa\v{r}ov\'{a}}, {N\v{e}mec}, {\v{S}ubrt}, {Mal\'{y}},
  {Ku\v{z}el}, Kadlec, {Ma\v{s}ek}, {N\v{e}mec} et~al.}}]{Jungwirth:2010_b}
\bibinfo{author}{\bibfnamefont{T.}~\bibnamefont{Jungwirth}},
  \bibinfo{author}{\bibfnamefont{P.}~\bibnamefont{{Horodysk\'{a}}}},
  \bibinfo{author}{\bibfnamefont{N.}~\bibnamefont{{Tesa\v{r}ov\'{a}}}},
  \bibinfo{author}{\bibfnamefont{P.}~\bibnamefont{{N\v{e}mec}}},
  \bibinfo{author}{\bibfnamefont{J.}~\bibnamefont{{\v{S}ubrt}}},
  \bibinfo{author}{\bibfnamefont{P.}~\bibnamefont{{Mal\'{y}}}},
  \bibinfo{author}{\bibfnamefont{P.}~\bibnamefont{{Ku\v{z}el}}},
  \bibinfo{author}{\bibfnamefont{C.}~\bibnamefont{Kadlec}},
  \bibinfo{author}{\bibfnamefont{J.}~\bibnamefont{{Ma\v{s}ek}}},
  \bibinfo{author}{\bibfnamefont{I.}~\bibnamefont{{N\v{e}mec}}},
  \bibnamefont{et~al.}, \bibinfo{journal}{Phys. Rev. Lett.}
  \textbf{\bibinfo{volume}{105}}, \bibinfo{pages}{227201}
  (\bibinfo{year}{2010}).

\bibitem[{\citenamefont{Moore et~al.}(2003)\citenamefont{Moore, Ferre, Mougin,
  Moreno, and Daweritz}}]{Moore:2003_a}
\bibinfo{author}{\bibfnamefont{G.~P.} \bibnamefont{Moore}},
  \bibinfo{author}{\bibfnamefont{J.}~\bibnamefont{Ferre}},
  \bibinfo{author}{\bibfnamefont{A.}~\bibnamefont{Mougin}},
  \bibinfo{author}{\bibfnamefont{M.}~\bibnamefont{Moreno}}, \bibnamefont{and}
  \bibinfo{author}{\bibfnamefont{L.}~\bibnamefont{Daweritz}},
  \bibinfo{journal}{J. Appl. Phys.} \textbf{\bibinfo{volume}{94}},
  \bibinfo{pages}{4530} (\bibinfo{year}{2003}).

\bibitem[{\citenamefont{Tesa\v{r}ov\'{a}
  et~al.}(2012)\citenamefont{Tesa\v{r}ov\'{a}, \v{S}ubrt, Mal\'{y}, N\v{e}mec,
  Ellis, Mukherjee, and Cerne}}]{Tesarova:2012_c}
\bibinfo{author}{\bibfnamefont{N.}~\bibnamefont{Tesa\v{r}ov\'{a}}},
  \bibinfo{author}{\bibfnamefont{J.}~\bibnamefont{\v{S}ubrt}},
  \bibinfo{author}{\bibfnamefont{P.}~\bibnamefont{Mal\'{y}}},
  \bibinfo{author}{\bibfnamefont{P.}~\bibnamefont{N\v{e}mec}},
  \bibinfo{author}{\bibfnamefont{C.~T.} \bibnamefont{Ellis}},
  \bibinfo{author}{\bibfnamefont{A.}~\bibnamefont{Mukherjee}},
  \bibnamefont{and} \bibinfo{author}{\bibfnamefont{J.}~\bibnamefont{Cerne}},
  \bibinfo{journal}{Rev. Sci. Inst.}
  {\bibinfo{volume}{83}}, \bibinfo{eid}{123108}
  (\bibinfo{year}{2012}).


\bibitem[{\citenamefont{Abolfath et~al.}(2001)\citenamefont{Abolfath,
  Jungwirth, Brum, and MacDonald}}]{Abolfath:2001_a}
\bibinfo{author}{\bibfnamefont{M.}~\bibnamefont{Abolfath}},
  \bibinfo{author}{\bibfnamefont{T.}~\bibnamefont{Jungwirth}},
  \bibinfo{author}{\bibfnamefont{J.}~\bibnamefont{Brum}}, \bibnamefont{and}
  \bibinfo{author}{\bibfnamefont{A.~H.} \bibnamefont{MacDonald}},
  \bibinfo{journal}{Phys. Rev. B} \textbf{\bibinfo{volume}{63}},
  \bibinfo{pages}{054418} (\bibinfo{year}{2001}).

\bibitem[{\citenamefont{Ebert}(1996)}]{Ebert:1996_a}
\bibinfo{author}{\bibfnamefont{H.}~\bibnamefont{Ebert}},
  \bibinfo{journal}{Rep. Prog. Phys.}
  \textbf{\bibinfo{volume}{59}}, \bibinfo{pages}{1665} (\bibinfo{year}{1996}).

\bibitem[{not({\natexlab{a}})}]{note1}
\bibinfo{note}{Cyclotron resonance (in non-magnetic materials) gives rise to a
  peak in absorption which shifts with increasing magnetic
  field.\cite{Orlita:2013_a} The term "magnetooptics" also commonly embraces
  this effect.}

\bibitem[{\citenamefont{Orlita et~al.}(2013)\citenamefont{Orlita, Escoffier,
  Plochocka, Raquet, and Zeitler}}]{Orlita:2013_a}
\bibinfo{author}{\bibfnamefont{M.}~\bibnamefont{Orlita}},
  \bibinfo{author}{\bibfnamefont{W.}~\bibnamefont{Escoffier}},
  \bibinfo{author}{\bibfnamefont{P.}~\bibnamefont{Plochocka}},
  \bibinfo{author}{\bibfnamefont{B.}~\bibnamefont{Raquet}}, \bibnamefont{and}
  \bibinfo{author}{\bibfnamefont{U.}~\bibnamefont{Zeitler}},
  \bibinfo{journal}{Comptes Rendus Physique} \textbf{\bibinfo{volume}{14}},
  \bibinfo{pages}{78 } (\bibinfo{year}{2013}). 

\bibitem[{\citenamefont{Buchmeier et~al.}(2009)\citenamefont{Buchmeier,
  Schreiber, B\"urgler, and Schneider}}]{Buchmeier:2009_a}
\bibinfo{author}{\bibfnamefont{M.}~\bibnamefont{Buchmeier}},
  \bibinfo{author}{\bibfnamefont{R.}~\bibnamefont{Schreiber}},
  \bibinfo{author}{\bibfnamefont{D.~E.} \bibnamefont{B\"urgler}},
  \bibnamefont{and} \bibinfo{author}{\bibfnamefont{C.~M.}
  \bibnamefont{Schneider}}, \bibinfo{journal}{Phys. Rev. B}
  \textbf{\bibinfo{volume}{79}}, \bibinfo{pages}{064402}
  (\bibinfo{year}{2009}).

\bibitem[{not({\natexlab{b}})}]{note9}
\bibinfo{note}{Hubert and Sch\"afer\cite{Schafer:1990_a} (HS) observed an
  effect related to magnetization gradient while studying in-plane
  magnetized domains of magnetic thin films. It was later named the HS
  effect\cite{Kambersky:1992_a} but since the original
  reference~\cite{Schafer:1990_a} also reports on the observation of
  the Voigt effect in reflection, terminology became somewhat
  confused. Boundary effects in systems with magnetic domains (and
  non-zero magnetic gradients) naturally receive the name HS
  effect\cite{Banno:2008_a} but x-ray Voigt effect in reflection off
  thin iron layers with homogeneous magnetization has also been called
  HS effect.\cite{Valencia:2010_a} In the present article, we adhere
  to the original convention\cite{Schafer:1990_a} and use the term
  ''Voigt effect in reflection''.}

\bibitem[{\citenamefont{Kambersk\'y}(1992)}]{Kambersky:1992_a}
\bibinfo{author}{\bibfnamefont{V.}~\bibnamefont{Kambersk\'y}},
  \bibinfo{journal}{J. Magn. Magn. Mat.}
  {\bibinfo{volume}{104--107}}, \bibinfo{pages}{311 }
  (\bibinfo{year}{1992}).

\bibitem[{\citenamefont{Banno}(2008)}]{Banno:2008_a}
\bibinfo{author}{\bibfnamefont{I.}~\bibnamefont{Banno}},
  \bibinfo{journal}{Phys. Rev. A} \textbf{\bibinfo{volume}{77}},
  \bibinfo{pages}{033818} (\bibinfo{year}{2008}).

\bibitem[{\citenamefont{Valencia et~al.}(2010)\citenamefont{Valencia, Kleibert,
  Gaupp, Rusz, Legut, Bansmann, Gudat, and Oppeneer}}]{Valencia:2010_a}
\bibinfo{author}{\bibfnamefont{S.}~\bibnamefont{Valencia}},
  \bibinfo{author}{\bibfnamefont{A.}~\bibnamefont{Kleibert}},
  \bibinfo{author}{\bibfnamefont{A.}~\bibnamefont{Gaupp}},
  \bibinfo{author}{\bibfnamefont{J.}~\bibnamefont{Rusz}},
  \bibinfo{author}{\bibfnamefont{D.}~\bibnamefont{Legut}},
  \bibinfo{author}{\bibfnamefont{J.}~\bibnamefont{Bansmann}},
  \bibinfo{author}{\bibfnamefont{W.}~\bibnamefont{Gudat}}, \bibnamefont{and}
  \bibinfo{author}{\bibfnamefont{P.~M.} \bibnamefont{Oppeneer}},
  \bibinfo{journal}{Phys. Rev. Lett.} \textbf{\bibinfo{volume}{104}},
  \bibinfo{pages}{187401} (\bibinfo{year}{2010}).

\bibitem[{\citenamefont{Postava et~al.}(1997)\citenamefont{Postava, Jaffres,
  Schuhl, Dau, Goiran, and Fert}}]{Postava:1997_a}
\bibinfo{author}{\bibfnamefont{K.}~\bibnamefont{Postava}},
  \bibinfo{author}{\bibfnamefont{H.}~\bibnamefont{Jaffres}},
  \bibinfo{author}{\bibfnamefont{A.}~\bibnamefont{Schuhl}},
  \bibinfo{author}{\bibfnamefont{F.~N.~V.} \bibnamefont{Dau}},
  \bibinfo{author}{\bibfnamefont{M.}~\bibnamefont{Goiran}}, \bibnamefont{and}
  \bibinfo{author}{\bibfnamefont{A.}~\bibnamefont{Fert}},
  \bibinfo{journal}{J. Magn. Magn. Mat.}
  \textbf{\bibinfo{volume}{172}}, \bibinfo{pages}{199 } (\bibinfo{year}{1997}).

\bibitem[{\citenamefont{Nemec et~al.}(2013)\citenamefont{Nemec, Novak,
  Tesarova, Rozkotova, Reichlova, Butkovicova, Trojanek, Olejnik, Maly, Campion
  et~al.}}]{Nemec:2012_b}
\bibinfo{author}{\bibfnamefont{P.}~\bibnamefont{N\v emec}},
  \bibinfo{author}{\bibfnamefont{V.}~\bibnamefont{Nov\'ak}},
  \bibinfo{author}{\bibfnamefont{N.}~\bibnamefont{Tesa\v rov\'a}},
  \bibinfo{author}{\bibfnamefont{E.}~\bibnamefont{Rozkotov\'a}},
  \bibinfo{author}{\bibfnamefont{H.}~\bibnamefont{Reichlov\'a}},
  \bibinfo{author}{\bibfnamefont{D.}~\bibnamefont{Butkovi\v cov\'a}},
  \bibinfo{author}{\bibfnamefont{F.}~\bibnamefont{Troj\'anek}},
  \bibinfo{author}{\bibfnamefont{K.}~\bibnamefont{Olejn\'\i{}k}},
  \bibinfo{author}{\bibfnamefont{P.}~\bibnamefont{Mal\'y}},
  \bibinfo{author}{\bibfnamefont{R.~P.} \bibnamefont{Campion}},
  \bibnamefont{et~al.}, \bibinfo{journal}{Nat. Commun.}
  \textbf{\bibinfo{volume}{4}}, \bibinfo{pages}{1422} (\bibinfo{year}{2013}).

\bibitem[{\citenamefont{Zemen et~al.}(2009)\citenamefont{Zemen, Kucera,
  Olejnik, and Jungwirth}}]{Zemen:2009_a}
\bibinfo{author}{\bibfnamefont{J.}~\bibnamefont{Zemen}},
  \bibinfo{author}{\bibfnamefont{J.}~\bibnamefont{Kucera}},
  \bibinfo{author}{\bibfnamefont{K.}~\bibnamefont{Olejnik}}, \bibnamefont{and}
  \bibinfo{author}{\bibfnamefont{T.}~\bibnamefont{Jungwirth}},
  \bibinfo{journal}{Phys. Rev.} \textbf{\bibinfo{volume}{B 80}}.

\bibitem[{\citenamefont{Kim et~al.}(2011)\citenamefont{Kim, Kurz, Acbas, Ellis,
  and Cerne}}]{Kim:2011_a}
\bibinfo{author}{\bibfnamefont{M.-H.} \bibnamefont{Kim}},
  \bibinfo{author}{\bibfnamefont{V.}~\bibnamefont{Kurz}},
  \bibinfo{author}{\bibfnamefont{G.}~\bibnamefont{Acbas}},
  \bibinfo{author}{\bibfnamefont{C.~T.} \bibnamefont{Ellis}}, \bibnamefont{and}
  \bibinfo{author}{\bibfnamefont{J.}~\bibnamefont{Cerne}}, \bibinfo{journal}{J.
  Opt. Soc. Am. B} \textbf{\bibinfo{volume}{28}}, \bibinfo{pages}{199}
  (\bibinfo{year}{2011}).

\bibitem[{\citenamefont{Sato}(1981)}]{Sato:1981_a}
\bibinfo{author}{\bibfnamefont{K.}~\bibnamefont{Sato}}, \bibinfo{journal}{Jpn.
  J. Appl. Phys.} \textbf{\bibinfo{volume}{20}}, \bibinfo{pages}{2403}
  (\bibinfo{year}{1981}).

\bibitem[{\citenamefont{Tesarova et~al.}(2012)\citenamefont{Tesarova, Nemec,
  Rozkotova, Subrt, Reichlova, Butkovicova, Trojanek, Maly, Novak, and
  Jungwirth}}]{Tesarova:2012_a}
\bibinfo{author}{\bibfnamefont{N.}~\bibnamefont{Tesa\v rov\'a}},
  \bibinfo{author}{\bibfnamefont{P.}~\bibnamefont{N\v emec}},
  \bibinfo{author}{\bibfnamefont{E.}~\bibnamefont{Rozkotov\'a}},
  \bibinfo{author}{\bibfnamefont{J.}~\bibnamefont{\v Subrt}},
  \bibinfo{author}{\bibfnamefont{H.}~\bibnamefont{Reichlov\'a}},
  \bibinfo{author}{\bibfnamefont{D.}~\bibnamefont{Butkovi\v cov\'a}},
  \bibinfo{author}{\bibfnamefont{F.}~\bibnamefont{Troj\'anek}},
  \bibinfo{author}{\bibfnamefont{P.}~\bibnamefont{Mal\'y}},
  \bibinfo{author}{\bibfnamefont{V.}~\bibnamefont{Nov\'ak}}, \bibnamefont{and}
  \bibinfo{author}{\bibfnamefont{T.}~\bibnamefont{Jungwirth}},
  \bibinfo{journal}{Appl. Phys. Lett.} \textbf{\bibinfo{volume}{100}},
  \bibinfo{pages}{102403} (\bibinfo{year}{2012}).

\bibitem[{\citenamefont{Hamrle et~al.}(2007)\citenamefont{Hamrle, Blomeier,
  Gaier, Hillebrands, Schneider, Jakob, Postava, and Felser}}]{Hamrle:2007_a}
\bibinfo{author}{\bibfnamefont{J.}~\bibnamefont{Hamrle}},
  \bibinfo{author}{\bibfnamefont{S.}~\bibnamefont{Blomeier}},
  \bibinfo{author}{\bibfnamefont{O.}~\bibnamefont{Gaier}},
  \bibinfo{author}{\bibfnamefont{B.}~\bibnamefont{Hillebrands}},
  \bibinfo{author}{\bibfnamefont{H.}~\bibnamefont{Schneider}},
  \bibinfo{author}{\bibfnamefont{G.}~\bibnamefont{Jakob}},
  \bibinfo{author}{\bibfnamefont{K.}~\bibnamefont{Postava}}, \bibnamefont{and}
  \bibinfo{author}{\bibfnamefont{C.}~\bibnamefont{Felser}},
  \bibinfo{journal}{J. Phys. D: Appl. Phys.}
  \textbf{\bibinfo{volume}{40}}, \bibinfo{pages}{1563} (\bibinfo{year}{2007}).

\bibitem[{\citenamefont{Oh et~al.}(1991)\citenamefont{Oh, Bartholomew, Ramdas,
  Furdyna, and Debska}}]{Oh:1991_a}
\bibinfo{author}{\bibfnamefont{E.}~\bibnamefont{Oh}},
  \bibinfo{author}{\bibfnamefont{D.~U.} \bibnamefont{Bartholomew}},
  \bibinfo{author}{\bibfnamefont{A.~K.} \bibnamefont{Ramdas}},
  \bibinfo{author}{\bibfnamefont{J.~K.} \bibnamefont{Furdyna}},
  \bibnamefont{and} \bibinfo{author}{\bibfnamefont{U.}~\bibnamefont{Debska}},
  \bibinfo{journal}{Phys. Rev. B} \textbf{\bibinfo{volume}{44}},
  \bibinfo{pages}{10551} (\bibinfo{year}{1991}).

\bibitem[{\citenamefont{Jancu et~al.}(1998)\citenamefont{Jancu, Scholz,
  Beltram, and Bassani}}]{Jancu:1998_a}
\bibinfo{author}{\bibfnamefont{J.-M.} \bibnamefont{Jancu}},
  \bibinfo{author}{\bibfnamefont{R.}~\bibnamefont{Scholz}},
  \bibinfo{author}{\bibfnamefont{F.}~\bibnamefont{Beltram}}, \bibnamefont{and}
  \bibinfo{author}{\bibfnamefont{F.}~\bibnamefont{Bassani}},
  \bibinfo{journal}{Phys. Rev. B} \textbf{\bibinfo{volume}{57}},
  \bibinfo{pages}{6493} (\bibinfo{year}{1998}).

\bibitem[{\citenamefont{Bebb}(1969)}]{Bebb:1969_a}
\bibinfo{author}{\bibfnamefont{H.~B.} \bibnamefont{Bebb}},
  \bibinfo{journal}{Phys. Rev.} \textbf{\bibinfo{volume}{185}},
  \bibinfo{pages}{1116} (\bibinfo{year}{1969}).


\bibitem[{\citenamefont{Chapler et~al.}(2013)\citenamefont{Chapler, Mack,
  Myers, Frenzel, Pursley, Burch, Dattelbaum, Samarth, Awschalom, and
  Basov}}]{Chapler:2013_a}
\bibinfo{author}{\bibfnamefont{B.~C.} \bibnamefont{Chapler}},
  \bibinfo{author}{\bibfnamefont{S.}~\bibnamefont{Mack}},
  \bibinfo{author}{\bibfnamefont{R.~C.} \bibnamefont{Myers}},
  \bibinfo{author}{\bibfnamefont{A.}~\bibnamefont{Frenzel}},
  \bibinfo{author}{\bibfnamefont{B.~C.} \bibnamefont{Pursley}},
  \bibinfo{author}{\bibfnamefont{K.~S.} \bibnamefont{Burch}},
  \bibinfo{author}{\bibfnamefont{A.~M.} \bibnamefont{Dattelbaum}},
  \bibinfo{author}{\bibfnamefont{N.}~\bibnamefont{Samarth}},
  \bibinfo{author}{\bibfnamefont{D.~D.} \bibnamefont{Awschalom}},
  \bibnamefont{and} \bibinfo{author}{\bibfnamefont{D.~N.} \bibnamefont{Basov}},
  \bibinfo{journal}{Phys. Rev. B} \textbf{\bibinfo{volume}{87}},
  \bibinfo{pages}{205314} (\bibinfo{year}{2013}).

\bibitem[{\citenamefont{Dietl et~al.}(2000)\citenamefont{Dietl, Ohno,
  Matsukura, Cibert, and Ferrand}}]{Dietl:2000_a}
\bibinfo{author}{\bibfnamefont{T.}~\bibnamefont{Dietl}},
  \bibinfo{author}{\bibfnamefont{H.}~\bibnamefont{Ohno}},
  \bibinfo{author}{\bibfnamefont{F.}~\bibnamefont{Matsukura}},
  \bibinfo{author}{\bibfnamefont{J.}~\bibnamefont{Cibert}}, \bibnamefont{and}
  \bibinfo{author}{\bibfnamefont{D.}~\bibnamefont{Ferrand}},
  \bibinfo{journal}{Science} \textbf{\bibinfo{volume}{287}},
  \bibinfo{pages}{1019} (\bibinfo{year}{2000}).


\bibitem[{not({\natexlab{c}})}]{note5}
\bibinfo{note}{Appendix~D discusses the smallness of $\ve_{yz}^2/\ve_{zz}$ from
  theoretical point of view. Alternatively, data in
  Ref.~\onlinecite{Acbas:2009_a} (Fig.~1a, $\theta_F\approx 10^{3}$~rad/cm)
  imply using Eq.~(13) of Ref.~\onlinecite{Kim:2007_a} $\sigma_{xy}\approx
  14\unit{(\Omega cm)^{-1}}$. This translates to $\ve_{xy}=0.05$ (relative) at
  $\hbar\omega=2$~eV, hence $\ve_{xy}^2/\ve_{xx}\approx 2\times 10^{-4}$ (for
  $\ve_{xx}\approx 13$). This is small compared to $\ve_{xx}-\ve_{zz}=0.02$
  (see Appendix~D).}

\bibitem[{\citenamefont{Lautenschlager
  et~al.}(1987)\citenamefont{Lautenschlager, Garriga, Logothetidis, and
  Cardona}}]{Lautenschlager:1987_a}
\bibinfo{author}{\bibfnamefont{P.}~\bibnamefont{Lautenschlager}},
  \bibinfo{author}{\bibfnamefont{M.}~\bibnamefont{Garriga}},
  \bibinfo{author}{\bibfnamefont{S.}~\bibnamefont{Logothetidis}},
  \bibnamefont{and} \bibinfo{author}{\bibfnamefont{M.}~\bibnamefont{Cardona}},
  \bibinfo{journal}{Phys. Rev. B} \textbf{\bibinfo{volume}{35}},
  \bibinfo{pages}{9174} (\bibinfo{year}{1987}).

\bibitem[{\citenamefont{Johnson et~al.}(1969)\citenamefont{Johnson, Sherman,
  and Weil}}]{Johnson:1969_a}
\bibinfo{author}{\bibfnamefont{C.~J.} \bibnamefont{Johnson}},
  \bibinfo{author}{\bibfnamefont{G.~H.} \bibnamefont{Sherman}},
  \bibnamefont{and} \bibinfo{author}{\bibfnamefont{R.}~\bibnamefont{Weil}},
  \bibinfo{journal}{Appl. Opt.} \textbf{\bibinfo{volume}{8}},
  \bibinfo{pages}{1667} (\bibinfo{year}{1969}).


\bibitem[{\citenamefont{da~Silva et~al.}(2004)\citenamefont{da~Silva,
  Campomanes, Leite, Orapunt, and O'Leary}}]{daSilva:2004_a}
\bibinfo{author}{\bibfnamefont{J.~H.~D.} \bibnamefont{da~Silva}},
  \bibinfo{author}{\bibfnamefont{R.~R.} \bibnamefont{Campomanes}},
  \bibinfo{author}{\bibfnamefont{D.~M.~G.} \bibnamefont{Leite}},
  \bibinfo{author}{\bibfnamefont{F.}~\bibnamefont{Orapunt}}, \bibnamefont{and}
  \bibinfo{author}{\bibfnamefont{S.~K.} \bibnamefont{O'Leary}},
  \bibinfo{journal}{J. Appl. Phys.} \textbf{\bibinfo{volume}{96}},
  \bibinfo{pages}{7052} (\bibinfo{year}{2004}).


\bibitem[{\citenamefont{Tauc et~al.}(1966)\citenamefont{Tauc, Grigorovici, and
  Vancu}}]{Tauc:1966_a}
\bibinfo{author}{\bibfnamefont{J.}~\bibnamefont{Tauc}},
  \bibinfo{author}{\bibfnamefont{R.}~\bibnamefont{Grigorovici}},
  \bibnamefont{and} \bibinfo{author}{\bibfnamefont{A.}~\bibnamefont{Vancu}},
  \bibinfo{journal}{physica status solidi (b)} \textbf{\bibinfo{volume}{15}},
  \bibinfo{pages}{627} (\bibinfo{year}{1966}). 


\bibitem[{not({\natexlab{d}})}]{note7}
\bibinfo{note}{Comparing (Ga,Mn)As to amorphous GaAs, it should be mentioned
  that chemical rather than structural disorder has also other ramifications
  than the inapplicability of Bloch theorem. The hybridization between $p$-like
  orbitals of GaAs with $d$-like Mn orbitals of course does change the orbital
  character of the valence band to some extent, so that the matrix elements in
  Eq.~(\ref{eq-19}) will be modified. In particular, the parameter
  $E_P=2m_0P^2/\hbar^2$ mentioned above Eq.~(\ref{eq-17}) will bear witness to
  the modified orbitals through $P=\hbar/m_0\langle S|\hat{p}_z|X\rangle$ and
  optical transition probabilities will naturally also be modified. This effect
  is "hidden" behind the Schrieffer-Wolff transformation included in the model
  of Ref.~\onlinecite{Abolfath:2001_a}. Consequently, heights of the peaks in
  $\theta(\omega)$ will change.}

\bibitem[{\citenamefont{Jungwirth et~al.}(2007)\citenamefont{Jungwirth, Sinova,
  MacDonald, Gallagher, Nov\'ak, Edmonds, Rushforth, Campion, Foxon, Eaves
  et~al.}}]{Jungwirth:2007_a}
\bibinfo{author}{\bibfnamefont{T.}~\bibnamefont{Jungwirth}},
  \bibinfo{author}{\bibfnamefont{J.}~\bibnamefont{Sinova}},
  \bibinfo{author}{\bibfnamefont{A.~H.} \bibnamefont{MacDonald}},
  \bibinfo{author}{\bibfnamefont{B.~L.} \bibnamefont{Gallagher}},
  \bibinfo{author}{\bibfnamefont{V.}~\bibnamefont{Nov\'ak}},
  \bibinfo{author}{\bibfnamefont{K.~W.} \bibnamefont{Edmonds}},
  \bibinfo{author}{\bibfnamefont{A.~W.} \bibnamefont{Rushforth}},
  \bibinfo{author}{\bibfnamefont{R.~P.} \bibnamefont{Campion}},
  \bibinfo{author}{\bibfnamefont{C.~T.} \bibnamefont{Foxon}},
  \bibinfo{author}{\bibfnamefont{L.}~\bibnamefont{Eaves}},
  \bibnamefont{et~al.}, \bibinfo{journal}{Phys. Rev. B}
  \textbf{\bibinfo{volume}{76}}, \bibinfo{pages}{125206}
  (\bibinfo{year}{2007}).

\bibitem[{\citenamefont{Moss}(1954)}]{Moss:1954_a}
\bibinfo{author}{\bibfnamefont{T.~S.} \bibnamefont{Moss}},
  \bibinfo{journal}{Proceedings of the Physical Society. Section B}
  \textbf{\bibinfo{volume}{67}}, \bibinfo{pages}{775} (\bibinfo{year}{1954}).

\bibitem[{not({\natexlab{e}})}]{note6}
\bibinfo{note}{Exchange energy as we describe it can lead to Bloch
  ferromagnetism at {\em low} electron densities as it is very clearly
  explained in Sec. II of Ref.~\onlinecite{Zhang:2005_c}). This transition
  results from the competition of the exchange energy with kinetic
  energy. At
  higher densities relevant for our system, the former does not exceed the
  latter so that the ferromagnetic state would not appear were it not for the
  additional interaction with manganese magnetic moments. However, since
  the exchange energy grows relative to $E_g$ with increasing carrier 
  density, it may be important to our considerations here.}

\bibitem{Zhang:2005_c} Y. Zhang and S. Das Sarma,
  Phys. Rev. B \textbf{72}, 115317 (2005).

\bibitem[{\citenamefont{Sipahi et~al.}(1996)\citenamefont{Sipahi, Enderlein,
  Scolfaro, and Leite}}]{Sipahi:1996_a}
\bibinfo{author}{\bibfnamefont{G.~M.} \bibnamefont{Sipahi}},
  \bibinfo{author}{\bibfnamefont{R.}~\bibnamefont{Enderlein}},
  \bibinfo{author}{\bibfnamefont{L.~M.~R.} \bibnamefont{Scolfaro}},
  \bibnamefont{and} \bibinfo{author}{\bibfnamefont{J.~R.} \bibnamefont{Leite}},
  \bibinfo{journal}{Phys. Rev. B} \textbf{\bibinfo{volume}{53}},
  \bibinfo{pages}{9930} (\bibinfo{year}{1996}).

\bibitem[{\citenamefont{Zhang and Das~Sarma}(2005)}]{Zhang:2005_a}
\bibinfo{author}{\bibfnamefont{Y.}~\bibnamefont{Zhang}} \bibnamefont{and}
  \bibinfo{author}{\bibfnamefont{S.}~\bibnamefont{Das~Sarma}},
  \bibinfo{journal}{Phys. Rev. B} \textbf{\bibinfo{volume}{72}},
  \bibinfo{pages}{125303} (\bibinfo{year}{2005}).

\bibitem[{\citenamefont{Yang et~al.}(2003)\citenamefont{Yang, Sinova,
  Jungwirth, Shim, and MacDonald}}]{Yang:2003_b}
\bibinfo{author}{\bibfnamefont{S.-R.~E.} \bibnamefont{Yang}},
  \bibinfo{author}{\bibfnamefont{J.}~\bibnamefont{Sinova}},
  \bibinfo{author}{\bibfnamefont{T.}~\bibnamefont{Jungwirth}},
  \bibinfo{author}{\bibfnamefont{Y.~P.} \bibnamefont{Shim}}, \bibnamefont{and}
  \bibinfo{author}{\bibfnamefont{A.~H.} \bibnamefont{MacDonald}},
  \bibinfo{journal}{Phys. Rev. B} \textbf{\bibinfo{volume}{67}},
  \bibinfo{pages}{045205} (\bibinfo{year}{2003}).

\bibitem[{\citenamefont{{Osgood III} et~al.}(1998)\citenamefont{{Osgood III},
  Bader, Clemens, White, and Matsuyama}}]{Osgood:1998_a}
\bibinfo{author}{\bibfnamefont{R.}~\bibnamefont{{Osgood III}}},
  \bibinfo{author}{\bibfnamefont{S.}~\bibnamefont{Bader}},
  \bibinfo{author}{\bibfnamefont{B.}~\bibnamefont{Clemens}},
  \bibinfo{author}{\bibfnamefont{R.}~\bibnamefont{White}}, \bibnamefont{and}
  \bibinfo{author}{\bibfnamefont{H.}~\bibnamefont{Matsuyama}},
  \bibinfo{journal}{J. Magn. Magn. Mat.}
  \textbf{\bibinfo{volume}{182}}, \bibinfo{pages}{297 } (\bibinfo{year}{1998}).


\bibitem[{not({\natexlab{f}})}]{note3}
\bibinfo{note}{A derivation based on the analogy between $\vec{M}$ and
  $\vec{B}$ is given in Ref.~\onlinecite{Osgood:1998_a}.}

\bibitem[{not({\natexlab{g}})}]{note4}
\bibinfo{note}{Onsager relations imply $\ve_{ij}(\vec{M})=\ve_{ji}(-\vec{M})$.
  For $\vec{M}||\hat{x}$, $\ve_{yz}(\vec{M})=-\ve_{yz}(-\vec{M})$ follows from
  applying mirror inversion with respect to the $xz$ plane to
  $J_y=\ve_{yz}(\vec{M}) E_z$.}

\bibitem[{\citenamefont{Birss}(1964)}]{Birss:1964}
\bibinfo{author}{\bibfnamefont{R.~R.} \bibnamefont{Birss}},
  \emph{\bibinfo{title}{Symmetry and magnetism}}
  (\bibinfo{publisher}{North-Holland Publishing Company},
  \bibinfo{address}{Amsterdam}, \bibinfo{year}{1964}).

\bibitem[{\citenamefont{Vi\v{s}\v{n}ovsk\'y}(1986)}]{Visnovsky:1986_a}
\bibinfo{author}{\bibfnamefont{{\v{S}}.}~\bibnamefont{Vi\v{s}\v{n}ovsk\'y}},
  \bibinfo{journal}{Czech. J. Phys. B} \textbf{\bibinfo{volume}{36}},
  \bibinfo{pages}{1424} (\bibinfo{year}{1986}).

\bibitem[{\citenamefont{Postava et~al.}(2002)\citenamefont{Postava, Hrabovsky,
  Pistora, Fert, Visnovsky, and Yamaguchi}}]{Postava:2002_a}
\bibinfo{author}{\bibfnamefont{K.}~\bibnamefont{Postava}},
  \bibinfo{author}{\bibfnamefont{D.}~\bibnamefont{Hrabovsky}},
  \bibinfo{author}{\bibfnamefont{J.}~\bibnamefont{Pistora}},
  \bibinfo{author}{\bibfnamefont{A.~R.} \bibnamefont{Fert}},
  \bibinfo{author}{\bibfnamefont{S.}~\bibnamefont{Visnovsky}},
  \bibnamefont{and}
  \bibinfo{author}{\bibfnamefont{T.}~\bibnamefont{Yamaguchi}},
  \bibinfo{journal}{J. Appl. Phys.} \textbf{\bibinfo{volume}{91}},
  \bibinfo{pages}{7293} (\bibinfo{year}{2002}).

\bibitem[{\citenamefont{Bhagavantam}(1966)}]{Bhagavantam:1966}
\bibinfo{author}{\bibfnamefont{S.}~\bibnamefont{Bhagavantam}},
  \emph{\bibinfo{title}{Crystal symmetry and physical properties}}
  (\bibinfo{publisher}{Academic Press}, \bibinfo{address}{London and New York},
  \bibinfo{year}{1966}).

\bibitem[{\citenamefont{Hamrlov{\'a} et~al.}(2013)\citenamefont{Hamrlov{\'a},
  Hamrle, Postava, and Pi\v{s}tora}}]{Hamrlova:2013_a}
\bibinfo{author}{\bibfnamefont{J.}~\bibnamefont{Hamrlov{\'a}}},
  \bibinfo{author}{\bibfnamefont{J.}~\bibnamefont{Hamrle}},
  \bibinfo{author}{\bibfnamefont{K.}~\bibnamefont{Postava}}, \bibnamefont{and}
  \bibinfo{author}{\bibfnamefont{J.}~\bibnamefont{Pi\v{s}tora}},
  \bibinfo{journal}{phys. stat. sol. (b)}, published on-line 
  (\bibinfo{year}{2013}), \bibinfo{note}{doi:
  10.1002/pssb.201349031}.

\bibitem[{\citenamefont{{de {Ranieri}} et~al.}(2008)\citenamefont{{de
  {Ranieri}}, Rushforth, {V\'{y}born\'{y}}, Rana, Ahmad, Campion, Foxon,
  Gallagher, Irvine, Wunderlich et~al.}}]{Ranieri:2008_a}
\bibinfo{author}{\bibfnamefont{E.}~\bibnamefont{{de {Ranieri}}}},
  \bibinfo{author}{\bibfnamefont{A.~W.} \bibnamefont{Rushforth}},
  \bibinfo{author}{\bibfnamefont{K.}~\bibnamefont{{V\'{y}born\'{y}}}},
  \bibinfo{author}{\bibfnamefont{U.}~\bibnamefont{Rana}},
  \bibinfo{author}{\bibfnamefont{E.}~\bibnamefont{Ahmad}},
  \bibinfo{author}{\bibfnamefont{R.~P.} \bibnamefont{Campion}},
  \bibinfo{author}{\bibfnamefont{C.~T.} \bibnamefont{Foxon}},
  \bibinfo{author}{\bibfnamefont{B.~L.} \bibnamefont{Gallagher}},
  \bibinfo{author}{\bibfnamefont{A.~C.} \bibnamefont{Irvine}},
  \bibinfo{author}{\bibfnamefont{J.}~\bibnamefont{Wunderlich}},
  \bibnamefont{et~al.}, \bibinfo{journal}{New J. Phys.}
  \textbf{\bibinfo{volume}{10}}, \bibinfo{pages}{065003}
  (\bibinfo{year}{2008}).

\bibitem[{\citenamefont{Piano et~al.}(2011)\citenamefont{Piano, Grein, Mellor,
  V\'yborn\'y, Campion, Wang, Eschrig, and Gallagher}}]{Piano:2010_a}
\bibinfo{author}{\bibfnamefont{S.}~\bibnamefont{Piano}},
  \bibinfo{author}{\bibfnamefont{R.}~\bibnamefont{Grein}},
  \bibinfo{author}{\bibfnamefont{C.~J.} \bibnamefont{Mellor}},
  \bibinfo{author}{\bibfnamefont{K.}~\bibnamefont{V\'yborn\'y}},
  \bibinfo{author}{\bibfnamefont{R.}~\bibnamefont{Campion}},
  \bibinfo{author}{\bibfnamefont{M.}~\bibnamefont{Wang}},
  \bibinfo{author}{\bibfnamefont{M.}~\bibnamefont{Eschrig}}, \bibnamefont{and}
  \bibinfo{author}{\bibfnamefont{B.~L.} \bibnamefont{Gallagher}},
  \bibinfo{journal}{Phys. Rev. B} \textbf{\bibinfo{volume}{83}},
  \bibinfo{pages}{081305} (\bibinfo{year}{2011}).

\bibitem[{\citenamefont{Jungwirth
  et~al.}(2006{\natexlab{b}})\citenamefont{Jungwirth, Ma\ifmmode~\check{s}\else
  \v{s}\fi{}ek, Wang, Edmonds, Sawicki, Polini, Sinova, MacDonald, Campion,
  Zhao et~al.}}]{Jungwirth:2005_a}
\bibinfo{author}{\bibfnamefont{T.}~\bibnamefont{Jungwirth}},
  \bibinfo{author}{\bibfnamefont{J.}~\bibnamefont{Ma\ifmmode~\check{s}\else
  \v{s}\fi{}ek}}, \bibinfo{author}{\bibfnamefont{K.~Y.} \bibnamefont{Wang}},
  \bibinfo{author}{\bibfnamefont{K.~W.} \bibnamefont{Edmonds}},
  \bibinfo{author}{\bibfnamefont{M.}~\bibnamefont{Sawicki}},
  \bibinfo{author}{\bibfnamefont{M.}~\bibnamefont{Polini}},
  \bibinfo{author}{\bibfnamefont{J.}~\bibnamefont{Sinova}},
  \bibinfo{author}{\bibfnamefont{A.~H.} \bibnamefont{MacDonald}},
  \bibinfo{author}{\bibfnamefont{R.~P.} \bibnamefont{Campion}},
  \bibinfo{author}{\bibfnamefont{L.~X.} \bibnamefont{Zhao}},
  \bibnamefont{et~al.}, \bibinfo{journal}{Phys. Rev. B}
  \textbf{\bibinfo{volume}{73}}, \bibinfo{pages}{165205}
  (\bibinfo{year}{2006}{\natexlab{b}}).


\bibitem[{\citenamefont{Hankiewicz et~al.}(2004)\citenamefont{Hankiewicz,
  Jungwirth, Dietl, Timm, and Sinova}}]{Hankiewicz:2004_a}
\bibinfo{author}{\bibfnamefont{E.~M.} \bibnamefont{Hankiewicz}},
  \bibinfo{author}{\bibfnamefont{T.}~\bibnamefont{Jungwirth}},
  \bibinfo{author}{\bibfnamefont{T.}~\bibnamefont{Dietl}},
  \bibinfo{author}{\bibfnamefont{C.}~\bibnamefont{Timm}}, \bibnamefont{and}
  \bibinfo{author}{\bibfnamefont{J.}~\bibnamefont{Sinova}},
  \bibinfo{journal}{Phys. Rev. B} \textbf{\bibinfo{volume}{70}},
  \bibinfo{pages}{245211} (\bibinfo{year}{2004}).


\bibitem[{\citenamefont{Koloren\ifmmode~\check{c}\else \v{c}\fi{}
  et~al.}(2002)\citenamefont{Koloren\ifmmode~\check{c}\else \v{c}\fi{},
  Smr\ifmmode~\check{c}\else \v{c}\fi{}ka, and St\ifmmode~\check{r}\else
  \v{r}\fi{}eda}}]{Kolorenc:2002_a}
\bibinfo{author}{\bibfnamefont{J.}~\bibnamefont{Koloren\ifmmode~\check{c}\else
  \v{c}\fi{}}},
  \bibinfo{author}{\bibfnamefont{L.}~\bibnamefont{Smr\ifmmode~\check{c}\else
  \v{c}\fi{}ka}}, \bibnamefont{and}
  \bibinfo{author}{\bibfnamefont{P.}~\bibnamefont{St\ifmmode~\check{r}\else
  \v{r}\fi{}eda}}, \bibinfo{journal}{Phys. Rev. B}
  \textbf{\bibinfo{volume}{66}}, \bibinfo{pages}{085301}
  (\bibinfo{year}{2002}).


\bibitem[{\citenamefont{Allen}(2006)}]{Allen:2006_a}
\bibinfo{author}{\bibfnamefont{P.}~\bibnamefont{Allen}}, in
  \emph{\bibinfo{booktitle}{Conceptual Foundations of Materials A Standard
  Model for Ground- and Excited-State Properties}}, edited by
  \bibinfo{editor}{\bibfnamefont{S.~G.} \bibnamefont{Louie}} \bibnamefont{and}
  \bibinfo{editor}{\bibfnamefont{M.~L.} \bibnamefont{Cohen}}
  (\bibinfo{publisher}{Elsevier}, \bibinfo{year}{2006}),
  vol.~\bibinfo{volume}{2} of \emph{\bibinfo{series}{Contemporary Concepts of
  Condensed Matter Science}}, pp. \bibinfo{pages}{165 -- 218}.


\bibitem[{not({\natexlab{h}})}]{note8}
\bibinfo{note}{The standard way to microscopically link $\Gamma$ to potential
  disorder is to include potential of randomly positioned impurities into
  $\hat{H}$ and average over the positions. This procedure is explained in Sec.
  8.1.2 of G. Mahan, Many-Particle physics (Kluwer Academic/Plenum Publishers,
  New York, 2000, third edition). Alternatively, $\Gamma/\hbar$ can also be
  introduced as inverse relaxation time when the electron is weakly interacting
  with its surroundings.\cite{Lax:1958_a}}

\bibitem[{\citenamefont{Lax}(1958)}]{Lax:1958_a}
\bibinfo{author}{\bibfnamefont{M.}~\bibnamefont{Lax}}, \bibinfo{journal}{Phys.
  Rev.} \textbf{\bibinfo{volume}{109}}, \bibinfo{pages}{1921}
  (\bibinfo{year}{1958}).

\bibitem[{\citenamefont{Al-Qadi et~al.}(2012)\citenamefont{Al-Qadi, Nishizawa,
  Nishibayashi, Kaneko, and Munekata}}]{Qadi:2012_a}
\bibinfo{author}{\bibfnamefont{B.}~\bibnamefont{Al-Qadi}},
  \bibinfo{author}{\bibfnamefont{N.}~\bibnamefont{Nishizawa}},
  \bibinfo{author}{\bibfnamefont{K.}~\bibnamefont{Nishibayashi}},
  \bibinfo{author}{\bibfnamefont{M.}~\bibnamefont{Kaneko}}, \bibnamefont{and}
  \bibinfo{author}{\bibfnamefont{H.}~\bibnamefont{Munekata}},
  \bibinfo{journal}{Applied Physics Letters} \textbf{\bibinfo{volume}{100}},
  \bibinfo{eid}{222410} (pages~\bibinfo{numpages}{4}) (\bibinfo{year}{2012}).



\end{thebibliography}
%% %\bibliography{dms-assorted}
%%%%%%%%%%%%\bibliographystyle{apsrev}
%\bibliographystyle{plain}

\end{document}